\documentclass[reprint,
superscriptaddress,
amsmath,amssymb,
aps,
prx,
]{revtex4-2}
\usepackage{snapshot}
\usepackage{graphicx}
\usepackage{bm}
\usepackage[normalem]{ulem}
\usepackage{hyperref}
\usepackage[capitalize]{cleveref} 
\hypersetup{
    breaklinks=true,
    colorlinks=true,
    linkcolor=blue,
    urlcolor=blue,
    citecolor=blue
}

\usepackage{subfigure}
\usepackage{amsthm}
\usepackage{amsfonts}
\usepackage{braket}
\usepackage{comment}
\usepackage{enumitem}
\usepackage{xcolor}
\usepackage{tikz} 
\usepackage{soul}
\usepackage{booktabs}

\makeatletter

\def\l@subsubsection#1#2{}
\makeatother
\begin{document}

	\title{Universal Freezing Transitions of Dipole-Conserving Chains}
	
	\author{Jonathan Classen-Howes}
	\thanks{These authors contributed equally to this work}
	\email[\\Contacts:\\]{jonathan.classen-howes@physics.ox.ac.uk}
	\email{\\riccardo.senese@physics.ox.ac.uk}
	\affiliation{Rudolf Peierls  Centre  for  Theoretical  Physics, University of Oxford, Oxford  OX1  3PU,  United  Kingdom}
	\author{Riccardo Senese}
	\thanks{These authors contributed equally to this work}
	\email[\\Contacts:\\]{jonathan.classen-howes@physics.ox.ac.uk}
	\email{\\riccardo.senese@physics.ox.ac.uk}
	\affiliation{Rudolf Peierls  Centre  for  Theoretical  Physics, University of Oxford, Oxford  OX1  3PU,  United  Kingdom}
	\author{Abhishodh Prakash}
	\affiliation{Rudolf Peierls  Centre  for  Theoretical  Physics, University of Oxford, Oxford  OX1  3PU,  United  Kingdom}
    \affiliation{Harish-Chandra Research Institute, Prayagraj (Allahabad) 211019, India}
	
	\date{\today}
	
	\begin{abstract}
	   We demonstrate the existence of a universal phase diagram of quantum chains with range-$k$ interactions subject to the conservation of a total charge and its dipole moment. These systems exhibit “freezing” transitions between strongly and weakly Hilbert-space-fragmented phases as the charge filling $\nu$ is varied. We show that these continuous phase transitions occur at a critical charge filling of $\nu_c=(k-2)^{-1}$ ~\emph{independently} of the on-site Hilbert space dimension $d$. To this end, we analytically prove that, for any $d$, any state with $\nu<\nu_c$ hosts a finite density of sites belonging to “blockages”, which we define as subregions of the chain across which transport of charge and dipole moment cannot occur. Some blockages arise from sequences of frozen sites, i.e.~sites with an unchanging on-site charge, while others do not involve frozen sites at all. We prove that the presence of blockages implies strong fragmentation of typical symmetry sectors into Krylov subspaces, each of which forms an exponentially vanishing fraction of the total sector. By studying the distribution of blockages we analytically characterise how typical states are subdivided into dynamically disconnected local ``active bubbles'', and prove that typical eigenstates at these charge fillings exhibit area-law entanglement entropy, while there exist rare eigenstates featuring non-area-law scaling. We also numerically show that for $\nu>\nu_c$ and arbitrary $d$, typical symmetry sectors are weakly fragmented, with their dominant Krylov sectors constituted of states that are free of blockages. We analytically derive some critical exponents characterising the transition and numerically determine the density of blockages at $\nu=\nu_c$, with clear implications for transport at the critical point. Finally, we investigate the properties of special-case models for which no phase transitions occur. 
	\end{abstract}

	\maketitle
	
	\tableofcontents

	\section{\label{sec:introduction}Introduction and summary of main results}
	
	When an interacting quantum many-body system undergoes dynamical evolution it generically \emph{thermalises}, i.e.~the long-time expectation value of local observables is found to coincide with the predictions of standard thermodynamic ensembles. The condition for thermalisation is formalised by the so-called eigenstate thermalisation hypothesis (ETH)~\cite{PhysRevA.43.2046, PhysRevE.50.888, RigolReview, Deutsch_ETH_Review}, which implies that the expectation value of local observables on eigenstates can be reproduced by appropriate thermal ensembles.

	A long-standing quest in the study of quantum dynamics is to find generic settings robust to perturbations where ETH fails. One prominent example is many-body localisation~\cite{Anderson_PhysRev.109.1492,BaskoAleinerAltschuler_MBL_20061126,Nandkishore_MBL_ETH,Abanin_MBLReview_RevModPhys.91.021001}, which arises in systems with strong quenched disorder. An alternative, promising route following the recent discovery of fractons~\cite{Chamon_Fracton_PhysRevLett.94.040402,Haah_fracton2,VijayHaahFu_Fracton_PhysRevB.92.235136,Nandkishore_fracton1,PretkoChenYou_2020fracton,GromovRadzihovsky2022fractonReview} are systems that exhibit Hilbert space fragmentation (HSF)~\cite{PhysRevX.9.021003, Sala_HSF, Khemani_HSF, Moudgalya_HSF, PhysRevB.101.125126, doi:10.1142/9789811231711_0009} in the presence of multipole conservation \footnote{We note that while the concept of HSF has been originally introduced in the context of dipole-conserving quantum chains, several mechanisms for generating HSF that do not rely on dipole (or multipole) conservation have been investigated \cite{PhysRevB.101.125126, Moudgalya_HSF, PhysRevLett.124.207602, PhysRevB.100.214313, PhysRevB.102.224303, PhysRevB.103.235133, 10.21468SciPostPhys.11.4.074, PhysRevResearch.3.033201, PhysRevB.103.L220304, PhysRevB.104.155117, PhysRevLett.129.090602, PhysRevB.106.L220301, 10.21468SciPostPhys.15.3.093, PhysRevLett.130.120401, PhysRevLett.132.040401, Classen-Howes_RSSY, PhysRevLett.134.010411},  and experimentally realised \cite{adler2024observation, PhysRevLett.133.196301}.}. In fractonic models, kinetic constraints that emerge through conservation laws combined with locality inhibit thermalisation by preventing the model from exploring all the states permitted by its conservation laws. One widely studied class of fractonic models, which will be the focus of this work, is that of 1D chains conserving a charge $N$ and its dipole moment $X$. These effectively arise in the context of the quantum Hall effect \cite{Wang_QH1,BERGHOLTZ_QH2} and
	in systems of particles exposed to strong tilt potentials \cite{Morong_StarkMBLnoDisorder, Nieuwenburg_Efield2, scherg2021observing, PhysRevLett.130.010201, PhysRevLett.127.240502, wang2024exploring}. In these systems, the combined effect of the two conserved charges and the locality of interactions greatly inhibits mobility and generates HSF. In particular, isolated charges cannot propagate on their own because they require other nearby charges to interact with. Hence, states in the same $(N,X)$ quantum number sector can be dynamically disconnected.

	Fractonic systems open up new possibilities for a rigorous characterization of ergodicity-breaking transitions in quantum many-body systems given that the underlying mechanisms rely only on symmetries and locality, which are easier to work with compared with, say, quenched disorder. In particular, there have been attempts to understand how systems undergo a continuous ``freezing'' phase transition between ``strong" and ``weak" HSF \cite{Sala_HSF} as a charge density $\nu$ is varied~\cite{Morningstar_HSF,Pozderac_DinfHSF,Wang_EastModel,PhysRevB.106.214426}. In the strongly fragmented phase, typical symmetry sectors can be subdivided into an exponentially large (in system size) number of dynamically disconnected ``Krylov subspaces", all of which represent a vanishing fraction of the total sector. This results in a strong violation of ETH~\cite{Sala_HSF, Khemani_HSF}. On the other hand, in the weakly fragmented phase ETH is only weakly violated. This is because in this phase almost all states in a symmetry sector belong to a single Krylov subspace, which dominates in size over the remaining (still exponentially many) Krylov sectors. It has been established that in the presence of weak fragmentation dipole-conserving models host subdiffusive transport \cite{Morningstar_HSF, PhysRevResearch.2.033124, PhysRevLett.125.245303, PhysRevE.103.022142}.

    Analytically characterising the strong-to-weak transition within dipole-conserving chains poses significant challenges \cite{Morningstar_HSF,Pozderac_DinfHSF}. Simplification is possible by switching to different 1D models with simpler kinetic constraints \cite{Wang_EastModel}, which appear to belong to the same universality class as dipole-conserving ones. This raises the question of whether HSF and its associated transition are inherently more complex in the paradigmatic dipole-conserving setup. In this work we show that actually, when approached with the right tools, the situation in such cases becomes in many respects elementary. This strongly substantiates the claim that HSF in 1D lattices represents a minimal tractable mechanism for ergodicity-breaking, \emph{irrespective} of the specific realization. We note that the richer setting intrinsic to dipole conservation, compared with the simpler model in \cite{Wang_EastModel}, enables us to understand the role that the local Hilbert space dimension $d$ plays within HSF, uncover fundamentally distinct types of transport obstructions, conjecture the existence of an inverse quantum many-body scar phenomenon, and analytically determine critical exponents, hence establishing on firm grounds the dipole-conserving universality class. We also remark that, compared with other types of dynamical constraints, dipole conservation represents a very simple mechanism from the point of view of experimental realizations, given that it naturally emerges (on prethermal scales) through the bare application of sufficiently strong linear potentials \cite{scherg2021observing,PhysRevLett.130.010201}.

	Quantum chains with any on-site dimension $d$, range-$k$ interactions and charge-dipole symmetry can be analysed by reformulating them as problems of hopping particles with maximal on-site occupation $d-1$ that conserve both total particle number $N$ and centre of mass $X/N$. Exploiting this, Ref.~\cite{Pozderac_DinfHSF} analytically argued that for unbounded maximal on-site occupation ($d=\infty$) the critical density is located at $\nu_c=(k-2)^{-1}$, where $k$ is the range of interactions. In this work we make considerable progress on characterising the transition in dipole-conserving systems with any on-site dimension $d$. Our main contributions can be briefly summarised as follows.

    \begin{figure}[t]
		\centering
		\includegraphics[width=0.9\linewidth]{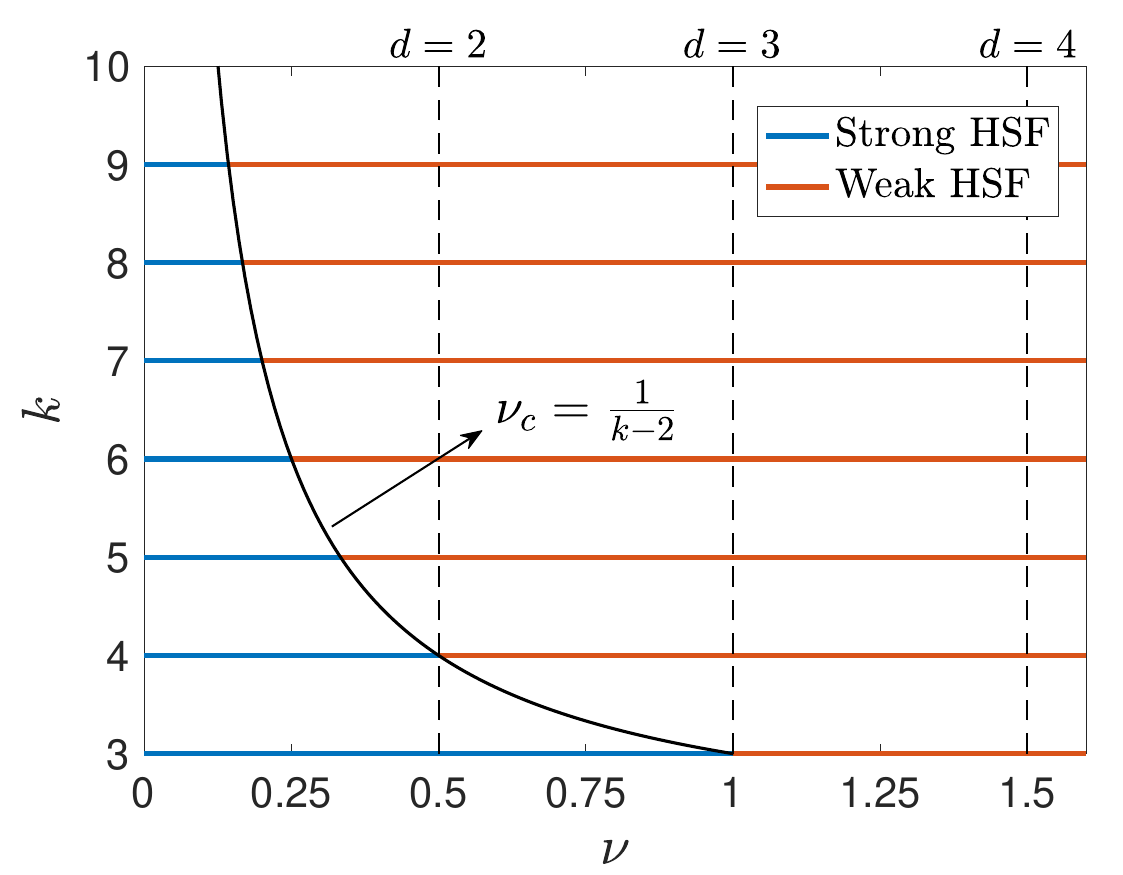}
		\caption{Conjectured universal phase diagram of charge- and dipole-conserving quantum chains with range-$k$ interactions, which feature Hilbert space fragmentation (HSF). The relevant variables are the continuous particle filling $\nu\ge0$ and the discrete range of interactions $k\ge 3$. A first strong-to-weak fragmentation transition occurs at $\nu_c=1/(k-2)$, a value independent of the on-site Hilbert space dimension $d$. Note that for each value of $d$ we depict the corresponding phase diagram only below the half filling $(d-1)/2$ (vertical dashed lines). Beyond half filling the phase diagram is mirrored by particle-hole symmetry, leading to a second opposite transition (not depicted here) at $d-1-\nu_c$. Also note that for $d=2$ it is only meaningful to consider $k\ge4$, as there can be no dynamical evolution at $k=3$ which respects the global conservation laws.}
		\label{fig:PhaseDiagram}
	\end{figure}

	\emph{i.} \ We show that the critical density $\nu_c=(k-2)^{-1}$ of Ref.~\cite{Pozderac_DinfHSF} is, in fact, “universal”: for any value of the on-site Hilbert space dimension $d$, there is a strong-to-weak transition at $\nu=\nu_c$, with a second weak-to-strong transition at $\nu=d-1-\nu_c$ by particle-hole symmetry. This leads to the universal $d$-independent phase diagram of \cref{fig:PhaseDiagram}. This result emerges from the findings presented in \cref{sec:FES,sec:strongfrag,sec:weakFragNumerics}, where we develop several new concepts and approaches (summarised in the next few points) for characterising strongly and weakly fragmented phases and the transition between them.
	
	\emph{ii.} \ In \cref{sec:FES} we introduce the concept of “blockages” to characterise the strongly fragmented phase, defining them as subregions of the system across which transport of particle number and dipole moment cannot occur. These constitute a generalisation of the ``bottlenecks'' introduced in Ref. \cite{Khemani_HSF}. One kind of blockage involves only “frozen” sites that have an unchanging particle number under dynamics. Remarkably, we also demonstrate the existence of a second kind of blockage that does not involve frozen sites at all.
    
	\emph{iii.} \ To identify and locate blockages we employ the concept of a ``fully extended state'' (FES), originally introduced in Ref.~\cite{Pozderac_DinfHSF} and whose scope we expand substantially. Intuitively, an FES is a configuration of particles obtained by starting from any state of the system and allowing the particles to maximally spread out, ignoring any local occupancy constraint introduced by $d$. In \cref{sec:FES} we prove for general interaction range $k$ that there exists a unique FES within each Krylov sector of a $d=\infty$ system, and furthermore that the \emph{local} structure of an FES can be used to determine the locations of certain types of blockages at any $d$, in spite of blockages reflecting \emph{global} properties of the system. We refer to this approach for identifying blockages as the ``FES picture''.
	
	\emph{iv.} \ Using the FES picture, we prove in \cref{sec:strongfrag} that for particle fillings $\nu<\nu_c$ and for \emph{any} $d$, any configuration of particles features a finite density of sites belonging to blockages, and that this implies that typical symmetry sectors at these fillings are strongly fragmented. In \cref{sec:StrongFragFurtherAnalysis} we use the FES picture to analytically lower bound the density of blockages in typical states, as well as the density of ``active bubbles'': local groups of interacting particles shielded from their surroundings by blockages. Additionally, we prove in \cref{sec:EEstrongfrag} that due to the presence of blockages, typical eigenstates in strongly fragmented symmetry sectors feature area-law entanglement entropy scaling, although there exist rare exceptions which may constitute inverse quantum many-body scars \cite{Srivatsa_IQMBS1,Chen_IQMBS2,Srivatsa_IQMBS3,Iversen_IQMBS4}.
	
	\emph{v.} \ To characterise the weakly fragmented phase, in \cref{sec:weakFragNumerics} we numerically show that for $\nu>\nu_c$ and arbitrary $d$, typical symmetry sectors feature a dominant Krylov sector which is blockage-free. We identify the dominant Krylov sector as the one containing a ``blockage-free extended state'', a kind of state similar in structure to an FES and which is unique to each symmetry sector. We develop an algorithm that maps arbitrary initial states to their corresponding blockage-free extended state with a high success rate for $\nu>\nu_c$, hence yielding strong numerical evidence for weak fragmentation.

    \emph{vi.} \ To further characterise the transition, in \cref{sec:critscalandexp} we analytically derive critical exponents for the density of blockages and average size of active bubbles, and find them to be in agreement with previous related results for dipole-conserving chains \cite{Morningstar_HSF, Pozderac_DinfHSF} and for other models in the same universality class \cite{Wang_EastModel}. We also numerically determine the scaling with system size of the density of blockages at $\nu=\nu_c$, and discuss the consequences it has on transport at the critical filling, which is at present poorly understood \cite{Morningstar_HSF, Pozderac_DinfHSF}.
	
	We note that, unlike previous analytical results on this phase transition \cite{Pozderac_DinfHSF}, our FES picture can be used to analytically characterise not just individual symmetry sectors, but whole families of typical symmetry sectors at arbitrary on-site dimensions $d$, while also providing several analytical insights into the dynamical impacts of strong fragmentation. Furthermore, the numerical algorithms we develop are not only efficient at large system sizes but also exactly simulate the dynamics of the systems, unlike the approximate numerical methods developed in Ref. \cite{Morningstar_HSF}.

    \begin{figure}[t]
\centering
\includegraphics[scale=0.28]{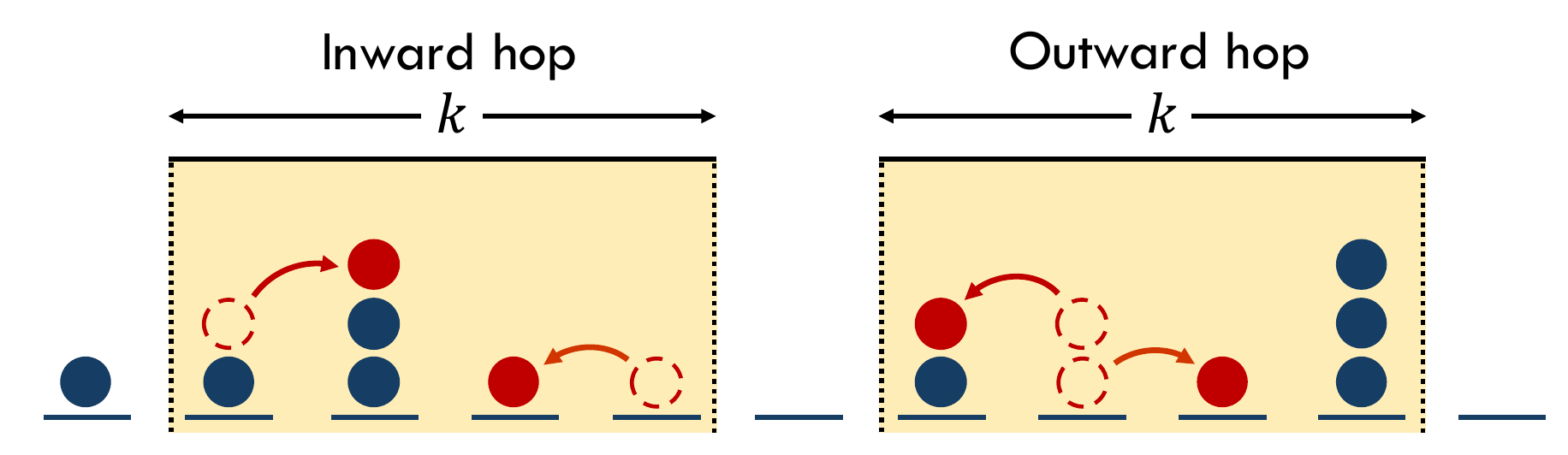}
\caption{Examples of inward hop and outward hop range-$k$ gates ($k=4$) that conserve $N$ and $X$, acting on a local region of a chain with $d=4$.}
\label{fig:HopsFig}	
\end{figure}

    The remainder of the paper is organised as follows. In Section \ref{sec:familyofmodels} we introduce our family of models and review various probes that can be used to distinguish between the strongly and weakly fragmented phases.  In Section \ref{sec:FES} we introduce the two distinct type of blockages and the FES picture, and derive several key results regarding these.  
    In \cref{sec:strongfrag} we define “typical” symmetry sectors and make use of the FES picture to prove that, for any on-site dimension $d$, they are strongly fragmented at fillings $\nu<\nu_c$. In Section \ref{sec:weakFragNumerics} we present numerical evidence at large system sizes that typical symmetry sectors are weakly fragmented for $\nu>\nu_c$ and arbitrary $d$.
    We make further use of the FES picture in Section \ref{sec:StrongFragFurtherAnalysis} to analyse various properties of the strongly fragmented phase, such as the average density of frozen sites and of different types of active bubbles.  In \cref{sec:critscalandexp} we derive critical exponents that describe the transition and discuss transport at the critical point. We also present further evidence that $\nu_c$ is the critical filling by numerically probing the critical scaling of the dimension of dominant Krylov sectors.  In Section \ref{sec:EEstrongfrag}  we discuss the implications of the presence of blockages for the scaling of eigenstate entanglement entropies and for quantum dynamics. In Section \ref{sec:StrongFragModels} we study special cases of the parameters $d$ and $k$ for which the model does not possess a phase transition. We provide concluding remarks in Section \ref{sec:conclusion}. Several appendices follow discussing technical details of our derivations.\\

\section{\label{sec:familyofmodels}Models and order parameters}

\subsection{Charge- and dipole-conserving local models}
\label{subsec:chargeandiplocmod}
In this section we introduce the systems that are the focus of our study \cite{Sala_HSF,Khemani_HSF, Morningstar_HSF}. 
While our main objective is the characterization of one-dimensional dipole-conserving \emph{quantum} systems, we will see that much progress can be made by considering states in a product state basis which can be given a classical interpretation \cite{Morningstar_HSF, Pozderac_DinfHSF, Wang_EastModel}.
For many purposes, it is hence sufficient to focus on an equivalent problem of classical particles hopping on one-dimensional lattices and subject to finite-range interactions that conserve a charge and its dipole moment. We consider a chain of $L$ sites with open boundary conditions (OBC), where each site can host from $0$ up to $d-1$ particles, with $2\le d \le \infty$. They dynamically evolve through sequences of hopping moves implemented via range-$k$ gates, as depicted in \cref{fig:HopsFig}. Letting $n_i$ indicate the number of particles on site $i$, every gate conserves the total number of particles $N$ and the dipole moment $X$ defined as follows
	\begin{equation}
		\label{eq:defNandX}
		N = \sum_{i=0}^{L-1} n_i \qquad \quad X = \sum_{i=0}^{L-1} i\, n_i\ .
	\end{equation}
	We define the filling $\nu$ and the intensive centre of mass $\nu_x$ as
	\begin{equation}
		\label{eq:nuandnux}
		\nu = \frac{N}{L} \qquad \quad \nu_x = \frac{X}{N \, L} \ ,
	\end{equation}
	where $X/N$ represents the centre of mass. To keep our discussion as general as possible, we require that the dynamics satisfies “gate-completeness”, i.e.~that the set of available gates (and their combinations) enables \emph{all} possible rearrangements of particles compatible with range-$k$ locality of interactions, maximal occupancy of $d-1$ and conservation of $N$ and $X$. This is a very natural requirement, because it implies that if two particles are within a distance $k$ from each other, they can perform any move compatible with the charge, dipole, and on-site dimension requirements such that the move  happens entirely within at most $k$ contiguous sites. Violations of gate-completeness, e.g.~due to additional symmetries, further reduce mobility. An example of this, for $d=4, k =4$, would be a chain in which the move on the left in \cref{fig:HopsFig} is available, but the move on the right is not allowed. We anticipate that, as a consequence of the additional hindrance to dynamics that a violation of gate-completeness entails, the main results for the strongly fragmented phase derived in this work remain unchanged. We discuss this point further in later sections.
    In the following, we will be mainly concerned with understanding which and how many configurations can be reached starting from a given initial configuration of particles and performing classical (stochastic) dynamics, e.g.~randomly drawing and applying sequences of range-$k$ gates. 
	
	All the results derived in this work via classical reasoning have direct consequences on the physics of 1D quantum lattice models with any on-site Hilbert space dimension $d$ and range-$k$ interactions that conserve a charge $\hat{N}$ and its dipole moment $\hat{X}$, with $[\hat{N},\hat{X}]=0$. In the case of Hamiltonian quantum dynamics, the Hamiltonian $\hat{H}$ is given by a sum of range-$k$ terms $\hat{h}_j^{(k)}$ such that
	\begin{equation}
		\label{eq:genericquantummodel}
		\hat{H} = \sum_j \hat{h}_j^{(k)} + \text{H.c.} \, , \  \quad \Big[\hat{h}_j^{(k)},\hat{N}\Big] = \Big[\hat{h}_j^{(k)}, \hat{X} \Big] = 0 \ \ \forall  \ j \ .
	\end{equation}
	Here any individual operator $\hat{h}_j^{(k)}$ (possibly non-Hermitian) can be chosen such that it maps product states in a properly defined “particle basis” to other product states in the same basis. Analogously to the classical case, we require gate-completeness of the set $\{\hat{h}_j^{(k)}\}$ and all their combinations. A standard example is a model of spin-$s$ variables with $k=4$ and Hamiltonian
	\begin{equation}
		\begin{aligned}
			\hat{H} &= \sum_{i=0}^{L-4} J_i \, \hat{S}_i^+ \hat{S}_{i+1}^- \hat{S}_{i+2}^- \hat{S}_{i+3}^+ \\ 
			& \qquad \quad + \sum_{i=0}^{L-3} K_i \, \hat{S}_i^+ (\hat{S}_{i+1}^-)^2 \hat{S}_{i+2}^+ + \text{H.c.} \ ,
		\end{aligned}
	\end{equation}
where $\hat{S}^\pm_i$ represent on-site raising and lowering operators and $J_i, K_i$ are arbitrary scalars. Indeed, the fragmentation phenomenon does not rely on fine tuning of couplings and holds for entire families of Hamiltonians built from the set $\{\hat{h}_j^{(k)}\}$ \cite{Moudgalya_HSF}. Note that any product state in the $\hat{S}^z$ basis is in one-to-one correspondence with a product state in a particle basis with $d = 2s +1$, where the on-site particle occupation $\hat{n}_i=(\hat{S}_i^z+s)$ allows one to define $\hat{N}$, $\hat{X}$ analogously to \cref{eq:defNandX}. Identical considerations can be applied to fermionic $(d=2)$ and bosonic $(d=\infty)$ models. Furthermore, instead of Hamiltonian dynamics, one can equally well consider (stochastic) quantum dynamics generated by unitary circuits built using range-$k$ quantum gates with charge and dipole conservation \cite{PhysRevX.9.021003,Khemani_HSF}.
	
	In the following, we often refer to single configurations of particles on the chain, or equivalently to product states in the particle basis, simply as “states”. In Section \ref{sec:EEstrongfrag}, where we discuss entanglement and quantum dynamics, we refer explicitly to “eigenstates” and “product states” to highlight the quantum nature of those states. From now on, we restrict our attention to fillings $0 \le \nu \le (d-1)/2$. The regime $(d-1)/2 \le \nu \le d-1$ can be exactly mapped to the former by particle-hole symmetry. Similarly, our conclusions for $\nu_x\le1/2$ can be automatically extended to $\nu_x\ge1/2$. 
    Finally, in what follows we often implicitly resolve the dynamics of particles in terms of sequences of \emph{pairwise one-site} hops. It is easy to verify that this is always possible, independently of the on-site dimension $d$, provided gate-completeness is satisfied.\footnote{Imagine a range-$k$ gate, like the one in \cref{fig:HopsFig}, and an overall hopping move implemented by it which conserves $N$ and $X$ and respects the constraint imposed by $d$. By definition, during the move no particle will exit the $k$-site region the gate acts on. Given the indistinguishability of the particles, if we were to imagine individually labeling them in the original configuration then there are many ways in which we can “interpret” the overall move, in terms of which particles moved where. We define a positive or negative dipole-moment quantum as the change in dipole moment associated with a single particle moving, respectively, one site to the right or to the left. The sum of the dipole-moment quanta gained must be equal to the sum of the quanta lost, by conservation of $X$. For $d=\infty$, it is hence obvious that irrespective of the interpretation chosen, we can always decompose the overall move into one-site hops of pairs in which one particle gains a dipole-moment quantum and the other loses it. For $d$ finite, it might happen that within a chosen interpretation, a pairwise one-site hop cannot be implemented because of the constraint induced by a fully occupied site. However, in such cases it is always possible to change interpretation and let one of the particles in the fully occupied site perform the one-site hop.}
	
	\subsection{Order parameters for the degree of fragmentation}
	\label{subsec:degreeoffragmentation}
	In the quantum setting, we refer to the dimension of a given $(N,X)$ symmetry sector as $D_{N,X}^{(d)}$. Symmetry sectors can be decomposed into ``Krylov sectors", defined as the subspaces spanned by quantum states that are connected by the unitary evolution $\hat{U}(t)$ generated by $\hat{H}$ from \cref{eq:genericquantummodel} (or its unitary circuit analog) when starting from a product state \cite{Moudgalya_HSF}. In a few simple cases, e.g.~for the special models discussed in \cref{sec:StrongFragModels}, Krylov sectors can be identified as the quantum number sectors of relatively simple nonlocal symmetries \cite{PhysRevB.101.125126, Moudgalya_HSF}, but this is not possible for the general class of models with arbitrary $d$ and $k$ considered in this work.
    
    To make contact with the classical picture discussed in the previous subsection, Krylov sectors in the classical case are defined to be the set of product states in the particle basis that span the sector's associated quantum Krylov subspace. Hence, classical Krylov sectors coincide with the subsets of states within each classical symmetry sector that are connected by the classical (stochastic) dynamics.
	
	The interplay between range-$k$ locality and conservation of $N$ and $X$ necessarily fractures $(N,X)$ sectors into exponentially many (in system size) Krylov sectors featuring a very wide range of dimensions \cite{Sala_HSF,Khemani_HSF}, in striking contrast with both generic and integrable quantum models \footnote{Resolving all quantum sectors associated with local conserved charges in integrable models generally leaves no or very little degeneracy.}. If we call $\mathcal{D}_{N,X}^{(d,k)}(i)\ge1$ the dimension of the $i$-th Krylov sector (for some choice of indexing $i$) of a given $(N,X)$ sector for on-site dimension $d$ and interaction range $k$, we have
	\begin{equation}
		D_{N,X}^{(d)} = \sum_{i=1}^{K^{(d,k)}_{N,X}} \mathcal{D}_{N,X}^{(d,k)}(i)\ ,
	\end{equation}
	where $K^{(d,k)}_{N,X}$ is the total number of Krylov sectors in the $(N,X)$ sector. In this work, we consider \emph{families} of $(N,X)$ sectors, where one family includes all the $(N,X)$ sectors characterised by $N=\nu L$ and $X = \nu_x \nu L^2$ as we let $L$ approach the thermodynamic limit, with $\nu$ and $\nu_x$ either fixed real numbers or fixed real functions of $L$. Then, following \cite{Sala_HSF,Khemani_HSF}, it is convenient to define the two phases of \emph{strong} and \emph{weak} fragmentation in terms of the ratio 
	\begin{align}
		\label{eq:ratio1} r^{(d,k)}_{N,X} &= \frac{\mathcal{D}_{\text{max}}^{(d,k)}}{D_{N,X}^{(d)}} \ ,  \\
		\label{eq:ratio2} \lim_{L\to\infty} r^{(d,k)}_{N,X}  &= \begin{cases} 
			0 \quad \ \text{strong fragmentation}\\
			1 \quad \ \text{weak fragmentation}
		\end{cases} ,
	\end{align}
	where $\mathcal{D}_{\text{max}}^{(d,k)}$ represents the dimension of the largest Krylov sector in a given $(N,X)$ sector. In the case of strong fragmentation each Krylov sector constitutes only a vanishingly small fraction of its corresponding symmetry sector, and numerical evidence has been given in Refs.~\cite{Sala_HSF,Khemani_HSF} that $r^{(d,k)}_{N,X}$ decays to zero exponentially fast with system size. In contrast, in weakly fragmented families the largest Krylov sector includes in the thermodynamic limit a measure-$1$ fraction of all configurations present in the corresponding $(N,X)$ sector.\\

A different probe for the nature of fragmentation was introduced in Ref. \cite{Morningstar_HSF}. Let $\rho_F^{(d,k)}(i)$ be the density of frozen sites in the $i$-th Krylov sector of a given $N$ sector, where a frozen site is a site whose occupation number is the same throughout the entire Krylov sector. Then it was numerically argued that in the thermodynamic limit the weighted average
	\begin{equation}
		\label{eq:orderparrhoF}
		\overline{\rho}^{(d,k)}_F(N,L) = \frac{\sum_{i=1}^{K^{(d,k)}_N}\rho_F^{(d,k)}(i)\, \mathcal{D}_{N,X}^{(d,k)}(i)}{ \sum_{i=1}^{K^{(d,k)}_N}\mathcal{D}_{N,X}^{(d,k)}(i)}
	\end{equation}
	is a non-increasing function of $\nu$ which is exactly zero beyond a critical point. Here we have defined $K^{(d,k)}_N$ to be the total number of Krylov sectors in fixed $N$ sectors, i.e. we are not resolving over the different dipole moments compatible with $N$. 

    \subsection{Connections with new results}

    Before delving into our formalism we present a brief overview of how our results connect with the order parameters just introduced and with previous work. In dipole-conserving chains much of the present understanding of the properties of $r_{N,X}^{(d,k)}$ and  $\overline{\rho}^{(d,k)}_F$ stems from numerical evidence involving only limited system sizes, due to the computational burden of exploring Krylov sectors in large Hilbert spaces. On the contrary, our approach yields both analytical results and efficient numerical methods capable of addressing large system sizes.

    In \cref{sec:strongfrag} we prove analytically that for any $d$, any filling $\nu < \nu_c = (k-2)^{-1}$, and any \emph{typical} family of $(N,X)$ sectors (i.e.~families characterised by $\nu_x=1/2 + o(L^0)$, see \cref{sec:sizesymmetrysec} or glossary in \cref{tab:glossary}), $r^{(d,k)}_{N,X}$ decays to zero exponentially with $L$, and hence that these families are strongly fragmented.

    Given the exponential decay of $r^{(d,k)}_{N,X}$ at $\nu<\nu_c$, it is convenient to define a function  $R^{(d,k)}(\nu)$ as
    \begin{equation}
		\label{eq:scalingrNX}
		\lim_{L\to\infty} \frac{1}{L}\ln r^{(d,k)}_{N,X} = R^{(d,k)}(\nu) \ . 
	\end{equation}
	The advantage of considering $R^{(d,k)}(\nu)$ compared with just probing the limit in \cref{eq:ratio2} is that we conjecture the former to represent a \emph{continuous} order parameter for the filling-induced phase transition. Indeed, our results imply $R^{(d,k)}(\nu)<0$ for $\nu< \nu_c$. We also note that the main analytic claim of Ref.~\cite{Pozderac_DinfHSF} coincides with $R^{(\infty,k)}(\nu_c)=0$. In \cref{sec:weakFragNumerics} (see also \cref{sec:criticalScalingRatio}) we provide strong numerical evidence for $d=2,3, 4$ and $\infty$, and large system sizes, that for $\nu > \nu_c$ the ratio $r^{(d,k)}_{N,X}$ goes to $1$ in the thermodynamic limit, i.e.~$R^{(d,k)}(\nu)=0 \ \forall \ \nu > \nu_c$. In Section \ref{sec:criticalScalingRatio} we present numerical results, for $d=2,3$ and small $L$ values, that are compatible with $R^{(d,k)}(\nu)$ being an increasing function of $\nu$ for $\nu \lesssim \nu_c$, and such that $R^{(d,k)}(\nu_c)=0$ (see  also data in Ref.~\cite{Wang_EastModel} for similar conclusions in a different context).  All this supports the fact that $\nu_c$ is the critical point for any $d$ and that the continuous nature of the transition in typical symmetry sectors can be probed by looking at the scaling in \cref{eq:scalingrNX} of $r^{(d,k)}_{N,X}$ with $L$.\\

    Turning to the average density of frozen sites $\overline{\rho}^{(d,k)}_F$, we note that the numerical results in Ref.~\cite{Morningstar_HSF} for $d=3, k=4$ are consistent with our claim that the critical density is $\nu_c=(k-2)^{-1}$. 
    In \cref{sec:strongfrag1} we analytically prove, for any $d$, that at the low fillings $\nu < (k-1)^{-1}$ (note that $(k-1)^{-1}<\nu_c$) the limit $\lim_{L \to \infty}\overline{\rho}^{(d,k)}_F(N,L)$ is nonzero.  In \cref{sec:weakFragNumerics} we provide strong numerical evidence for large system sizes and $d=2,3,4,\infty$ that $\overline{\rho}_F^{(d,k)}$ vanishes for $\nu > \nu_c$. In \cref{sec:StrongFragFurtherAnalysis} we analytically prove for $d=2$ and $\infty$ that $\lim_{L \to \infty}\overline{\rho}^{(d,k)}_F(N,L)$ is also nonzero for fillings $(k-1)^{-1}\le\nu<\nu_c$, and numerically show the same for other finite $d$ values (and large system sizes). All this supports the fact that $\lim_{L \to \infty}\overline{\rho}^{(d,k)}_F(N,L)$ is an alternative ordere parameter, different from $R^{(d,k)}(\nu)$ in Eq.~(\ref{eq:scalingrNX}), for the continuous filling-induced phase transition. We note that both order parameters are highly nonlocal; for example, one cannot recognize a site as frozen without having first globally resolved the Krylov sector. Despite this powerful connection between the strongly fragmented phase and the presence of finite densities of frozen sites, it is quite easy to verify that in the regime $(k-1)^{-1}\le\nu<\nu_c$ rare Krylov sectors devoid of frozen sites exist for any $d$ (see \cref{sec:strongfrag2}).

	\section{Blockages and fully extended states}
	\label{sec:FES}

We present in this section a framework that allows us to analytically characterise the strongly fragmented phase of models with any on-site dimension $d$. It also forms the basis of our analytical and numerical procedures for addressing the weakly fragmented phase. 

    The main idea behind the framework is that the degree of “dynamical connectivity” in the system is directly related to the presence or absence of \emph{blockages} along the chain. These have been defined in Section \ref{sec:introduction} as subregions of the chain across which transport of particle number and dipole moment cannot occur, i.e.~blockages prevent the regions to their left and right from exchanging particles and dipole moment quanta. We prove the existence of two distinct types of blockages, which we classify as \emph{type-1} and \emph{type-2}. The former involve $k-1$ or more consecutive empty sites that are frozen, and hence create a disconnection in the system due to the finite range $k$ of interactions. Type-2 blockages are more subtle because they do not involve frozen sites but still prevent transport. For $d=\infty$, we show that type-1 and type-2 blockages can be efficiently located by analysing simple \emph{local} properties of a special state called a ``fully extended state'' (FES) that is present in each Krylov sector, despite the fact that blockages reflect \emph{global} properties of the sector. By leveraging these $d=\infty$ methods, we then generalise our results to finite on-site dimension $d$ by a procedure we refer to as the “FES picture”. Finally, we highlight the existence and discuss some features of Krylov sectors that possess a high degree of connectivity due to their lack of blockages. As is easily anticipated, the existence of these special sectors is responsible for the onset of the weakly fragmented phase. 

    As several new concepts are introduced in the remainder of the work, we present a glossary of specialised terms and abbreviations in \cref{tab:glossary} as a guide to the reader. 

    \renewcommand{\arraystretch}{1.5} 
\begin{table*}[!t]
    \centering
    \textbf{Glossary\vspace{0.2cm}}
    \scriptsize
    \begin{tabular}{|p{4.2cm}|p{12cm}|}
    \hline
    \textbf{Blockage} & Subregion of the chain across which transport of particles or dipole quanta cannot occur. \\
    \hline
    \textbf{Fully extended state (FES)} & State of a $d=\infty$ system to which no outward hop can be applied. \\
    \hline
    \textbf{Type-1 blockage} & Sequence of $k-1$ or more consecutive holes in an FES. It results in a frozen blockage. \\
    \hline
    \textbf{Type-2 edge}& Sequence of $k-2$ consecutive holes enclosed by two particles in an FES.\\ 
    \hline
    \textbf{Type-2 blockage} & Subregion of an FES devoid of type-1 blockages and which includes two type-2 edges and the sites they enclose. This structure results in an active blockage. \\
    \hline
    \textbf{2-colour connectivity} & Property that underlies the reason why the regions to the left and right of a type-2 blockage cannot exchange particles and dipole moment quanta. Defined in \cref{sec:BlockagesFES2colour}. \\
    \hline
    \textbf{FES picture} & Process of mapping a state in a finite-$d$ system $S$ to the FES of an auxiliary system $\tilde{S}$ with $\tilde{d}=\infty$. \\
    \hline
    \textbf{Particle-connected (PC) FES}& FES with no type-1 or type-2 blockages between the leftmost and rightmost particles.  \\ 
    \hline
    \textbf{Blockage-free FES}& PC FES with no type-1 or type-2 blockages along the entire chain.   \\ 
    \hline
    \textbf{Typical $(N,X)$ sectors} & For any value of $\nu$ which sets a family of $N = \nu L$ sectors, $(\nu L,X)$ sectors characterised by the intensive centre of mass condition $X/(N L) - 1/2 = o(L^0)$ . \\
    \hline
    \textbf{PC string} & Subregion of the chain with at most 1 particle on each site and separations between particles (or between particles and string boundaries) of exactly $k-3$ holes, with the exception of at most one separation of $k-2$ holes. \\
    \hline
    \textbf{Active bubble} & Contiguous subregion of the chain which contains at least two ``active'' particles, is devoid of frozen blockages (i.e.~$k-1$ or more consecutive frozen sites), and is enclosed by a frozen blockage on each side. \\
    \hline
    \textbf{PC extended state} & Finite-$d$ analog of a PC FES. No outward hops can be applied to this state and the bulk structure is identical to that of a PC FES. The two differ at the boundaries, as a PC extended state can host boundary stacks of several fully occupied sites.\\
    \hline
    \textbf{Blockage-free extended state}& PC extended state with no blockages along the entire chain.\\
    \hline
    \textbf{PC contracted state}& Particle-hole conjugate of PC extended state. \\
    \hline
    \end{tabular}
    \caption{Glossary of specialised terms and abbreviations used in this work.}
    \label{tab:glossary}
\end{table*}

	\subsection{Blockages as local features of fully extended states}
	\label{sec:BlockagesFES2colour}
    In our framework for identifying blockages, the concept of a “fully extended state” (FES) originally introduced in Ref. \cite{Pozderac_DinfHSF} plays a crucial role. An FES is a configuration of particles in a $d=\infty$ system in which no pair of particles can perform an outward hop, i.e.~a hop in which they move away from each other (\emph{cf}.~\cref{fig:HopsFig}). A useful fact to remember is that \emph{two particles separated by $k-3$ or more empty sites (holes) cannot perform an outward hop}. Therefore, an FES should be visualised as follows: any site can host at most $1$ particle, and different particles must be separated by sequences of at least $k-3$ holes; otherwise it would still be possible to perform further outward hops. On the two boundary sites of the open chain, however, an indefinite number of particles can be stacked and there is no lower bound on the number of holes separating each of these stacks from the next closest particle. An example of an FES with $k=6,N=10,X=130$ is
\begin{align*}
    \begin{tikzpicture}[scale=0.7, transform shape]
    \def\siteLength{0.3} 
    \def\spacing{0.45}  
    \def\particleSize{2.5pt} 
    \def\adjust{0.03}
    \foreach \i in {0, 1, 2, 3, 4, 5, 6, 7, 8, 9, 10, 11, 12, 13, 14, 15, 16, 17, 18,19,20,21,22,23,24} {
        \draw (\i * \spacing, 0) -- (\i * \spacing + \siteLength, 0);
    }
    \foreach \i in {0, 2, 7, 11, 17, 21, 24} {
        \fill (\i * \spacing + \siteLength/2, 0.2) circle (\particleSize);
    }
    \fill (0+\siteLength/2,0.45) circle (\particleSize);
    \fill (24*\spacing +\siteLength/2,0.45) circle (\particleSize);
    \fill (24*\spacing +\siteLength/2,0.7) circle (\particleSize);
    \node at (4.5 * \spacing +\siteLength/2, -0.3) {\(k-2\)};
    \node at (9 * \spacing +\siteLength/2, -0.3) {\(k-3\)};
    \node at (14 * \spacing +\siteLength/2, -0.3) {\(k-1\)};
    \node at (19 * \spacing +\siteLength/2, -0.3) {\(k-3\)};
    \node at (25 * \spacing, 0) {.};
\end{tikzpicture}
\end{align*}
    
    In a $d=\infty$ system, an FES can always be reached by starting from any initial configuration of particles and applying outward hops (irrespective of the order of application) until it is no longer possible to proceed further \cite{Pozderac_DinfHSF} (see also \cref{appendix:uniqueFES}). Clearly, this statement relies on the natural assumption of gate-completeness discussed in \cref{subsec:chargeandiplocmod}. In Appendix \ref{appendix:uniqueFES} we prove the  following “uniqueness” property of FESs for generic range $k$:\\
	
	\emph{In a $d=\infty$ system, each Krylov sector possesses a unique FES, i.e.~an FES cannot be dynamically connected to a different FES.} \\
	
	We now show that in $d=\infty$ systems this uniqueness result enables us to easily identify blockages from \emph{local} features of FESs, despite the fact that blockages reflect \emph{global} properties of Krylov sectors. This will later also allow us to readily identify certain types of blockages in systems with finite $d$. 
	An obvious but important property for the following derivation is that any subsystem composed of adjacent sites of an FES also represents an FES (with associated uniqueness) if considered in isolation. We call any such contiguous subregion of an FES a “sub-FES”. We are now ready to introduce two types of blockages that can be located by analysing the local structure of FESs. \\

	\textbf{Type-1 blockages}. \ Consider a $d=\infty$ system and an FES that possesses a sequence of $k-1$ or more consecutive holes somewhere along the chain. 
	\begin{align*}
		\begin{tikzpicture}[scale=0.8, transform shape]
			\def\siteLength{0.3} 
			\def\spacing{0.45}  
			\def\particleSize{2.5pt} 
			\def\adjust{0.03}
			\foreach \i in {0, 1, 3, 4} {
				\draw (\i * \spacing, 0) -- (\i * \spacing + \siteLength, 0);
			}
			\foreach \i in {0, 4} {
				\fill (\i * \spacing + \siteLength/2, 0.2) circle (\particleSize);
			}
			\node at (-1 * \spacing+\siteLength/2+\adjust, 0.15) {\dots};
			\node at (2 * \spacing+\siteLength/2+\adjust, 0.0) {\dots};
			\node at (5 * \spacing+\siteLength/2+\adjust, 0.15) {\dots};
			\node at (2 * \spacing +\siteLength/2, -0.3) {\(\ge k-1\)};
            \node at (6 * \spacing, 0) {.};
		\end{tikzpicture}
	\end{align*}
	Since it is not possible for $2$ particles separated by $k-1$ or more empty sites to interact, the subsystems on either side of these holes can only be dynamically connected if one of them expands further to bridge the divide. However, this would violate the uniqueness of FESs, given that both these subsystems represent sub-FESs. This proves that the $k-1$ empty sites are frozen, that is, they are empty in the entire Krylov sector, and hence the regions on their left and right are dynamically disconnected. A consequence of this is that any dynamics will conserve $N$ and $X$ separately on the left and right of the hole sequence. Thus, the hole sequence represents a blockage composed of empty frozen sites, i.e.~an example of a frozen blockage. \\
	
	\begin{figure}[t]
		\centering
		\begin{tikzpicture}[scale=0.85, transform shape]
			\def\height{0.0} 
			\def\length{0.2}
			\definecolor{darkgreen}{rgb}{0.0, 0.5, 0.0}
			\coordinate (A) at (0,\height);
			\coordinate (B) at (1,\height);
			\coordinate (C) at (3.2,\height);
			\coordinate (D) at (4.5,\height);
			\coordinate (E) at (5.1,\height);
			\coordinate (F) at (6,\height);
			\coordinate (G) at (7.1,\height);
			
			\draw[line width=1.8pt, brown] (A) -- (B);
			\draw[line width=1.8pt, darkgreen] (B) -- (C);
			\draw[line width=1.8pt, blue] (C) -- (D);
			\draw[line width=1.8pt, red] (D) -- (E);
			\draw[line width=1.8pt, orange] (E) -- (F);
			\draw[line width=1.8pt, purple] (F) -- (G);
			\draw[white] (7.1,\height) -- (8.3,\height);

			\draw[line width=5pt] (A) -- ++(0,-\length) -- ++(0,2*\length);
			\draw[line width=1.2pt] (B) -- ++(0,-\length) -- ++(0,2*\length);
			\draw[line width=1.2pt] (C) -- ++(0,-\length) -- ++(0,2*\length);
			\draw[line width=1.2pt] (D) -- ++(0,-\length) -- ++(0,2*\length);
			\draw[line width=1.2pt] (E) -- ++(0,-\length) -- ++(0,2*\length);
			\draw[line width=1.2pt] (F) -- ++(0,-\length) -- ++(0,2*\length);
			\draw[line width=5pt] (G) -- ++(0,-\length) -- ++(0,2*\length);
			
			\draw[line width=5pt] (2,\height+1.0) -- ++(0,-1*\length) -- ++(0,2*\length);
			\node at (3.5,\height+1.0) {\(=\) type-1 blockage};
			\draw[line width=1.2pt] (2,\height+1.6) -- ++(0,-1*\length) -- ++(0,2*\length);
			\node at (3.2,\height+1.6) {\(=\) type-2 edge};
			\node at (0.5,\height+0.5) {\(\mathcal{A}\)};
			\node at (0.5,\height+0.5) {\(\mathcal{A}\)};
			\node at (-1.05,1.7) {\large (a)};
			
			\draw[line width=1.2pt, dashed, gray!90] (2.6,-2*\length) rectangle (4.8,2*\length);
		\end{tikzpicture}\\
		\begin{tikzpicture}[scale=0.85, transform shape]
			\def\siteLength{0.3} 
			\def\spacing{0.45}  
			\def\particleSize{2.5pt} 
			\def\adjust{0.03}
			\foreach \i in {0, 1, 3, 4, 5, 7, 8, 10, 11, 13, 14, 15, 17, 18} {
				\draw (\i * \spacing, 0) -- (\i * \spacing + \siteLength, 0);
			}
			
			\definecolor{darkgreen}{rgb}{0.0, 0.5, 0.0}
			\newcommand{\getcolor}[1]{%
				\ifnum#1=0 darkgreen\fi
				\ifnum#1=1 blue\fi
				\ifnum#1=2 blue\fi
				\ifnum#1=3 blue\fi
				\ifnum#1=4 blue\fi
				\ifnum#1=5 red\fi
			}
			\foreach \i [count=\j from 0] in {0, 4, 8, 10, 14, 18} {
				\fill[fill=\getcolor{\j}] (\i * \spacing + \siteLength/2, 0.2) circle (\particleSize);
			}
			\node[text=darkgreen] at (-1 * \spacing+\siteLength/2+\adjust, 0.15) {\textbf{\dots}};
			\node at (2 * \spacing+\siteLength/2+\adjust, 0.0) {\dots};
			\node at (6 * \spacing+\siteLength/2+\adjust, 0.0) {\dots};
			\node at (9 * \spacing+\siteLength/2+\adjust, 0.0) {\dots};
			\node at (12 * \spacing+\siteLength/2+\adjust, 0.0) {\dots};
			\node at (16 * \spacing+\siteLength/2+\adjust, 0.0) {\dots};
			\node[text=red] at (19 * \spacing+\siteLength/2+\adjust, 0.15) {\textbf{\dots}};
			\node at (2 * \spacing +\siteLength/2, -0.3) {\(k-2\)};
			\node at (6 * \spacing +\siteLength/2, -0.3) {\(k-3\)};
			\node at (12 * \spacing +\siteLength/2, -0.3) {\(k-3\)};
			\node at (16 * \spacing +\siteLength/2, -0.3) {\(k-2\)};
			\node at (-0.3,1) {\large (b)};
		\end{tikzpicture}\\
		\begin{tikzpicture}[scale=0.85, transform shape]
			\def\siteLength{0.3} 
			\def\spacing{0.45}  
			\def\particleSize{2.5pt} 
			\def\adjust{0.03}
			\foreach \i in {0, 1, 3, 4, 5, 7, 8, 10, 11, 13, 14, 15, 17, 18} {
				\draw (\i * \spacing, 0) -- (\i * \spacing + \siteLength, 0);
			}
			
			\definecolor{darkgreen}{rgb}{0.0, 0.5, 0.0}
			\newcommand{\getcolor}[1]{%
				\ifnum#1=0 darkgreen\fi
				\ifnum#1=1 blue\fi
				\ifnum#1=2 blue\fi
				\ifnum#1=3 blue\fi
				\ifnum#1=4 blue\fi
				\ifnum#1=5 red\fi
			}
			\foreach \i [count=\j from 0] in {0, 4, 8, 10, 14, 18} {
				\fill[fill=\getcolor{\j}] (\i * \spacing + \siteLength/2, 0.2) circle (\particleSize);
			}
			\node[text=darkgreen] at (-1 * \spacing+\siteLength/2+\adjust, 0.15) {\textbf{\dots}};
			\node at (2 * \spacing+\siteLength/2+\adjust, 0.0) {\dots};
			\node at (6 * \spacing+\siteLength/2+\adjust, 0.0) {\dots};
			\node at (9 * \spacing+\siteLength/2+\adjust, 0.0) {\dots};
			\node at (12 * \spacing+\siteLength/2+\adjust, 0.0) {\dots};
			\node at (16 * \spacing+\siteLength/2+\adjust, 0.0) {\dots};
			\node[text=red] at (19 * \spacing+\siteLength/2+\adjust, 0.15) {\textbf{\dots}};
			\draw (-0.6,-0.4) rectangle (2.17,0.6);
			\draw (2.17,-0.4) rectangle (4.87,0.6);
			\draw (4.87,-0.4) rectangle (9,0.6);
			\node at (-0.3,1) {\large (c)};
		\end{tikzpicture}
		\caption{Schematic representation of type-2 blockages, as identified from the sub-FES that occupies a region $\mathcal{A}$ of the chain. (a) Example of region $\mathcal{A}$ enclosed by two type-1 blockages. Subregions separated by a type-2 edge (i.e., a sequence of $k-2$ holes in an FES) are associated with different colours. Any region consisting of two type-2 edges and the sites enclosed by them represents a type-2 blockage. (b) Close-up of the portion of region $\mathcal{A}$ highlighted by the dashed gray rectangle in panel (a), where two sequences of $k-2$ holes (type-2 edges) appear in the FES. In between the two type-2 edges, particles are separated only by sequences of $k-3$ holes. According to the colour scheme, regions separated by one type-2 edge have particles of different colours. (c) Example of a possible partition of the portion of $\mathcal{A}$ from panel (b) into disjoint subregions that can at most host two different colours. Note that it is irrelevant where exactly in-between two chosen particles of the FES we place a given partition cut.\label{fig:colorscheme}}
	\end{figure}

    \textbf{Type-2 blockages}. \ Consider a $d=\infty$ system in an FES and a subregion $\mathcal{A}$ of the chain that is enclosed by two type-1 blockages, so that $\mathcal{A}$ is dynamically disconnected from the rest of the chain. We also require that within $\mathcal{A}$ there are no type-1 blockages. In cases where there are no type-1 blockages in the entire system, $\mathcal{A}$ coincides with the whole chain. In the following we call any sequence of $k-2$ holes enclosed between two particles in an FES a “type-2 edge”. An example of such a region $\mathcal{A}$ is presented in Fig.~\ref{fig:colorscheme}(a). Within $\mathcal{A}$ we employ a useful colour scheme according to which regions separated by a type-2 edge have particles of different colours. An example is shown in Fig.~\ref{fig:colorscheme}(b), where two type-2 edges are separated by a nonzero number of sequences of $k-3$ holes. Note that two type-2 edges can also be next to each other, i.e., enclose just one particle and no sequence of $k-3$ holes.

	Consider the local Krylov sector composed of all the particle configurations in $\mathcal{A}$ that can be reached starting from the sub-FES that originally occupies $\mathcal{A}$. In Appendix \ref{sec:proof2colour} we prove the following:\\
	
	\emph{Any configuration in this local Krylov sector can be dynamically reached by partitioning $\mathcal{A}$ into disjoint subregions that contain at most two different colours each and, starting from the sub-FES, performing independent series of hops within each of these subregions.} \\
	
	Note that different particle configurations in the local Krylov sector might be associated with different partitions. In the following, we refer to the statement above as “2-colour connectivity”. An example of such a partition for the local subregion in Fig.~\ref{fig:colorscheme}(b) is represented in Fig.~\ref{fig:colorscheme}(c).

	With reference to the colours and particle configurations of Fig.~\ref{fig:colorscheme}(b)-(c), we refer to the region consisting of the two type-2 edges and all the sites enclosed by them as the ``central region''. An elementary consequence of the 2-colour connectivity is that the green and red regions cannot exchange particles or dipole moment quanta, and hence that the central region represents a blockage. Indeed, the 2-colour connectivity ensures that given any particle configuration in the local Krylov sector, we can find a cut somewhere along the central region such that the particle number and dipole moment to the left and right of the cut in the chosen configuration are the same as in the sub-FES. 
	Contrary to type-1 blockages, here no frozen site is involved and the green and red regions can exchange particles and dipole quanta with the blockage. Identical conclusions hold for all the other colours in $\mathcal{A}$, \emph{cf.} Fig.~\ref{fig:colorscheme}(a). 
	
	In the following we refer to any subregion of an FES devoid of type-1 blockages and consisting of two type-2 edges and the sites enclosed by them as a “type-2” blockage. These constitute an example of active blockages as they do not involve frozen sites. For a summary of the concepts just introduced see the glossary in \cref{tab:glossary}.

	\subsection{The FES picture}
    \label{sec:FESpicsec}
	We have just seen how certain types of local configurations of particles and holes in the FES of a $d=\infty$ system allow us to identify blockages and frozen sites that characterise the entire Krylov sector. A key realisation is that this method based on FESs allows us to also draw important conclusions about systems with finite $d$. Consider a configuration of particles on the chain in a system $S$ with $d$ finite. Starting from this, we imagine performing dynamics in an auxiliary system $\tilde{S}$ which is identical to $S$ aside from having no upper bound on the maximal on-site occupancy, i.e., $\tilde{d}=\infty$. We can hence apply outward hops until we reach the corresponding unique FES. If in this FES some sequences of empty sites constitute type-1 blockages, the same sites will also be empty and frozen in the finite-$d$ system $S$, thus representing frozen blockages there as well. This is because the Krylov sector associated with the chosen particle configuration in the finite-$d$ system $S$ is included in the Krylov sector associated with the $\tilde{d}=\infty$ system $\tilde{S}$. Similarly, type-2 blockages in $\tilde{S}$ constitute (or are part of) blockages also in $S$, in that they prevent the propagation of particles and dipole moment between the regions on their left and right. In the following, we refer to this approach that maps a configuration of the system $S$ to its corresponding FES in the auxiliary system $\tilde{S}$ as the “FES picture”.
	
	Note that in general $S$ might possess frozen sites and blockages in addition to those it has in common with $\tilde{S}$, due to the maximal occupation constraint of $d-1$ particles per site. By particle-hole symmetry, some of these are trivial generalisations of those already derived, and typically arise in $N$-sectors with filling $\nu$ beyond the half filling $(d-1)/2$. For example, a configuration of particles with some sites forming a type-1 blockage can be mapped to its particle-hole symmetric configuration, and in the latter the same sites will be frozen and have maximal occupation $d-1$. Far more interesting are blockages in finite-$d$ systems $S$ that cannot be identified using the FES picture. The simplest example is $k-1$ or more contiguous sites with maximal occupation $d-1$ embedded in an otherwise empty chain. These sites are clearly frozen, but when using the FES picture no type-1 blockage emerges to signal it, and hence this finite-$d$ feature is invisible to the FES picture. The possible presence of these types of “finite-$d$” blockages is automatically accounted for by the methods we present in \cref{sec:weakFragNumerics} when discussing weak fragmentation.

	\subsection{Blockage-free FESs}
	\label{sec:ergodicFES}
	
	A particularly interesting class of FESs is constituted of those devoid of any type-1 and type-2 blockages. Defining $n_1$ and $n_2$ respectively to be the number of type-1 blockages and of type-2 edges in an FES, we define a “blockage-free FES” to be a fully extended state characterised by $n_1=0$ and $n_2 = 0$ or $1$. Thus, if one considers all the separations between neighbouring particles in a blockage-free FES, there can be at most one separation of $k-2$ holes while all the other separations must be of $k-3$ holes, with the only exception being the separations next to the two boundary sites which can be composed of fewer than $k-3$ holes. Note that stacking remains allowed at the two boundary sites and that the single sequence of $k-2$ holes, if present, can also be in-between one of the boundary sites and the next closest particle. An example of a blockage-free FES for $k=6$, $L=21$, $N=13$, $X=112$ is given by
	\begin{align*}
		\begin{tikzpicture}[scale=0.8, transform shape]
			\def\siteLength{0.3} 
			\def\spacing{0.45}  
			\def\particleSize{2.5pt} 
			\def\adjust{0.03}
			\foreach \i in {0, 1, 2, 3, 4, 5, 6, 7, 8, 9, 10, 11, 12, 13, 14, 15, 16, 17, 18,19,20} {
				\draw (\i * \spacing, 0) -- (\i * \spacing + \siteLength, 0);
			}
			\foreach \i in {0, 2, 6, 10, 15, 19, 20} {
				\fill (\i * \spacing + \siteLength/2, 0.2) circle (\particleSize);
			}
			\fill (0+\siteLength/2,0.45) circle (\particleSize);
			\fill (0+\siteLength/2,0.7) circle (\particleSize);
			\fill (0+\siteLength/2,0.95) circle (\particleSize);
			\fill (0+\siteLength/2,1.2) circle (\particleSize);
			\fill (20*\spacing +\siteLength/2,0.45) circle (\particleSize);
			\fill (20*\spacing +\siteLength/2,0.7) circle (\particleSize);
			\node at (4 * \spacing +\siteLength/2, -0.3) {\(k-3\)};
			\node at (8 * \spacing +\siteLength/2, -0.3) {\(k-3\)};
			\node[blue] at (12.5 * \spacing +\siteLength/2, -0.3) {\(k-2\)};
			\node at (17 * \spacing +\siteLength/2, -0.3) {\(k-3\)};
            \node at (21 * \spacing, 0) {.};            
		\end{tikzpicture}
	\end{align*}
	Importantly, blockage-free FESs can only exist for states with filling $\nu\ge\nu_c$, \emph{cf}.~\cref{sec:strongfrag}. Fillings $\nu < \nu_c$ are compatible with another special class of FESs, composed of a sub-FES which hosts no blockages and is surrounded by empty sites (which constitute blockages) up to the chain boundaries (note that the sub-FES can overlap with one of the two boundaries). We refer to the latter type of FESs and to blockage-free FESs collectively as “particle-connected” (PC) FESs, given that both types host no disconnections between the leftmost and rightmost particle.
	Appendices \ref{appendix:uniqueFES} and \ref{app:Eoftype1FESforNandX} prove the following:\\
	
	\emph{Given an interaction range $k$, there exists one and only one PC FES within each $(N,X)$ sector.}\\

    Cases in which the unique PC FES is not a blockage-free FES arise when $\nu < \nu_c$ or when the centre of mass $X/N$ is significantly far from the centre of the chain. 
	From the analysis in \cref{sec:strongfrag} it will become apparent that, for any $d$, in weakly fragmented and typical $(N,X)$ sectors (see glossary in \cref{tab:glossary}) a \emph{necessary} condition for a Krylov sector to be the dominant one, i.e.~the one appearing in the limit of $r^{(d,k)}_{N,X}$ going to $1$ in \eqref{eq:ratio2}, is for it to be mapped by the FES picture to the unique blockage-free FES associated with the chosen $(N,X)$ sector. The intuition behind this is that a dominant Krylov sector cannot possess the strong dynamical disconnections induced by the presence of type-1 and type-2 blockages, which would otherwise further splinter it.  \\
	
	In Section \ref{sec:weakFragNumerics} we define “blockage-free extended states” as generalisations of blockage-free FESs that take into account the maximal on-site occupation of $d-1$ for finite-$d$ systems. These have the same blockage-free FES structure in the bulk of the chain but differ in proximity of the boundaries, given that the maximal on-site occupation prevents stacking of an indefinite number of particles on the two boundary sites. Blockage-free extended states, for which an existence and uniqueness property holds in typical symmetry sectors for $\nu>\nu_c$, will allow us to numerically argue for any $d$ that a \emph{necessary and sufficient} condition for a Krylov sector to be the dominant one is for it to contain the unique blockage-free extended state associated with the chosen $(N,X)$ sector. This perspective offers an interesting interpretation of the uniqueness of dominant Krylov sectors in the weakly fragmented phase, attributing it to the uniqueness of blockage-free extended states.
	
	Given the identical structure of blockage-free FESs and blockage-free extended states in the bulk of the chain, for the following it will be useful to define a “particle-connected (PC) string” to be any subregion of the chain that has
	\begin{enumerate}
		\item At most one particle on each site. 
		\item Separations between neighboring particles, or between particles and string boundaries, of exactly $k-3$ holes, with the exception of at most one separation of $k-2$ holes.
	\end{enumerate}
	Note that, in connection with the 2-colour connectivity introduced in \cref{sec:BlockagesFES2colour}, each contiguous subregion in the bulk of an FES hosting at most two colours coincides with a PC string.

	\section{Analytic characterisation of strong fragmentation}
	\label{sec:strongfrag}

	In this section we prove that, given any on-site Hilbert space dimension $d$, any \emph{typical} (see below) family of symmetry sectors with $\nu<\nu_c=(k-2)^{-1}$ is strongly fragmented. In particular, we show that the decay of the ratio $r_{N,X}^{(d,k)}$ in \cref{eq:ratio1} is exponential with system size. The main idea formalised in the proofs is that, for fillings $\nu < \nu_c$, there is an \emph{extensive} presence of blockages along the chain, of both type-1 and type-2. Their appearance implies a strong reduction in the connectivity of the system, i.e.~a significant restriction to the number of configurations that can be reached from a given initial state via range-$k$ dipole-conserving dynamics. 

    We start by discussing the scaling of the sizes of symmetry sectors, and defining what we mean by a “typical” symmetry sector. This, combined with the FES picture, allows us to very easily prove that the region $\nu<(k-1)^{-1}$ is strongly fragmented. We then present the full proof, which addresses the less trivial region $(k-1)^{-1}\le \nu <\nu_c$. Some of the technical details for this part are discussed in Appendix \ref{appendix:detailsoffullproof}. 

\subsection{\label{sec:sizesymmetrysec}Size of symmetry sectors and typicality}

Call $D_N^{(d)}$ the total number of configurations in the symmetry sector of $N=\nu L$ particles, without fixing the dipole moment $X$. In Appendix \ref{appendix:scalingofsizeofss} we show that for $0 < \nu < d - 1$ and large $L$ one has
\begin{equation}
\label{eq:generaldDN}
    \ln D_N^{(d)}(L) = L \, \eta_d(\nu) - \frac{1}{2}\ln L + \mathcal{O}(L^0) \ . 
\end{equation}
Here $\eta_d(\nu)$ is a strictly concave function of $\nu$ with a unique global maximum at half filling $\nu^*=(d-1)/2$, where it takes the value $\eta_d(\nu^*)=\ln d$. 
As before, we call $D_{N,X}^{(d)}$ the size of the symmetry sector of $N$ particles and $X=\nu \, \nu_x L^2$ dipole moment. Leveraging results from \cite{MelczerPanovaPemantle}  we find its asymptotic form to be (see Appendix \ref{appendix:scalingofsizeofss}) 
\begin{equation}
		\begin{aligned}
        \label{eq:DNXscaling}
			\ln D_{N,X}^{(d)}(L) &= L  \Big(\eta_d(\nu)  - \Lambda_d(\nu,\nu_x) \Big) \\
			& \qquad \quad  - 2 \ln L + \mathcal{O}(L^0) \ ,
		\end{aligned}
	\end{equation}
where $\Lambda_d(\nu,\nu_x) \ge 0$. For $\nu_x$ sufficiently close to $1/2$ the function $\Lambda_d$ can be expanded as
\begin{equation}
\label{eq:Lambdadnux0}
    \Lambda_d(\nu,\nu_x) = \lambda_d(\nu) \, \nu_{x_0}^2 + o(\nu_{x_0}^2) \ ,
\end{equation}
where $\lambda_d$ is a positive function and $\nu_{x_0}=\nu_x - 1/2$ is the intensive centre of mass when the origin is set in the middle of the chain. If we fix $\nu$,  \cref{{eq:DNXscaling},{eq:Lambdadnux0}} imply that for any family of $(N=\nu L,X)$ sectors characterised by $\nu_{x_0}=o(L^0)$, i.e.~vanishing (in the thermodynamic limit) intensive centre of mass with respect to the middle of the chain, one gets
\begin{equation}
		\label{eq:Xfamilynux0zero}
		\frac{1}{L}\ln D_{N,X}^{(d)} = \eta_d(\nu) + o(L^0) \quad \quad \nu_{x_0} = o(L^0).
	\end{equation}
    By comparing with \cref{eq:generaldDN} we see that $\nu_{x_0}=o(L^0)$ families are \emph{typical}, i.e., together they contain a fraction that tends to $1$ exponentially with $L$ of all the configurations in the chosen $N=\nu L$ sector. 

\begin{figure}[t]
    \centering
    \includegraphics[scale=0.23]{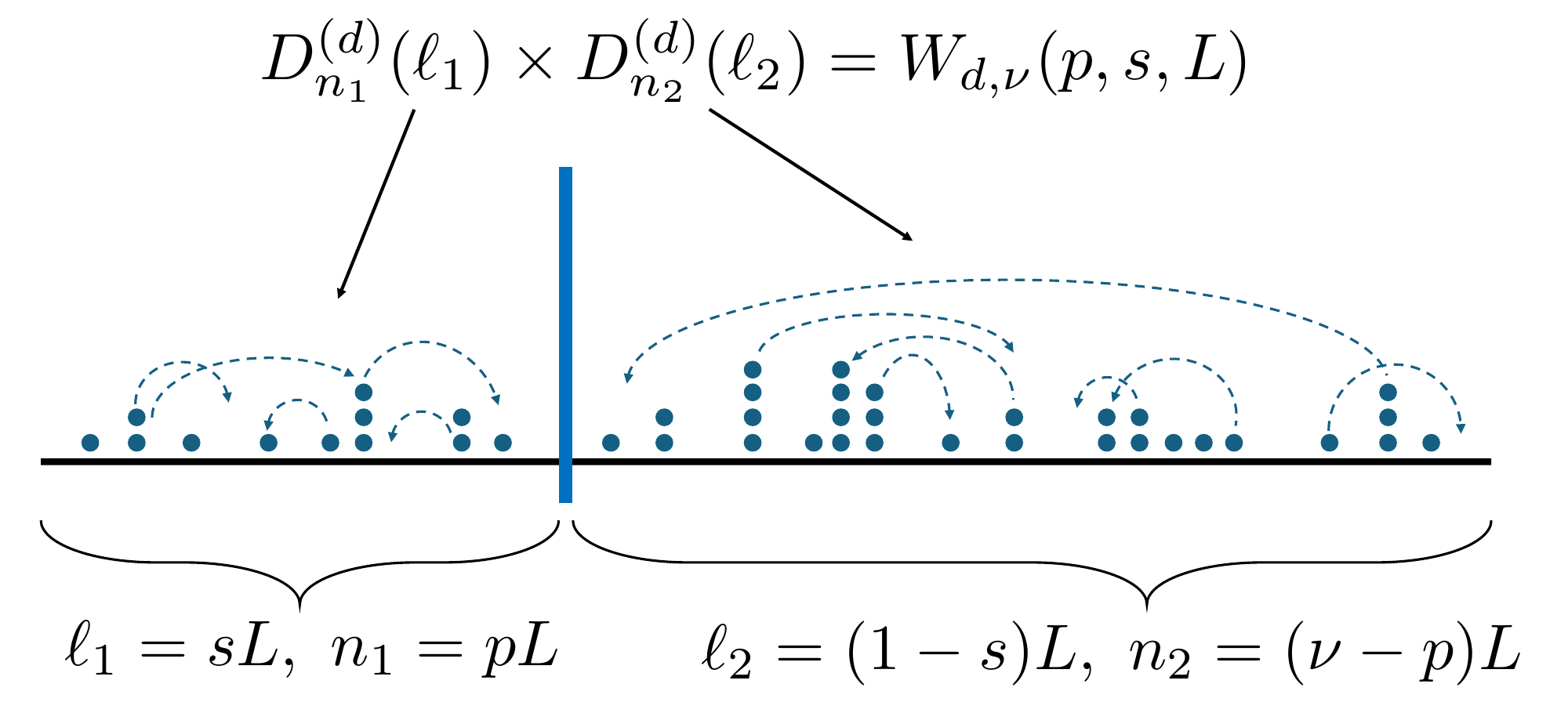}
    \caption{Example of a bipartition of the chain with $L$ sites and $N=\nu L$ particles. $W_{d,\nu}$ indicates the total number of configurations compatible with it.}
    \label{fig:exampleBipart}
\end{figure}

    Another important property needed in the following is related to “spatial” typicality, which is a natural consequence of the strict concavity of $\eta_d(\nu)$: \\ 

    \emph{Typical states in a given family of $N=\nu L$ sectors are homogeneous over extensive length scales, i.e.~the “local” filling of any extensive subregion equals the global $\nu$.}\\

This is nothing but the familiar \footnote{Since the work of Boltzmann on the microscopic definition of entropy.} combinatorial property that homogeneous arrangements dominate in number the configuration space. To formalize this, consider a chain of length $L$ and $N=\nu L$ particles, and imagine bipartitioning it in such a way that one of the two subregions contains $s L$ sites and $p L$ particles, with $0 < s < 1$. We are interested in the total number $W_{d,\nu}(p,s,L)$ of different particle configurations compatible with such a bipartition (see \cref{fig:exampleBipart}). We can extract the asymptotic dependence of $W_{d,\nu}(p,s,L)$ directly from \eqref{eq:generaldDN}
	\begin{align}
		\label{eq:bipartitionchain} 
		g_{d,\nu}(p, s) &=\lim_{L\to\infty}\frac{1}{L} \ln W_{d,\nu}(p,s,L)\ \\
		&= s \, \eta_d\left(\frac{p}{s} \right)+(1-s) \, \eta_d\left(\frac{\nu-p}{1-s} \right) \ . \nonumber
	\end{align}
It then easily follows from the strict concavity of $\eta_d$ that 
    \begin{equation}
    \label{eq:ineqforgdnu}
        g_{d,\nu}(p, s) < \eta_d(\nu) \quad \ \forall \ p \neq p^* \ ,
    \end{equation}
    where $p^* = \nu \, s$ is the unique global maximum of $g_{d,\nu}$, as a consequence of the identity $g_{d,\nu}(p^*,s)=\eta_d(\nu)$. As there are only $\mathcal{O}(L)$ ways to distribute the total number of particles $N=\nu L$ between the two subregions (i.e., to choose $p$ once $s$ is fixed), we see by comparing \eqref{eq:ineqforgdnu} with \eqref{eq:generaldDN} that “homogeneous” configurations in which both subregions have local filling equal to the global $\nu$ are typical, i.e.~their number exponentially dominates in system size over all other configurations in the $N=\nu L$ sector. It is straightforward to generalise this argument to the case in which we partition the system into any $G=\mathcal{O}(L^0)$ number of extensively large subregions of size $s_i L$, with $s_1+\ldots+s_G=1$. Altogether, this proves the statement above. \footnote{This homogeneity over extensive subregions is expected from simple intuitive arguments. Indeed, for asymptotically large values of $L$ we can imagine generating a random state by randomly associating a site $i=0, \ldots, L-1$ with each particle, avoiding occupations beyond $d-1$. In the vast majority of states obtained in this way, any extensive subregion is expected to have a filling identical to the global filling $\nu$.} Note that this spatial typicality condition is consistent with the typicality of $(N,X)$ sectors with vanishingly small $\nu_{x_0}$.

	\subsection{Warm-up: strong fragmentation for $\nu < (k-1)^{-1}$}
	\label{sec:strongfrag1}

Before turning to the proof of strong fragmentation in the entire region $\nu < \nu_c = (k-2)^{-1}$, it is instructive to consider the proof for  $\nu<(k-1)^{-1}$. This is the simplest case, as a consequence of the presence of an extensive number of frozen sites. 

Consider a chain $S$ with $2 \le d \le \infty$ and global filling $\nu < (k-1)^{-1} $. Starting from any configuration of particles on the chain, we employ the FES picture of Section \ref{sec:FES} and reach an FES in the auxiliary system $\tilde{S}$. We call $f$ the fraction of sites in the chosen Krylov sector of $S$ that are part of a frozen blockage. Given the current upper bound on $\nu$, we automatically know from the FES picture that
	\begin{equation}
		\label{eq:fnonzerofornu}
		f \ge 1 - \nu (k-1) > 0 \ . 
	\end{equation}
	This lower bound is obtained by noticing that in the FES of $\tilde{S}$ the lowest number of sites involved in a type-1 blockage is obtained when each one of the $N$ particles is separated from its neighbours by exactly $k-2$ holes. 
	We now prove that as a consequence of Eq.~(\ref{eq:fnonzerofornu}), any typical $(N,X)$ family (\emph{cf.}~Section \ref{sec:sizesymmetrysec} or \cref{tab:glossary}) in $S$ is strongly fragmented according to the ratio test in \cref{eq:ratio2}. For any typical $(N,X)$ family, the exponential scaling of $D^{(d)}_{N,X}$ is given by Eq.~(\ref{eq:Xfamilynux0zero}). 
	On the other hand, the $f \, L $ empty frozen sites induce a natural bipartition of the chain in which they represent one (possibly noncontiguous) of two independent subregions that cannot exchange particles or dipole moment. The total number $W_{d,\nu}(p=0, s=f, L)$ of particle configurations of any dipole moment compatible with this bipartition is clearly larger than the dimension of any Krylov subspace in the chosen $(N,X)$ sector, i.e., 
	\begin{equation}
		\mathcal{D}^{(d,k)}_\text{max} < W_{d,\nu}(p=0, s=f, L) \ . 
	\end{equation}
	From the uniqueness of the maximum of $g_{d,\nu}(p,s)$ discussed in Section \ref{sec:sizesymmetrysec} and the fact that here $p=0 \neq \nu \, f = p^*$, we automatically see that
	\begin{equation}
		\label{eq:useofg1}
		\lim_{L\to\infty} \frac{1}{L} \ln r^{(d,k)}_{N,X} \le g_{d,\nu}(0, f) - \eta_d(\nu) < 0 \ . 
	\end{equation}
	This proves, for any $d$, that all the $(N,X)$ families characterised by $\nu<(k-1)^{-1}$ and $\nu_{x_0}=o(L^0)$ (typical) are strongly fragmented according to the ratio test, given that the ratio in Eq.~(\ref{eq:ratio1}) decays to zero exponentially with $L$.

	\subsection{The full proof}
	\label{sec:strongfrag2}
    To conclude the proof we need to address the interval $(k-1)^{-1} \le \nu < (k-2)^{-1} = \nu_c$, to which we refer for convenience as $\mathcal{C}_{k-2}$. Unlike the $\nu$ region discussed in the previous subsection, for $\nu \in \mathcal{C}_{k-2}$  it is possible to have Krylov sectors with no frozen sites. 

    To prove strong fragmentation in $\mathcal{C}_{k-2}$ it is not necessary to understand the specific features of the largest Krylov subspaces in a given family of $(N,X)$ sectors, i.e.~the one entering the definition \eqref{eq:ratio1} of $r^{(d,k)}_{N,X}$. Indeed, our strategy is simpler: we aim to show that for any $\nu \in \mathcal{C}_{k-2}$, the largest Krylov sectors occupy an exponentially vanishing fraction of the corresponding symmetry sectors $(N,X)$, irrespective of their characteristics. 

    We start by noticing that in $\mathcal{C}_{k-2}$ there still exist exponentially many FESs of the auxiliary chain $\tilde{S}$ that possess a nonvanishing fraction $f>0$ of frozen sites that are part of a type-1 blockage. This leads us to\\

    \textbf{Case 1:} The largest Krylov subspaces within a typical family of $(N,X)$ sectors in $S$ possess a nonzero fraction $f$ of frozen sites that are part of a type-1 blockage.\\

    By reasoning identical to the previous section we see that, in Case 1, $r_{N,X}^{(d,k)}$ is exponentially suppressed in $L$. \\
    
    Now we turn to the scenario in which $f=0$, which is possible because the  inequality in Eq.~(\ref{eq:fnonzerofornu}) does not hold in $\mathcal{C}_{k-2}$. Assume that, in addition to $f=0$, there are no particles stacked on the two boundary sites of the FES of $\tilde{S}$ to which the Krylov subspace of $S$ in question is connected. If we call $\rho_2$ and $\rho_3$ the densities of particles that are separated from the next closest particle to their right by  $k-2$ and $k-3$ holes respectively, then in the thermodynamic limit we have
	\begin{equation}
		\rho_2 + \rho_3 = \nu \qquad \quad \rho_2(k-1)+\rho_3(k-2) = 1 \ , 
	\end{equation}
	from which we find
	\begin{equation}
		\label{eq:solforrho23}
		\rho_2=1-\nu(k-2) \qquad \quad \rho_3 = \nu(k-1)-1 \ . 
	\end{equation}
	This proves that for fillings $\nu$ within $\mathcal{C}_{k-2}$, one has $\rho_2 > 0$ and $\rho_3\ge0$. If we keep $f=0$ but allow stacking of particles at the two boundary sites, then $\rho_2$ necessarily increases with respect to the case \eqref{eq:solforrho23} with no stacking. This means that every Krylov sector with $f=0$ must possess an extensive number of type-2 edges, and thus, of type-2 blockages. Say we select one among the $\mathcal{O}(L)$ many type-2 blockages. Given the extensive number of possible choices, we can always select a type-2 blockage ($\mathcal{B}$) of $L$-independent size such that the regions to its left ($\mathcal{A}_1$) and right ($\mathcal{A}_2$), up to the boundaries, are both extensively large. Given that $\mathcal{A}_1$ and $\mathcal{A}_2$ are prevented from exchanging particles and dipole moment, we can upper bound the dimension of the largest Krylov sector as
	\begin{equation}
		\label{eq:dimABCbound}
		\mathcal{D}^{(d,k)}_\text{max} < D_{N_{\mathcal{A}_1+\mathcal{B}},X_{\mathcal{A}_1+\mathcal{B}}}^{(d)} \, D_{N_{\mathcal{A}_2+\mathcal{B}},X_{\mathcal{A}_2+\mathcal{B}}}^{(d)} \ . 
	\end{equation}
	We notice that starting from a chain with $L$ sites and $N$ particles that possesses $D_N^{(d)}$ configurations, if we add to it $\ell=\mathcal{O}(L^0)$ sites which contain $n=\mathcal{O}(L^0)$ particles, then $\lim_{L\to\infty}D^{(d)}_{N+n}(L+\ell)/D^{(d)}_{N}(L)$ is a finite $\mathcal{O}(L^0)$ number, as seen by applying Eq.~(\ref{eq:generaldDN}). The same is true when we consider the sizes $D_{N,X}^{(d)}(L)$ of $(N,X)$ sectors. This implies that there always exists an $L$-independent constant $Q$ such that from Eq.~(\ref{eq:dimABCbound}) we can arrive to 
	\begin{equation}
		\label{eq:QboundDmax}
		\mathcal{D}^{(d,k)}_\text{max} < Q \, D_{N_{\mathcal{A}_1+\mathcal{B}},X_{\mathcal{A}_1+\mathcal{B}}}^{(d)} \, D_{N_{\mathcal{A}_2},X_{\mathcal{A}_2}}^{(d)} \ . 
	\end{equation}
    
From this we can consider\\

 \textbf{Case 2:} The largest Krylov subspaces within a typical family of $(N,X)$ sectors have $f=0$, and the regions $\mathcal{A}_1+\mathcal{B}$ and $\mathcal{A}_2$ can be chosen in such a way that their local filling is different from $\nu$ in the limit of large $L$.\\

 Remembering that extensive subregions with local filling different from the global one can characterise only atypical states (see Section \ref{sec:sizesymmetrysec}), using \eqref{eq:ineqforgdnu} we obtain again that $\lim_{L\to\infty}\ln r_{N,X}^{(d,k)}/L<0$, i.e.~$r^{(d,k)}_{N,X}$ decays exponentially with $L$.

 Let us now consider the final scenario, given by\\

 \textbf{Case 3:} The largest Krylov subspaces within a typical family of $(N,X)$ sectors have $f=0$, but the “inhomogeneous” assumption on $\mathcal{A}_1+\mathcal{B}$ and $\mathcal{A}_2$ of Case 2 does not hold.\\

	Case 3 implies that any extensive subregion of the chain has local filling approaching the global $\nu$ for large $L$ values. As we show in Appendix \ref{appendix:detailsoffullproof}, this provides us with an algorithm to partition the system into an \emph{extensive} number $G$ of subregions $\mathcal{A}_i$ that are separated by type-2 blockages $\mathcal{B}_i$ of $L$-independent size, as pictured in Fig.~\ref{fig:type2partition}. The length $\ell_i$ of each subregion $\mathcal{A}_i$ is guaranteed to be subextensive. Given such a partition of the chain, and given the disconnecting effect induced by type-2 blockages $\mathcal{B}_i$, one arrives at the inequality (see Appendix \ref{appendix:detailsoffullproof})
    \begin{equation}
		\label{eq:finalbound}
		r^{(d,k)}_{N,X} < L^2 \prod_{i=1}^G \frac{\widetilde{Q}}{\ell_i^2} \ . 
	\end{equation} 
Here $\widetilde{Q}$ is an $L$-indepdendent positive constant such that $ \widetilde{Q} <\ell_i^2  \ \ \forall \ i$. Given that $G = \mathcal{O}(L)$, \cref{eq:finalbound} shows that also in this case $r_{N,X}^{(d,k)}$ decays exponentially with $L$, hence concluding the proof. \\

\begin{figure}[t]
		\begin{align*}
			\begin{tikzpicture}[scale=0.8, transform shape]
				\def\hlength{1.8cm}
				\def\vheight{0.5cm}
				\foreach \i in {0,1,3,4} {
					\draw (\i * \hlength, 0.25) -- (\i * \hlength + 0.9 *\hlength, 0.25);
				}
				\foreach \i in {0,1,2,3} {
					\draw (\i * \hlength + 0.9 *\hlength, 0) -- (\i * \hlength + 0.9 *\hlength, \vheight);
				}
				\foreach \i in {0,1,2,3} {
					\draw (\i * \hlength + \hlength, 0) -- (\i * \hlength + \hlength, \vheight);
				}
				\node at (2*\hlength+0.8cm, 0.25) { . . .};
				\node at (0*\hlength+0.45*\hlength, -0.1) {\( \mathcal{A}_1 \)};
				\node at (1*\hlength+0.45*\hlength, -0.1) {\( \mathcal{A}_2 \)};
				\node at (3*\hlength+0.45*\hlength, -0.1) {\( \mathcal{A}_{G-1} \)};
				\node at (4*\hlength+0.45*\hlength, -0.1) {\( \mathcal{A}_G \)};
				\node at (0*\hlength+0.95*\hlength, 0.75) {\( \mathcal{B}_1 \)};
				\node at (1*\hlength+0.95*\hlength, 0.75) {\( \mathcal{B}_2 \)};
				\node at (2*\hlength+0.95*\hlength, 0.75) {\( \mathcal{B}_{G-2} \)};
				\node at (3*\hlength+0.95*\hlength, 0.75) {\( \mathcal{B}_{G-1} \)};
			\end{tikzpicture}
		\end{align*}
		\caption{Partition of the chain into contiguous subregions ($\mathcal{A}_i$) separated by type-2 blockages of $L$-independent size ($\mathcal{B}_i$). Each vertical line represents a type-2 edge.\label{fig:type2partition}}
	\end{figure}

A few remarks are in order. In Appendix \ref{appendix:detailsoffullproof} we briefly discuss the fate of strong fragmentation in \emph{atypical} families of $(N,X)$ sectors, i.e.~those with a centre of mass significantly far from the centre of the chain.

In \cref{sec:StrongFragFurtherAnalysis}, we shall prove for $d=2$ and $d=\infty$ that \emph{typical} states in a given $N=\nu L$ sector with $\nu < \nu_c$ host an extensive number of type-1 blockages. We note that if one could easily prove the same for any $d$, then the previous proof could be limited to the case of $f>0$, which is trivial. 

Another route to a proof could make explicit use of the structure of the largest Krylov sectors in the strongly fragmented phase. For example, a plausible assumption would be that even for $\nu < \nu_c$ the particle-connected (PC) FESs (for $d=\infty$) and PC extended states (for $d$ finite, see glossary in \cref{tab:glossary}), are always part of the largest Krylov subspaces in typical symmetry sectors, exactly as happens for $\nu>\nu_c$ (see \cref{sec:weakFragNumerics}). However, an analytical justification of such an assumption appears highly nontrivial to obtain, and robust numerical verification poses challenges beyond those tackled by the approach of \cref{sec:weakFragNumerics}. We note that Ref.~\cite{Pozderac_DinfHSF}, by making use of similar assumptions (verified numerically for small $L$), argued for an exponential decay of the ratio of dimensions for $d=\infty$, $\nu < \nu_c$ and $\nu_{x_0}=1/2$.

Finally, we remark that even if the assumption of gate-completeness (see \cref{sec:familyofmodels}) does not hold, our results still prove that $r_{N,X}^{(d,k)}$ decays to zero exponentially with $L$ in typical symmetry sectors for $\nu<\nu_c$, and hence that strong fragmentation holds. This is a trivial consequence of the fact that violations of gate-completeness further suppress dynamics and reduce the connectivity in the system, and hence all the blockages identified by our method remain such. It is in principle possible, however, that in some models that violate gate-completeness the critical point of the strong-to-weak transition is higher than $\nu_c=(k-2)^{-1}$. We discuss this point further at the end of the next section, which numerically addresses weak fragmentation.

    	\section{Numerically identifying the weakly fragmented phase}
	\label{sec:weakFragNumerics}

	In this section, we provide numerical evidence that, for $\nu>\nu_c$ and generic $d$, the model is in its weakly fragmented phase. By this, we mean that typical $(N,X)$ sectors at these particle densities possess a dominant Krylov sector to which almost all states in the $(N,X)$ sector belong in the thermodynamic limit. We show that the dominant Krylov sectors are blockage-free and can be characterised by the fact that they each contain a ``blockage-free extended state''. These states are a natural finite-$d$ generalisation of the blockage-free FESs defined in Section \ref{sec:ergodicFES} (see \cref{tab:glossary}), and their presence in a Krylov sector implies that the latter is blockage-free. We demonstrate that there is at most one blockage-free extended state per $(N,X)$ sector, and that these states only exist  for $\nu\geq\nu_c$. This explains why the critical density is $d$-independent: for any value of $d$, the density $\nu_c=(k-2)^{-1}$ is the smallest density at which it is possible to obtain blockage-free Krylov sectors. Our numerical results directly imply that typical states  with $\nu>\nu_c$ are blockage-free; hence, the absence of blockages in typical states is not only \emph{necessary} for the onset of the weakly fragmented phase, it is also \emph{sufficient}.

	\subsection{Blockage-free extended states }
	\label{sec:ESandCSdef}
	
    Identically to how, in $d=\infty$ systems, blockage-free FESs are special cases of particle-connected (PC) FESs (see \cref{tab:glossary} and \cref{sec:ergodicFES}), the blockage-free extended states in finite-$d$ systems are special cases of what we call ``particle-connected (PC) extended states''. These are a finite-$d$ generalisation of PC FESs, and have a very similar structure in which no particles can perform outward hops. In particular, at most one pair of particles are separated by $k-2$ holes, and almost all others are separated by $k-3$ holes. The only exceptions are at the boundaries, and it is here that the difference at finite $d$ becomes important. In $d=\infty$ systems, indefinitely many particles can be stacked on the boundary sites; in a finite-$d$ system, such stacks of particles must be spread over several sites. For example, the PC extended state with $d=4$, $k=5$, $L=16$, $N=15$, $X=67$ is   
    
\begin{equation*}
    \begin{tikzpicture}[scale=0.8, transform shape]
    \def\siteLength{0.3} 
    \def\spacing{0.45}  
    \def\particleSize{2.5pt} 
    \def\adjust{0.03}
    \foreach \i in {-1,0,1,2,3,4,5,6,7,8,9,10,11,12,13,14} {
        \draw (\i * \spacing, 0) -- (\i * \spacing + \siteLength, 0);
    }
    \foreach \i in {-1,0,1,2,4,7,11,14} {
        \fill (\i * \spacing + \siteLength/2, 0.2) circle (\particleSize);
    }
    \fill (-1*\spacing + \siteLength/2, 0.65) circle (\particleSize);
    \fill (-1*\spacing + \siteLength/2, 0.42) circle (\particleSize);
    \fill (0*\spacing + \siteLength/2, 0.65) circle (\particleSize);
    \fill (0*\spacing + \siteLength/2, 0.42) circle (\particleSize);
    \fill (1*\spacing + \siteLength/2, 0.65) circle (\particleSize);
    \fill (1*\spacing + \siteLength/2, 0.42) circle (\particleSize);
      \fill (14*\spacing + \siteLength/2, 0.42) circle (\particleSize);
        \node at (5.5 * \spacing +\siteLength/2, -0.3) {\(k-3\)};
            \node at (9 * \spacing +\siteLength/2, -0.3) {\(k-2\)};
                \node at (12.5 * \spacing +\siteLength/2, -0.3) {\(k-3\)};
    \node at (15 * \spacing, 0) {,};                
    \end{tikzpicture}
\end{equation*}
where we note the stack of particles spread over several sites at the left boundary. Hence, when stacks  are present at the boundaries of a PC extended state, we define them to be composed of several sites with $d-1$ particles, followed by at most one site with fewer than $d-1$ particles. We also have that, same as the $d=\infty$ case, fewer than $k-3$ holes can separate a stack from the next particle over. It is apparent that the PC extended state above follows these rules.

We then define a blockage-free extended state to be a PC extended state that is devoid of blockages along the entire chain. It is clear from the structure of a PC extended state that any blockages must arise at its boundaries and be composed of strings of frozen sites. 

A final thing to note is that in finite-$d$ systems, there is a unique PC extended state corresponding to each $(N,X)$ sector. The proof of this statement follows identical logic to the $d=\infty$ case presented in Appendix \ref{app:Eoftype1FESforNandX}. This furthermore implies that there can be \emph{at most one} blockage-free extended state per $(N,X)$ sector.

	\subsection{Proving absence of blockages}
	\label{sec:CStoES}

    Here we prove that for finite $d$, PC extended states in typical $(N,X)$ sectors at fillings $\nu>\nu_c$ do not have any frozen sites at their boundaries and hence these states (and more broadly the Krylov sectors that contain them) are blockage-free. To show this, we make use of the particle-hole conjugates of PC extended states, which we refer to as ``particle-connected (PC) contracted states''. These states are obtained by taking a PC extended state and flipping the particle occupation number $n_i$ at each site $i$ to $d-1-n_i$. An example of a PC contracted state with $d=4,k=5,L=16,N=29,$ and $X=258$ is given by
\begin{equation*}
    \begin{tikzpicture}[scale=0.8, transform shape]
    \def\siteLength{0.3} 
    \def\spacing{0.45}  
    \def\particleSize{2.5pt} 
    \def\adjust{0.03}
    \foreach \i in {-1,0,1,2,3,4,5,6,7,8,9,10,11,12,13,14} {
        \draw (\i * \spacing, 0) -- (\i * \spacing + \siteLength, 0);
    }
    \foreach \i in {2,...,13} {
        \fill (\i * \spacing + \siteLength/2, 0.2) circle (\particleSize);
    }
    \foreach \i in {3,...,13} {
        \fill (\i * \spacing + \siteLength/2, 0.42) circle (\particleSize);
    }
    \foreach \i in {3,4,6,7,9,10,11} {
        \fill (\i * \spacing + \siteLength/2, 0.65) circle (\particleSize);
    }
        \node at (3.5 * \spacing +\siteLength/2, -0.3) {\(k-3\)};
            \node at (6.5 * \spacing +\siteLength/2, -0.3) {\(k-3\)};
                \node at (10 * \spacing +\siteLength/2, -0.3) {\(k-2\)};

        \node at (15 * \spacing, 0) {.};
    
    \end{tikzpicture}
\end{equation*}
Just as it is impossible to perform any outward hops on a PC extended state, it is also impossible to perform any inward hops on a PC contracted state. Also, as an immediate corollary of the result in \cref{app:Eoftype1FESforNandX}, there is exactly one PC contracted state in each $(N,X)$ sector.

In \cref{app:ProofCStoES}, we derive an algorithm for mapping from the PC contracted state to the PC extended state in a given $(N,X)$ sector (and vice versa) via a series of pairwise hops. These hops are implemented by range-$k$ gates which preserve dipole moment and particle number, and hence are compatible with the dynamics of the system. An immediate corollary of this algorithm is that the PC extended state and PC contracted state in each $(N,X)$ sector are in the same Krylov sector. This implies that the PC extended states in typical ($N,X$) sectors with densities $\nu>\nu_c$ (and, implicitly, $\nu<d-1-\nu_c$) are blockage-free extended states. This is because, as demonstrated in \cref{sec:sizesymmetrysec}, typical ($N,X$) sectors at large $L$ have intensive centres of mass $X/(LN)$ infinitesimally close to 1/2 (the centre of the chain). Thus, the PC extended states in these ($N,X$) sectors will have stacks of particles at their boundaries for $\nu>\nu_c$, whereas the PC contracted states  will have strings of holes at their boundaries. This implies that these PC extended states have \emph{no frozen sites}: all sites involved in the stacks of particles at the boundaries will have had their particle numbers change as part of the algorithm mapping to the PC contracted state.  Furthermore,  since the algorithm in \cref{app:ProofCStoES} involves the exchange of particle number and dipole quanta between  distant regions of the state, we see that \emph{no blockages} are present in the rest of the state either. Hence the PC extended state is blockage-free. This absence of blockages in certain Krylov sectors already suggests that, for typical $(N,X)$ sectors with $\nu>\nu_c$, ergodic behaviour can occur and the system can thermalise.

	\subsection{Numerical evidence of weak fragmentation}
	\label{sec:WeakFragNumericsSubsec}

\begin{figure*}[th]
\centering
{\includegraphics[scale = 0.42]{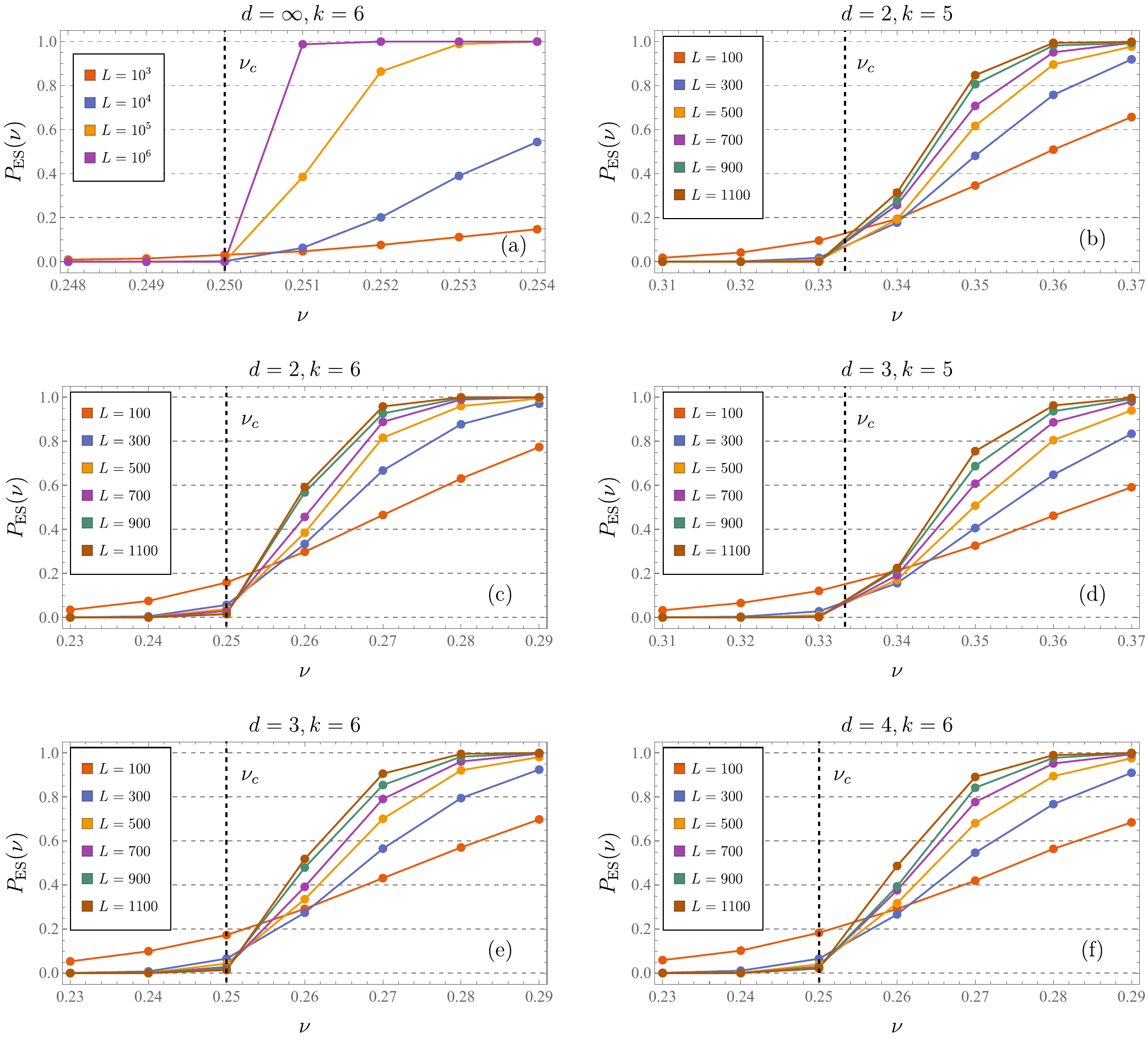}}
\caption{Success rate $P_{\text{ES}}(\nu)$ of the algorithms defined in \cref{sec:WeakFragNumericsSubsec} reaching a PC extended state (or equivalently a PC contracted state) from a random initial state  as a function of the filling $\nu$ for various choices of $d$ and $k$. For plot (a) with $d=\infty$, states were mapped to their corresponding FESs from which it was directly determined if the PC FES had been reached. The numbers of states sampled per data point were, from lowest $L$ to highest, 9000, 3000, 2700, and 400. For plots, (b)-(f), the algorithm using CS-to-ES expansions and ES-to-CS contractions was used. Sample sizes for (b) and (c), from lowest $L$ to highest, were 15000, 6000, 3000, 1500, 1000, and 1000. The sample sizes for (d), (e), and (f) were twice as much. The standard error, obtained by dividing the standard deviation associated with each data point by the square root of the sample size, is mostly too small to see.\label{fig:eFESprob}}

\renewcommand{\figurename}{Numerical evidence of weak fragmentation}
\end{figure*}

    \begin{figure}[th]
\centering
\includegraphics[scale=0.3]{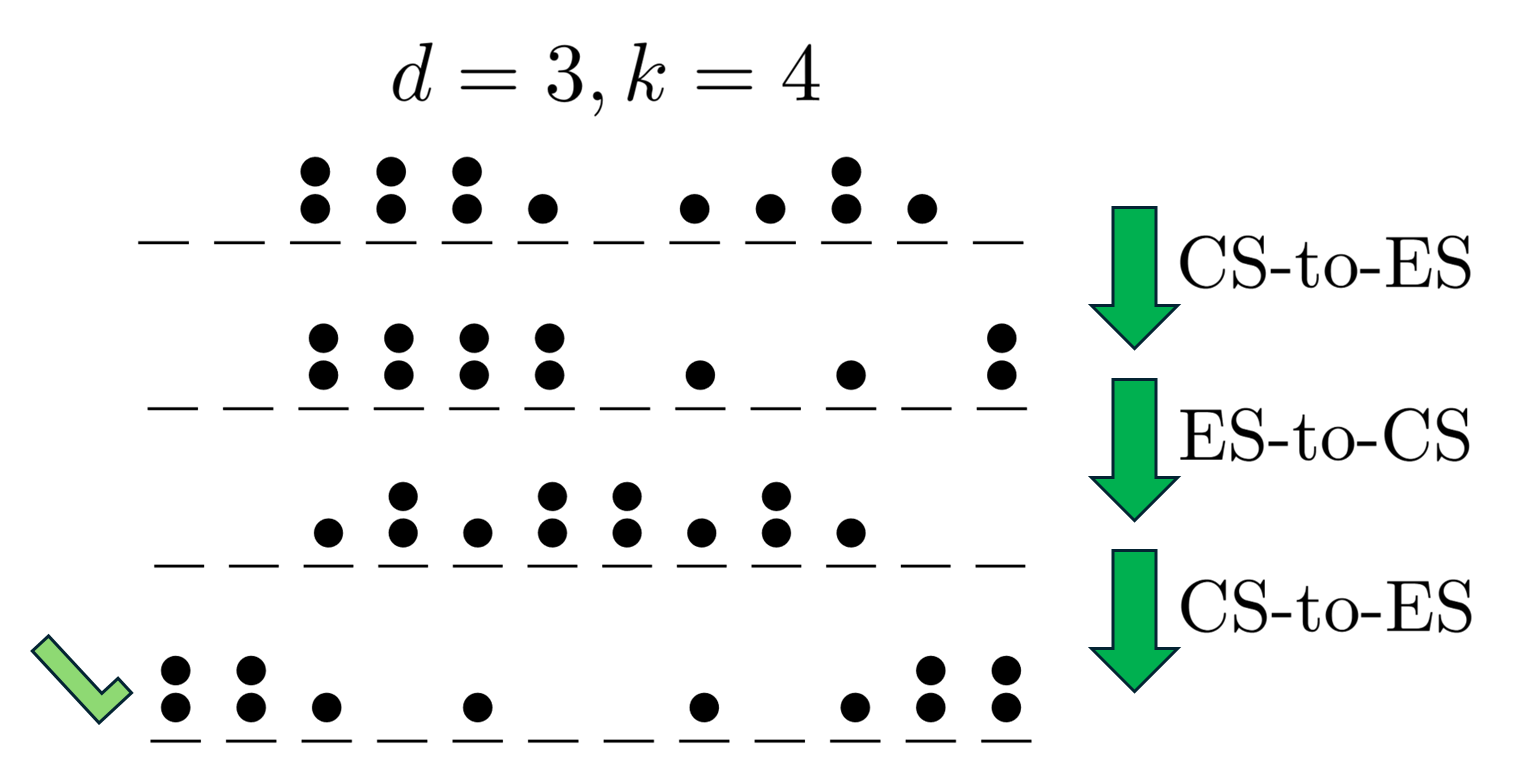}
\caption{The algorithm for testing if a state shares a Krylov subspace with a PC extended state, or equivalently with a PC contracted state, as applied to an arbitrary state of a $d=3$ system with $k=4$ range interactions. In the first step, the PC CS-to-ES algorithm defined in \cref{app:ProofCStoES} is applied; however, as part of the expansion, a set of 4 contiguous sites on the left becomes fully occupied. This prevents further expansion. This dense region is melted in the subsequent step, where the PC ES-to-CS algorithm is applied. If we were to again apply the PC CS-to-ES algorithm (as depicted above), we would obtain a PC extended state. In fact, the algorithm does not perform this last step: it recognises the before-last state as a PC contracted state and registers a success at that point.}
\label{fig:WHSFAlgorithm}
\end{figure}

We have established that for $\nu>\nu_c$ and in typical $(N,X)$ sectors, the unique Krylov sectors which contain PC extended states are blockage-free. In this section, we numerically show that for particle densities $\nu>\nu_c$,  the fraction of states which belong to these blockage-free Krylov sectors goes to 1 as system size $L$ increases. Hence, the blockage-free Krylov sectors dominate their respective $(N,X)$ sectors and the model is in its weakly fragmented phase. 

This result explains why the critical density $\nu_c$ is independent of the on-site dimension $d$. Indeed, for any value of $d$ including $d=\infty$, the particle density $\nu_c=(k-2)^{-1}$ is the smallest density at which it is possible to construct PC extended states which span the whole system and are hence blockage-free. Thus, $\nu_c$ is the lowest density at any $d$ at which ergodic behaviour can occur.

For systems with $d=\infty$, it is easy to numerically determine the percentage of states at a given filling $\nu$ that belong to Krylov subspaces with PC FESs. One simply generates a large number of states at filling $\nu$, maps them to their corresponding FESs via the algorithm in \cref{app:FESalgorithm}, and computes the percentage of states which were mapped to a PC FES. We present the results of such numerics in \cref{fig:eFESprob}(a), and we note the percentage tends to 1 for $\nu>\nu_c$, whereas it vanishes for $\nu<\nu_c$. This shows that the Krylov subspaces with blockage-free FESs dominate for $\nu>\nu_c$ and that the system is in its weakly fragmented phase. However, we cannot use the FES picture to draw similar conclusions for finite $d$, as there can be blockages present in finite-$d$ states which the FES picture does not detect. Furthermore, simply applying outward hops to a finite-$d$ state will not necessarily map it to the most extended particle configuration in its Krylov subspace, as often the system will become stuck in an intermediate state with a large buildup of particles somewhere in the bulk of the chain. Instead, a combination of inward and outward hops is often required.

We here present an algorithm that attempts to map a given initial state at finite $d$ via a series of hops to either the PC extended state or the PC contracted state in its $(N,X)$ sector. The success rate of this algorithm lower bounds the probability that a given state belongs to the same Krylov subspace as the PC extended state (and, equivalently, the PC contracted state). It hence also lower bounds the probability that a given state is devoid of \emph{any} blockages (including finite-$d$ blockages invisible to the FES picture). The success rate of the algorithm as a function of $\nu$ is plotted in \cref{fig:eFESprob}(b)-(f). We see that for densities $\nu>\nu_c$, the success rate tends to 1, indicating that typical states belong to blockage-free Krylov subspaces and the model is in the weakly fragmented phase. For $\nu<\nu_c$, the success rate tends to 0, which is consistent with the strongly fragmented phase where no Krylov subspace dominates.

The algorithm itself makes use of the algorithm derived in \cref{app:ProofCStoES} for mapping between PC extended and contracted states. As shorthand, let us refer to the algorithm for mapping from a PC contracted state to a PC extended state as the CS-to-ES algorithm, and the one mapping from an extended to a contracted state as the ES-to-CS algorithm. Our main algorithm for attempting to map an arbitrary initial state to a PC extended or contracted state is then as follows.
	
	\begin{itemize}
		\item Apply the CS-to-ES expansion algorithm to the state. If a PC extended state is obtained, terminate the algorithm and register a success; if not, proceed to the next step.
		\item  Apply the ES-to-CS contraction algorithm. Register a success if a PC contracted state is obtained, and proceed to the next step if not.
		\item Apply the previous two steps in sequence an arbitrary number $Q$ times in total. If the algorithm still has not terminated, then register an ambiguous result.
	\end{itemize}
	
	To generate the numerics in \cref{fig:eFESprob}, we set $Q=4$. The rationale behind the above algorithm is as follows. After the first application of the PC CS-to-ES algorithm, if the state does not reach a PC extended state, it was often observed that the resulting state was found to be composed of individual PC strings. These were often separated from each other by sequences of $k-1$ or more fully occupied sites  or sequences of $k-1$ or more holes. By applying a subsequent  ES-to-CS contraction, the sequences of $k-1$ or more fully occupied sites  are melted by the contraction process as their edges are peeled off, but the sequences of $k-1$ or more holes remain. These hole sequences are then potentially bridged during the following CS-to-ES expansion step, although this may introduce the presence of new dense regions of $k-1$ or more maximally occupied sites. In this alternating fashion, the aim of the algorithm is to melt the dense regions and bridge the holes separating active regions. We present an example of the algorithm in action in \cref{fig:WHSFAlgorithm}.\\

    Together with the results of the preceding sections, we have now fully established the universal phase diagram of \cref{fig:PhaseDiagram},  with a $d$-independent critical density of $\nu_c=(k-2)^{-1}$. 

    We note that the algorithm derived in \cref{app:ProofCStoES}, and hence most of the results derived in this section, generally rely on gate-completeness (see \cref{sec:familyofmodels}) to hold true. The extent to which they hold in systems that violate gate-completeness, but which still possess a weakly fragmented phase, is dependent on context. In such cases, the system might possess a higher critical density than $\nu_c$ because a higher filling is needed in order to overcome the hindrances introduced by the violations of gate-completeness and give rise to weak fragmentation. For example, a system with $k=5$ interactions which is only gate-complete at the $k=4$ level may have a phase-transition point somewhere in between $\nu=1/3$ and $\nu=1/2$.

	\section{Distribution of blockages and active bubbles}
	\label{sec:StrongFragFurtherAnalysis}

    In this section, we make use of the FES picture developed in \cref{sec:FES} to obtain further analytical and numerical results characterising the strongly fragmented phase. In particular, we demonstrate (analytically for $d=2$ and $\infty$ and numerically for other values of $d$) that typical states at fillings $\nu<\nu_c$ have a finite density of frozen sites. This shows that the mean density of frozen sites  defined in \cref{eq:orderparrhoF}  is a valid order parameter for identifying the strongly fragmented phase, confirming a hypothesis set out in \cite{Morningstar_HSF}. To arrive at this result we show that typical states at fillings $\nu<\nu_c$ have a finite density of both type-1 blockages and type-2 edges; hence, although states with only type-2 blockages (such as those discussed in the proof in \cref{sec:strongfrag2}) can exist for $\nu<\nu_c$, they are vanishingly rare.

    We begin in \cref{sec:nopropcondition} by using the FES picture to derive a fractonic constraint on particle mobility for $\nu<\nu_c$. We then make use of this constraint in \cref{subsec:MeanB1B2andp1} to prove a lower bound on the mean density of type-1 blockages and type-2 edges in  systems with $d=2$ and $d=\infty$ and fillings $\nu<\nu_c$. This also proves the existence of a finite density of frozen sites at these fillings. We derive a related analytic result in \cref{subsec:ActiveBubbleDensityFunction}, where we show that the FES picture can also be used to derive lower bounds on the densities of “active bubbles” of different sizes. These are local active regions of the chain and were first defined in \cite{Morningstar_HSF}. In \cref{subsec:type1distributionNumerics}, we use the FES picture to obtain strong numerical evidence that typical states with $\nu<\nu_c$ host a finite density of \emph{both} type-1 blockages and type-2 edges for other values of $d$ as well.

	\subsection{A sufficient condition for constrained particle mobility}
    \label{sec:nopropcondition}

	We here derive a sufficient condition which, when satisfied by a group of particles, places strong bounds on how far they can propagate from their point of origin. It can only generally be satisfied for densities $\nu<\nu_c$ and illustrates how fractonic restrictions on particle mobility lead to the presence of blockages in the strongly fragmented phase.  Furthermore, the condition  can be used to efficiently identify many (though not all) of the type-1 blockages in a system, and it is in some respects more analytically tractable than the FES picture (which on the other hand identifies all type-1 blockages). We use this condition in the following subsections to analytically lower bound the average density of type-1 blockages and type-2 edges, as well as the average density of active bubbles.
	
	We consider a system $S$ with on-site dimension $d$. Pick a site of the chain as the origin, that is, call it site $0$. Assume that there are $N_R$ particles in the region $\mathcal{A}_R$ that goes from site $0$ to the right boundary of the chain, and let $L_R$ denote the total number of sites in $\mathcal{A}_R$. Assume furthermore that either the rest of the system is empty or that the particles to the left of site 0 are sufficiently far away that they never interact with the particles in $\mathcal{A}_R$.  Label the initial positions of the particles  in $\mathcal{A}_R$, from leftmost to rightmost, as $\{i^0_n\}_{n=1}^{N_R}$, and say those positions satisfy
	\begin{equation}
		i^0_n\geq (n-1)(k-2).
		\label{eq:NoPropCond}
	\end{equation}
	We note that this constraint requires $L_R\geq (N_R-1)(k-2)+1$, which implies that in the thermodynamic limit $\lim_{L\rightarrow\infty}N_R/L_R\leq\nu_c$. We will show that condition \eqref{eq:NoPropCond} ensures that, under dynamical evolution of the system $S$, none of those $N_R$ particles ever leaves the region $\mathcal{A}_R$.\\

	Assume, for the sake of contradiction, that it is possible to start from an initial state satisfying \eqref{eq:NoPropCond} and to propagate at least one particle originally in $\mathcal{A}_R$ to the left of site $0$. From the point of view of the FES picture, this implies that in the auxiliary system $\tilde{S}$,  the leftmost of the particles initially in $\mathcal{A}_R$ is mapped to a position left of site 0 in the FES. Due to the 2-color connectivity of Section \ref{sec:FES}, we know that the expansion from the initial state in $\mathcal{A}_R$ to the final auxiliary FES will occur independently within different subregions of the system. As stressed before, the particles in each of these subregions form a particle-connected (PC) string (see glossary in \cref{tab:glossary}) in the final FES. Consider the leftmost of these PC strings, to which the leftmost particle belongs, and say that it is constituted of $m$ particles in total. Since these particles only interacted amongst themselves during the expansion to an FES, they must have conserved their initial local dipole moment $X_R$, which by \cref{eq:NoPropCond}  must satisfy
	\begin{equation}
		X_R = \sum_{j=1}^m i^0_j \geq \sum_{j=1}^m (j-1) (k-2) = \frac{m(m-1)}{2} (k-2).
		\label{eq:lowerBound}
	\end{equation}
	We next consider what PC string made of $m$ particles would have the highest possible dipole moment while having a particle left of site 0. If we label the final positions of the particles as $i_1,\dots,i_m$, then clearly this PC string is given by $i_1=-1$ and $i_j=(j-1)(k-2)$ for $j=2,\dots,m$. Hence, the leftmost PC string in the FES must have a dipole moment of at most
	\begin{equation}
		-1+ \sum_{j=2}^m (j-1) (k-2) = -1+\frac{m(m-1)}{2} (k-2) \, .
	\end{equation}
	This is less than the lower bound in \cref{eq:lowerBound}, and so this propagation cannot happen. Hence, regardless of the value of $d$, particles that have initial positions satisfying \cref{eq:NoPropCond} cannot propagate left of site 0 by interacting solely among themselves.

	\subsection{Average density of type-1 blockages and type-2 edges}
	\label{subsec:MeanB1B2andp1}

 Using the sufficient condition in \cref{eq:NoPropCond}, we now rigorously derive a lower bound on the average density of type-1 blockages and type-2 edges for states in $d=2$ and $d=\infty$ systems. This also implies a lower bound on the density of frozen sites. The average is taken over all particle configurations in a specified $N$ sector. We define the average densities of type-1 blockages and type-2 edges $\overline{\rho}_t^{(d,k)}$ (with $t=1,2$ respectively) in a system with on-site dimension $d$ and interaction range $k$ as
	\begin{equation}
    \label{eq:avedensitywt}
		\overline{\rho}_t^{(d,k)}(N,L) = \frac{1}{L}\frac{\beta_t^{(d,k)}(N,L)}{D_N^{(d)}(L)},
	\end{equation}
	where $\beta_t^{(d,k)}$ is the total number of type-1 blockages or type-2 edges  across all states in the $N$ sector.

	We first consider the case of type-1 blockages. For simplicity, we only compute a lower bound on the density  of type-1 blockages of a particular format. In particular, we consider the following configuration of holes and particles:
	\begin{align}
		\begin{tikzpicture}[scale=0.8, transform shape]
			\def\siteLength{0.3} 
			\def\spacing{0.45}  
			\def\particleSize{2.5pt} 
			\def\adjust{0.03}
			\foreach \i in {0, 2, 3, 4, 6} {
				\draw (\i * \spacing, 0) -- (\i * \spacing + \siteLength, 0);
			}
			\definecolor{darkgreen}{rgb}{0.0, 0.5, 0.0}
			\newcommand{\getcolor}[1]{%
				\ifnum#1=0 black\fi
				\ifnum#1=1 blue\fi
				\ifnum#1=2 blue\fi
				\ifnum#1=3 black\fi
				\ifnum#1=4 blue\fi
				\ifnum#1=5 blue\fi
			}
			\foreach \i [count=\j from 0] in {3} {
				\fill[fill=\getcolor{\j}] (\i * \spacing + \siteLength/2, 0.2) circle (\particleSize);
			}
			\node at (-1.5 * \spacing+\siteLength/2+\adjust, 0.15) {\dots};
			\node at (1 * \spacing+\siteLength/2+\adjust, 0.0) {\dots};
			\node at (5 * \spacing+\siteLength/2+\adjust, 0.0) {\dots};
			\node at (7.5 * \spacing+\siteLength/2+\adjust, 0.15) {\dots};
			\draw [dashed,very thick] (7*\spacing+0.02,-0.5) -- (7*\spacing+0.02,0.8);
			\draw [dashed,very thick] (0*\spacing-0.15,-0.5) -- (0*\spacing-0.15,0.8);
			\node at (1 * \spacing +\siteLength/2, -0.3) {\(k-3\)};
			\node at (5 * \spacing +\siteLength/2, -0.3) {\(k-1\)};
            \node at (9 * \spacing, 0) {.};
		\end{tikzpicture}
		\label{eq:type1bottleneck}
	\end{align}
	
    We impose that the particles right and left of, respectively, the right and left dashed lines satisfy mirrored versions of the no-propagation constraints in \eqref{eq:NoPropCond}. As a result, the $k-1$ holes on the right in \eqref{eq:type1bottleneck} are frozen and constitute a type-1 blockage. The inclusion of a particle in \eqref{eq:type1bottleneck} ensures that we do not double-count type-1 blockages.

	We lower bound $\beta^{(d,k)}_1(N,L)$ by counting the total number of occurrences of \eqref{eq:type1bottleneck} across all possible states with $N$ particles and $L$ sites. This  is computed by summing over the $\mathcal{O}(L)$ possible positions in the chain  at which the blockage in \eqref{eq:type1bottleneck} can be situated, and then for each such position summing over all possible configurations compatible with \eqref{eq:NoPropCond} for the remaining particles in the system. This yields
	\begin{widetext}
		\begin{equation}
			\label{eq:boundB1}
			\beta^{(d,k)}_1(N,L)\geq \sum_{L_R=0}^{L-2k+3}\sum_{N_R=N_\text{min}}^{N_\text{max}} b^{(d,k)}_1(N_R,L_R)\,b^{(d,k)}_1(N-N_R-1,L-L_R-2k+3),
		\end{equation}
	\end{widetext}
	where $b^{(d,k)}_1(N_R,L_R)$ denotes the total number of ways to arrange $N_R$ particles over $L_R$ sites such that the particles obey the constraint in \cref{eq:NoPropCond} and $N_{\text{min}},N_{\text{max}}$ represent the smallest and biggest, respectively, number of particles $N_R$ compatible with a given $L_R$.

    For $d=2$ and $d=\infty$, it is possible to obtain exact formulas for the function $b^{(d,k)}_1(N_R,L_R)$, and the summation in \cref{eq:boundB1} can be performed analytically in the thermodynamic limit. We perform these calculations explicitly in \cref{app:DSandP}, and present only the final result here. The average densities of type-1 blockages for $d=2$ and $d=\infty$ are lower bounded respectively  by
    \begin{align}
    \lim_{L\rightarrow\infty}\overline{\rho}^{(2,k)}_{1}(\nu\, L, L) &\geq \nu (1-\nu)  (1-\nu/\nu_c)^2\label{eq:lowerbavdend2}, \\
	\lim_{L\to \infty}\overline{\rho}^{(\infty,k)}_{1}(\nu L,L) &\geq \nu (1+\nu)^{-2}(1-\nu/\nu_c)^2, \label{eq:LowerBound_pb1_dinf}
	\end{align}
    where $\nu_c=(k-2)^{-1}$ as previously defined. We note that these lower bounds only hold for $\nu \le\nu_c$. The fact that they vanish at $\nu_c$ is consistent with the onset of weak fragmentation numerically established in \cref{sec:weakFragNumerics}. 
    Since each type-1 blockage contains at least $k-1$ frozen sites, the above result immediately implies a lower bound on the average density of frozen sites $\overline{\rho}^{(d,k)}_{F}(N,L)$ (defined in \cref{subsec:degreeoffragmentation}) for $d=2$ and $d=\infty$, namely  $\overline{\rho}^{(d,k)}_{F}(N,L)\geq (k-1)\overline{\rho}^{(d,k)}_{1}(N, L)$. Hence, we have analytically shown that  $\overline{\rho}^{(d,k)}_{F}$ is nonzero for $\nu<\nu_c$ and $d=2$ and $d=\infty$ whereas the results of \cref{sec:weakFragNumerics} numerically show that it vanishes for $\nu\geq\nu_c$. Hence the average density of frozen sites is a valid order parameter.

    \begin{figure*}[th]
		\centering
		{\includegraphics[scale = 0.4]{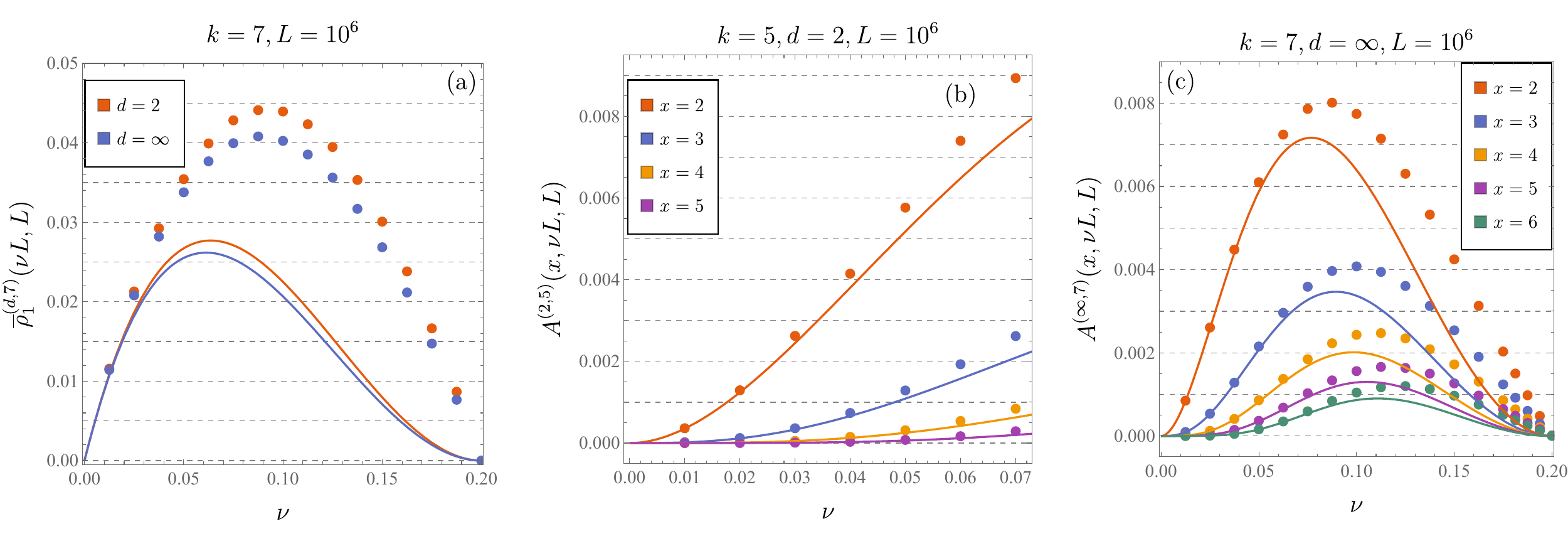}}
		\caption{
        Comparisons between numerical results and analytic lower bounds. In (a), the analytic lower bounds on the density of type-1 blockages in $d=2$ and $d=\infty$ systems with $k=7$ ($\nu_c=1/5$) are compared with numerical densities obtained using the FES picture (see \cref{subsec:type1distributionNumerics} for details on the numerical method). 100 randomly drawn states were averaged over per data point. In (b),   the  analytic lower bounds on the densities of active bubbles with different numbers of particles are compared with numerical results for a system with $d=2$ and $k=5$ ($\nu_c=1/3$). 10 random states were averaged over per data point, and the active bubbles were identified by mapping out each state's corresponding Krylov subspace. In (c), the same plot is shown for a system with $d=\infty$ and $k=7$. 100 states were averaged over per data point, and active bubbles were identified by mapping the states to their corresponding FESs.  In all plots, the standard error on the data is too small to see.}
        \label{fig:IslB1Dist}
	\end{figure*}

    Similarly, one can also use the no-propagation constraints of \cref{eq:NoPropCond} to lower bound the average density of type-2 edges. We focus on type-2 edges of the format
    \begin{align}
		\begin{tikzpicture}[scale=0.8, transform shape]
			\def\siteLength{0.3} 
			\def\spacing{0.45}  
			\def\particleSize{2.5pt} 
			\def\adjust{0.03}
			\foreach \i in {0, 2, 3, 4, 6,7,8,10} {
				\draw (\i * \spacing, 0) -- (\i * \spacing + \siteLength, 0);
			}
			\definecolor{darkgreen}{rgb}{0.0, 0.5, 0.0}
			\newcommand{\getcolor}[1]{%
				\ifnum#1=0 black\fi
				\ifnum#1=1 blue\fi
				\ifnum#1=2 blue\fi
				\ifnum#1=3 black\fi
				\ifnum#1=4 black\fi
				\ifnum#1=5 black\fi
				\ifnum#1=7 black\fi
			}
			\foreach \i [count=\j from 0] in {3,7} {
				\fill[fill=black] (\i * \spacing + \siteLength/2, 0.2) circle (\particleSize);
			}
			\node at (-1.5 * \spacing+\siteLength/2+\adjust, 0.15) {\dots};
			\node at (1 * \spacing+\siteLength/2+\adjust, 0.0) {\dots};
			\node at (5 * \spacing+\siteLength/2+\adjust, 0.0) {\dots};
			\node at (11.5 * \spacing+\siteLength/2+\adjust, 0.15) {\dots};
			\node at (9 * \spacing+\siteLength/2+\adjust, 0.0) {\dots};
			\draw [dashed,very thick] (11*\spacing+0.02,-0.5) -- (11*\spacing+0.02,0.8);
			\draw [dashed,very thick] (0*\spacing-0.15,-0.5) -- (0*\spacing-0.15,0.8);
			\node at (1 * \spacing +\siteLength/2, -0.3) {\(k-3\)};
			\node at (5 * \spacing +\siteLength/2, -0.3) {\(k-2\)};
			\node at (9 * \spacing +\siteLength/2, -0.3) {\(k-3\)};
            \node at (13 * \spacing, 0) {,};
		\end{tikzpicture}
		\label{eq:type2edge}
	\end{align}
	where again the particles to the right and left of the dashed vertical lines follow the no-propagation constraints \eqref{eq:NoPropCond}. Again using the results of \cref{app:DSandP}, we get that the average density $\overline{\rho}_2^{(d,k)} (N,L)$ of type-2 edges is lower bounded by
    \begin{align}
    \lim_{L\rightarrow\infty}\overline{\rho}^{(2,k)}_{2}(\nu L,L) \geq \nu^2 (1-\nu)^{k-2}(1-\nu/\nu_c)^2, \label{eq:LowerBound_t2_d2} \\
	\lim_{L\to \infty}\overline{\rho}^{(\infty,k)}_{2}(\nu L,L)\geq \nu^2 (1+\nu)^{-k}(1-\nu/\nu_c)^2. \label{eq:LowerBound_t2_dinf}
	\end{align}
    From this we see that the mean density of type-2 edges is also nonvanishing throughout the strongly fragmented phase. An almost identical analysis shows the density of type-2 blockages, defined in \cref{sec:BlockagesFES2colour} as two type-2 edges with no type-1 blockage between them, to be nonvanishing.

    We plot the analytic lower bounds on type-1 blockages in \cref{fig:IslB1Dist}(a), where we compare them with the numerically computed average densities. We note there is close agreement for $\nu\ll\nu_c$ and that the lower bounds continue to account for a significant percentage  of the total number of blockages throughout most of the strongly fragmented phase. The main discrepancy between the analytic lower bounds and the full numerical results lies in the scaling close to the critical point, where the analytic lower bound vanishes quadratically whereas the numerical average vanishes linearly. We discuss critical scaling further in \cref{sec:critscalandexp}.

    By self-averaging \cite{Wiseman_SelfAveraging}, one would expect that the lower bounds we computed on the average densities of type-1 blockages, frozen sites, and type-2 edges also apply to typical \emph{individual} states in the thermodynamic limit. In other words, we expect that for a typical state at sufficiently large $L$, the density of type-1 blockages, frozen sites, and type-2 edges will also be lower bounded by the equations derived above, up to corrections that are vanishingly small in system size $L$. We make this argument precise in \cref{app:SelfAveraging}, where we show that this is indeed the case and that type-1 blockages and type-2 edges are uniformly distributed in typical states. Hence, typical states in the strongly fragmented phase feature a finite density both of type-1 blockages and type-2 edges. Importantly, this shows that even though Krylov subspaces without any frozen sites exist in the strongly fragmented phase, they form a vanishingly small part of the overall spectrum at any given $\nu$.

	\subsection{Active bubble densities}
	\label{subsec:ActiveBubbleDensityFunction}

	Since typical states in the strongly fragmented phase contain a finite density of fairly uniformly distributed type-1 blockages, dynamical evolution for these states must be confined to dynamically disconnected local regions that are $L$-independent in size. Adapting a concept from Ref.~\cite{Morningstar_HSF}, we refer to these regions as ``active bubbles''.  More specifically, we define an active bubble to be a set of consecutive  sites containing at least 2 particles and such that:
	\begin{itemize}
		\item Nowhere in the active bubble is there a sequence of $k-1$ or more frozen sites.
		\item The leftmost and rightmost sites of the active bubble are not frozen sites.
		\item The first $k-1$ sites to the left and right of the active bubble are frozen sites and therefore constitute a frozen blockage.
	\end{itemize}
	Since the active bubbles are dynamically disconnected from each other, we may independently map out the local Krylov sectors associated with each one of them, where a local Krylov sector consists of all particle configurations available to an active bubble under dynamical evolution. A simple example in a $d=2$, $k=5$ system is an active bubble consisting of $x=2$ particles spread out over $\ell=4$ sites, which would have two states in its local Krylov sector:
	\begin{equation}
		\left\{\begin{tikzpicture}[scale=0.8, transform shape]
			\def\siteLength{0.3} 
			\def\spacing{0.45}  
			\def\particleSize{2.5pt} 
			\def\adjust{0.03}
			\foreach \i in {0,1,2, 3} {
				\draw (\i * \spacing, 0) -- (\i * \spacing + \siteLength, 0);
			}
			\definecolor{darkgreen}{rgb}{0.0, 0.5, 0.0}
			\newcommand{\getcolor}[1]{%
				\ifnum#1=0 black\fi
				\ifnum#1=1 black\fi
				\ifnum#1=2 black\fi
				\ifnum#1=3 black\fi
				\ifnum#1=4 black\fi
				\ifnum#1=5 black\fi
			}
			\foreach \i [count=\j from 0] in {1,2} {
				\fill[fill=\getcolor{\j}] (\i * \spacing + \siteLength/2, 0.2) circle (\particleSize);
			}
		\end{tikzpicture}\ , \ \begin{tikzpicture}[scale=0.8, transform shape]
			\def\siteLength{0.3} 
			\def\spacing{0.45}  
			\def\particleSize{2.5pt} 
			\def\adjust{0.03}
			\foreach \i in {0,1, 2,3} {
				\draw (\i * \spacing, 0) -- (\i * \spacing + \siteLength, 0);
			}
			\definecolor{darkgreen}{rgb}{0.0, 0.5, 0.0}
			\newcommand{\getcolor}[1]{%
				\ifnum#1=0 black\fi
				\ifnum#1=1 black\fi
				\ifnum#1=2 black\fi
				\ifnum#1=3 black\fi
				\ifnum#1=4 black\fi
				\ifnum#1=5 black\fi
			}
			\foreach \i [count=\j from 0] in {0,3} {
				\fill[fill=\getcolor{\j}] (\i * \spacing + \siteLength/2, 0.2) circle (\particleSize);
			}
			
		\end{tikzpicture}\right\}.
		\label{eq:x2l4krylov}
	\end{equation}
	An active bubble with $x=2,\ell=5$ would also have two states:
	\begin{equation}
		\left\{\begin{tikzpicture}[scale=0.8, transform shape]
			\def\siteLength{0.3} 
			\def\spacing{0.45}  
			\def\particleSize{2.5pt} 
			\def\adjust{0.03}
			\foreach \i in {0,1,2, 3,4} {
				\draw (\i * \spacing, 0) -- (\i * \spacing + \siteLength, 0);
			}
			\definecolor{darkgreen}{rgb}{0.0, 0.5, 0.0}
			\newcommand{\getcolor}[1]{%
				\ifnum#1=0 black\fi
				\ifnum#1=1 black\fi
				\ifnum#1=2 black\fi
				\ifnum#1=3 black\fi
				\ifnum#1=4 black\fi
				\ifnum#1=5 black\fi
			}
			\foreach \i [count=\j from 0] in {1,3} {
				\fill[fill=\getcolor{\j}] (\i * \spacing + \siteLength/2, 0.2) circle (\particleSize);
			}
		\end{tikzpicture}\ , \ \begin{tikzpicture}[scale=0.8, transform shape]
			\def\siteLength{0.3} 
			\def\spacing{0.45}  
			\def\particleSize{2.5pt} 
			\def\adjust{0.03}
			\foreach \i in {0,1, 2,3,4} {
				\draw (\i * \spacing, 0) -- (\i * \spacing + \siteLength, 0);
			}
			\definecolor{darkgreen}{rgb}{0.0, 0.5, 0.0}
			\newcommand{\getcolor}[1]{%
				\ifnum#1=0 black\fi
				\ifnum#1=1 black\fi
				\ifnum#1=2 black\fi
				\ifnum#1=3 black\fi
				\ifnum#1=4 black\fi
				\ifnum#1=5 black\fi
			}
			\foreach \i [count=\j from 0] in {0,4} {
				\fill[fill=\getcolor{\j}] (\i * \spacing + \siteLength/2, 0.2) circle (\particleSize);
			}
		\end{tikzpicture}\right\}.
		\label{eq:x2l5krylov}
	\end{equation}
	We refer to the different particle configurations within local Krylov sectors as ``active bubble configurations''.\\
	
	Similarly to the mean density of type-1 blockages and type-2 edges, we can lower bound the average densities of different types of active bubbles over an $N$ sector. We expect the distribution of bubbles in typical thermodynamically large states to reflect these averages by arguments similar to those made in \cref{app:SelfAveraging}. We denote the average density of occurrences of active bubbles containing $x$ particles in a system with on-site dimension $d$ as $A^{(d,k)}(x,N,L)$. We may expand $A^{(d,k)}(x,N,L)$ as a sum over $\ell$, where $\ell$ is the total number of sites a given active bubble occupies:
	\begin{equation}
		A^{(d,k)}(x,N,L) = \sum_{\ell=\ell_{d,\text{min}}(x,k)}^{\ell_{\text{max}}(x,k)} m^{(d,k)}(x,\ell) a^{(d,k)}(x,\ell,N,L).
		\label{eq:MasterD2ActiveBubble}
	\end{equation}
	In the above expression, the multiplicity $m^{(d,k)}(x,\ell)$ gives the total number of active bubble configurations compatible with a particular $x$, $\ell$, and $k$; the function $a^{(d,k)}(x,\ell,N,L)$ represents the average density of occurrences along the chain of any such active bubble configuration (we note that this density is particle-configuration independent); and $\ell_{d,\text{min}}(x,k)$ and $\ell_{\text{max}}(x,k)$ are, respectively, the smallest and largest number of sites compatible with an active bubble with $x$ particles for interaction range $k$. 
    As a simple example, we consider the case of $d=2$, $k=5$ and $x=2$. We see from the configurations in Eqs. \eqref{eq:x2l4krylov} and \eqref{eq:x2l5krylov} that
	\begin{equation}
		\begin{aligned}
			\ell_{2,\text{min}}(2,5)&=4,\, \ell_{\text{max}}(2,5)=5,\\
			m_2(2,4,5)&=m_2(2,5,5)=2. 
		\end{aligned}
	\end{equation}
    We may lower bound the function $a^{(2,k)}(x,\ell,N,L)$ analytically in the thermodynamic limit for $d=2,\infty$, similarly to how we lower bounded the densities of type-1 blockages and type-2 edges. We carry out the calculation explicitly in \cref{app:DSandP},  yielding the lower bounds
	\begin{align}
		\lim_{L\rightarrow\infty} a^{(2,k)}(x,\ell,\nu L,L)&\geq\nu^{x} (1-\nu)^{\ell+2} (1-\nu/\nu_c)^2,\label{eq:AB_d2}\\
        \lim_{L\rightarrow\infty}a^{(\infty,k)}(x,\ell,\nu L, L)&\geq \nu^{x} (1+\nu)^{-(x+\ell+2)}(1-\nu/\nu_c)^2 .\label{eq:AB_dinf}
	\end{align}
	Applying this to the $x=2,k=5,d=2$ example, we find
	\begin{equation}
		\begin{aligned}
			\lim_{L\rightarrow\infty}A^{(2,5)}(x=2,&\nu L, L)\\ &\geq 2  \nu^2 ((1-\nu)^6+(1-\nu)^7)(1-3\nu)^2.
		\end{aligned}
	\end{equation}
	In Appendix \ref{app:activebubbleDistLowerBounds}, we also work out $\lim_{L\rightarrow\infty}A^{(2,5)}(x,\nu L, L)$ for $x=3$ to $x=5$, as well as $\lim_{L\rightarrow\infty}A^{(\infty,7)}(x,\nu L, L)$ for $x=2$ to $x=6$. We plot the derived lower bounds in Fig. \ref{fig:IslB1Dist}(b-c), where we compare them to the true numerically computed densities. The $d=2$ numerical results were obtained by mapping out the entire Krylov subspaces of random initial states and using this to identify active bubbles, and the $d=\infty$ ones were obtained by mapping random initial states to their corresponding FES (see \cref{subsec:type1distributionNumerics} for details). 
    \begin{figure*}[!t]
		\centering
		\includegraphics[scale=0.36]{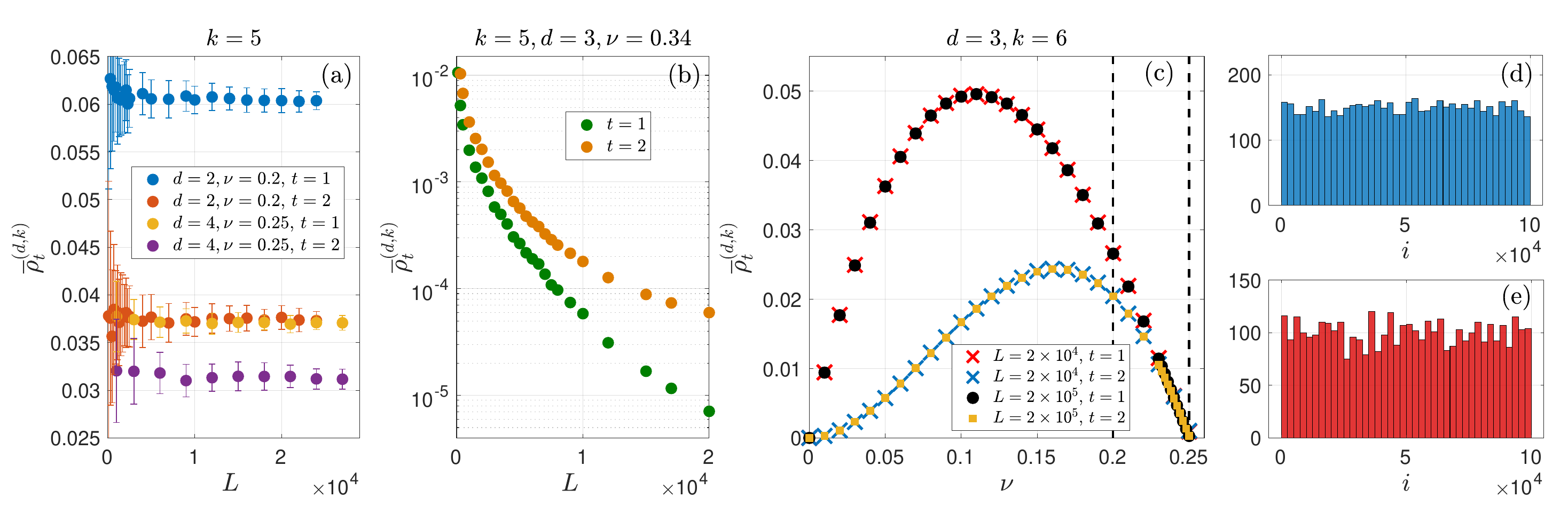}
		\caption{(a) Numerical estimates for $\overline{\rho}_t^{(d,k)}$ (from random sampling) as a function of $L$ for $100$ randomly generated states with local on-site dimensions $d=2,4$, for $k=5$ and fillings $\nu<\nu_c=1/3$. The error bars show the standard deviation associated with the set of $100$ random states at each $L$ value. (b) $\overline{\rho}_t^{(d,k)}$ as a function of $L$ for $1000$ randomly generated states at $\nu=0.34>1/3=\nu_c$, $k=5$ and $d=3$. (c) $\overline{\rho}_t^{(d,k)}$ for $d=3, \, k=6$ as a function of $\nu<\nu_c=0.25$ for $L=2 \times 10^{4}$ ($1000$ randomly generated states) and $L= 2 \times 10^{5}$ ($100$ randomly generated states). The leftmost vertical line coincides with $(k-1)^{-1}$ and the rightmost with $\nu_c=(k-2)^{-1}$. (d) Spatial distribution for $k=5$ of type-1 blockages in a single randomly drawn state with $L=10^5$, $d=2$, $\nu=0.2 <\nu_c$. Here, $i=1, \ldots, L$ labels the sites of the chain. The histogram shows the number of type-1 blockages that occurred in each bin, with bin width equal to 250. (e) Analogous histogram for $k=5$, $L=10^5$, $d=4$, $\nu=0.25 <\nu_c$.}
        \label{fig:FESdistribution}
    \end{figure*}
    We note a close match for $\nu\ll\nu_c$, and that the bound captures the majority of active bubbles up to intermediate values of $\nu/\nu_c \sim 1/2$, following which the analytic lower bound vanishes quadratically whereas the exact numerical density of bubbles vanishes linearly (in accordance with the results on critical scaling in \cref{sec:crscblockandab}). The vanishing density at the critical point reflects the fact that (typically) the whole chain represents a single active bubble at $\nu>\nu_c$, as we showed in \cref{sec:weakFragNumerics}. Therefore, the density of \emph{distinct} bubbles tends to zero in the thermodynamic limit.

    In addition to the active bubble density function $A^{(d,k)}(x,N,L)$ defined in \eqref{eq:MasterD2ActiveBubble}, we note that one can introduce an equally meaningful quantity $\tilde{A}^{(d,k)}(\ell,N,L)$, defined similarly to \eqref{eq:MasterD2ActiveBubble} but with the sum over $\ell$ replaced with a sum over the number of particles $x$. This yields the average density of occurrences of active bubbles of a given length $\ell$, irrespective of how many particles are involved in them. The methods described above to characterise $A^{(d,k)}(x,N,L)$ can be trivially generalised to address $\tilde{A}^{(d,k)}(\ell,N,L)$. The critical scaling of the latter is discussed in  \cref{sec:critscalandexp}.

	\subsection{Additional numerical results}
	\label{subsec:type1distributionNumerics}
	
	We next provide additional numerics supporting the results of the previous subsections and showing that several of the properties analytically demonstrated for systems with $d=2$ and $d=\infty$ also hold for other finite values of $d$. These numerics support the claim that typical states in the strongly fragmented phase feature an extensive number of evenly distributed type-1 blockages and type-2 edges. States with blockages then become vanishingly rare for $\nu>\nu_c$.\\

	We can numerically determine the location of type-1 and type-2 blockages associated with a given state by using the FES picture as defined in \cref{sec:FESpicsec}, i.e.~by mapping a state of a finite-$d$ system $S$ to its associated FES in the $\tilde{d}=\infty$ auxiliary system $\tilde{S}$. We can then directly identify the type-1 blockages and type-2 edges from the FES. An efficient algorithm for performing this mapping, which makes use of the properties of particle-connected (PC) strings (see \cref{tab:glossary}), is presented in \cref{app:FESalgorithm}.

	In Fig.~\ref{fig:FESdistribution}(a)-(c) we plot numerical results for the average densities $\overline{\rho}_1^{(d,k)}$ of type-1 blockages and $\overline{\rho}_2^{(d,k)}$ of type-2 edges, as defined in \cref{eq:avedensitywt}. These results were obtained by uniformly sampling states in a given $N$ sector and identifying blockages in each of them by the procedure outlined above. We note that these states were found to have vanishingly small intensive centres of mass $\nu_{x_0}$ as expected for typical states. In Fig.~\ref{fig:FESdistribution}(a), where the system is in its strongly fragmented phase given that $\nu<\nu_c$, we see that for various values of $d$ the average densities tend to a constant with increasing $L$. The corresponding standard deviations also appear to decrease with $L$, indicating the onset of self-averaging.  In Fig.~\ref{fig:FESdistribution}(b) type-1 blockages and type-2 edges are seen to become vanishingly rare with increasing $L$ for $\nu>\nu_c$, which is consistent with the onset of the weakly fragmented phase. In Fig.~\ref{fig:FESdistribution}(c), we fix $L$ and study $\overline{\rho}_t^{(d,k)}$ as a function of $\nu$. Below $\nu_c$ we find $\overline{\rho}_t^{(d,k)}$ to be a smooth positive function that tends to zero linearly as $\nu \to \nu_c$, in accordance with the critical scalings derived in \cref{sec:crscblockandab}. Given that above the critical filling the average density $\overline{\rho}_t^{(d,k)}$ is zero in the thermodynamic limit, the latter is seen to have a discontinuous first derivative with respect to $\nu$ at $\nu_c$, signaling a phase transition. We also note that there is no evidence of a discontinuity at $\nu=(k-1)^{-1}$, reinforcing our earlier result that even if it becomes possible for densities $\nu>(k-1)^{-1}$ to generate states without type-1 blockages, these states are extremely rare and do not lead to an additional thermodynamic phase transition. Note also that $\overline{\rho}_t^{(d,k)}$ has a peak at some finite value of $\nu$, and decreases both for fillings above and below it, as expected.  Indeed, even if more sites are involved in type-1 blockages at a given filling $\nu_1$ compared to any other filling $\nu_2>\nu_1$, the number of distinct type-1 blockages is strongly suppressed at small $\nu$ given that many different type-1 blockages merge, thus forming fewer but longer type-1 blockages. Similar reasoning can be applied to type-2 edges.
    In Figs.~\ref{fig:FESdistribution}(d)-(e) we present the spatial distribution of type-1 blockages along the chain for two randomly generated states at large $L$ with particle densities $\nu<\nu_c$ and $d=2,4$. The distribution of blockages is seen to be quite homogeneous along the entire chain. 
	This was found to be the case for all other randomly generated states tested, and the spatial distribution of type-2 edges was also found to be generally uniform. This numerically confirms the analytical predictions of \cref{app:SelfAveraging}, based on self-averaging.

   Finally, it is worth noticing that the equivalence of the critical filling for the weak-to-strong transition (see \cref{sec:WeakFragNumericsSubsec}) and the $\overline{\rho}_t^{(d,k)}$ transition leaves space for conjecturing that type-1 and type-2 blockages represent the main cause of dynamical disconnections also in finite-$d$ systems for $\nu<\nu_c$, with other types of blockages, intuitively expected to be rare at low fillings $\nu <\nu_c\le1$, playing a less central role.

\section{Critical scaling and exponents}
\label{sec:critscalandexp}
In this section, we focus on the critical scaling of various quantities. In \cref{sec:crscblockandab} we consider the scaling with $(\nu_c - \nu)$ of the total number of blockages, of the average length of active bubbles, and of a characteristic length scale $\xi$. In \cref{sec:blockandPCstratnuc} we numerically determine at the critical point the scaling with system size of the average density of blockages and of the average size of the largest blockage-free subregion. We discuss the implications of these results for transport at $\nu=\nu_c$. In \cref{sec:criticalScalingRatio} we present exact numerical results for small $L$ and $d=2,3$ of the dependence on $L$ of the ratio $r^{(d,k)}_{N,X}$ defined in \cref{eq:ratio1}, for $\nu<\nu_c$, $\nu=\nu_c$ and $\nu>\nu_c$. These represent additional evidence that the critical filling $\nu_c=(k-2)^{-1}$ does not depend on $d$. 

Our discussion implicitly addresses models that possess a weakly fragmented phase (and hence a transition), i.e.~we exclude the special cases $d=3,k=3$ and $d=2,k=4$, which are considered separately in \cref{sec:StrongFragModels}. 

\subsection{Critical scaling of blockages and active bubbles}
\label{sec:crscblockandab}

\begin{figure*}[th]
    \centering
    \includegraphics[scale=0.37]{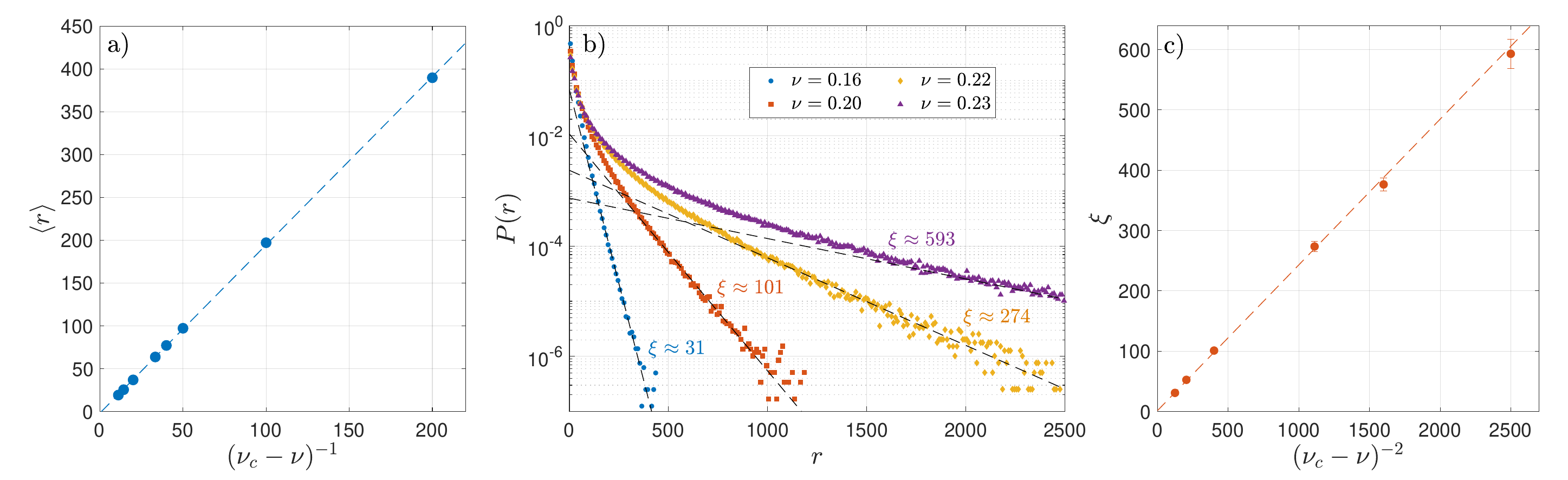}
    \caption{Numerical results for active regions length, for $d=2, k=6$ ($\nu_c = 0.25$). Data are obtained by applying the FES picture to 3000 randomly drawn initial states at a given $\nu$ for $L=10^5$. (a) Average length $\braket r$ as a function of $(\nu_c-\nu)^{-1}$. Dashed line is a linear fit. (b) Probability distribution $P(r)$ for a few values of $\nu$. To smooth out fluctuations, the lengths $r$ have been grouped into bins of length 10. From the linear fits on the exponentially decaying tails we extract the length scale $\xi$. (c) Length scale $\xi$ as a function of $(\nu_c-\nu)^{-2}$, and associated linear fit (dashed line). }
    \label{fig:xi_length_scale}
\end{figure*}

Consider a typical initial state at $\nu < \nu_c$ for any $d$, and apply the FES picture to it. Let $\rho_{1,\ell}$ designate the density of particles in the FES that have to their right a type-1 blockage of length $\ell$, i.e., exactly $\ell$ holes before the next particle. Similarly, let $\rho_2$ indicate the density of particles that have to their right a type-2 edge, i.e., exactly $k-2$ holes before the next particle; and $\rho_3$ the density of particles that have to their right exactly $k-3$ holes. Clearly, $\rho_2$ defined in this way coincides with the density of type-2 edges, where for notational simplicity we are dropping the explicit $d$- and $k$-dependence used in previous sections. Similarly, the sum over $\ell$ of $\rho_{1,\ell}$ coincides with the total density of type-1 blockages, i.e.~$\rho_1=\sum_\ell \rho_{1,\ell}$. 

Given that in the following we consider only intensive quantities, we can ignore the presence of possible stacks of particles at the boundaries. Indeed, for typical initial states such stacks can host at most a subextensive number of particles for $\nu < \nu_c$, \emph{cf.}~\cref{app:SelfAveraging}. We hence obtain in the limit of large $L$
\begin{align}
\label{eq:rho123_1}
\rho_3+\rho_2+\sum_{\ell=k-1}^{\ell_\text{max}} \rho_{1,\ell}  &= \nu \\
\label{eq:rho123_2}
(k-2)\rho_3+(k-1)\rho_2+\sum_{\ell=k-1}^{\ell_\text{max}} (\ell+1) \rho_{1,\ell} &= 1 \ .
\end{align}
In principle $\ell_\text{max}$ can grow with $L$ (subextensively). However, at any finite $\nu$ type-1 blockages with $\ell = \mathcal{O}(L^{\gamma})$ and $0 < \gamma < 1$ are extremely unlikely to arise. \footnote{Instead of the typicality class characterised by homogeneity over extensive length scales, one can consider smaller but still dominant classes of states that have local filling equal to the global $\nu$ over any region of length $\mathcal{O}(L^\gamma)$, for any $0<\gamma < 1$.} This means that for any finite $\nu$ there exists a large but finite cutoff length $\ell_\text{cut}=\mathcal{O}(L^0)$ such that $\sum_{\ell=\ell_\text{cut}+1}^{\ell_\text{max}}\ell \, \rho_{1,\ell} \ll 1 $. Combining \eqref{eq:rho123_1} and \eqref{eq:rho123_2}, and neglecting the vanishingly small error in replacing $\ell_\text{max}$ with $\ell_\text{cut}$, we obtain
\begin{equation}
\label{eq:rho23proptocrit}
\rho_2 + \sum_{\ell=k-1}^{\ell_\text{cut}} (\ell-k +3) \rho_{1,\ell} = \frac{1}{\nu_c}(\nu_c - \nu) \ ,
\end{equation}
where we have used $\nu_c = (k-2)^{-1}$. Note that the above equation is also valid when the densities $\rho_t$ (associated with a single typical state) are replaced by average densities over a given $N = \nu L$ sector, i.e.~those defined in \cref{eq:avedensitywt} from \cref{sec:StrongFragFurtherAnalysis}. From \eqref{eq:rho23proptocrit} we obtain
\begin{equation}
\label{eq:criticalboundrho12}
\frac{(\nu_c - \nu)}{\ell_\text{cut} \, \nu_c} < \rho_2 + \sum_{\ell=k-1}^{\ell_\text{cut}} \rho_{1,\ell} <\frac{(\nu_c - \nu)}{\nu_c} \ .
\end{equation}
Given the finiteness of $\ell_\text{cut}$ we obtain from \eqref{eq:criticalboundrho12} that $\rho_2 + \rho_1 \propto (\nu_c - \nu)$. It is our expectation that the previous scaling applies separately to the two terms. Indeed, the numerical results from \cref{subsec:type1distributionNumerics} indicate that these densities have comparable magnitude in the entire regime $\nu < \nu_c$. Hence
\begin{equation}
\label{eq:finalscalingrho12}
\rho_t \propto (\nu_c - \nu)^\beta \qquad \beta = 1, \quad \qquad t = 1, 2 \ .
\end{equation}
Notice that from the scaling of $\rho_1$ and the finiteness of $\ell_\text{cut}$ we also obtain $\rho_F \propto (\nu_c - \nu)^{\beta}$ with $\beta =1$, where $\rho_F$ is the average density of frozen sites in the chosen $\nu L$ sector at any given $d$. This is because at $d=\infty$ type-1 blockages are the only source of frozen sites \footnote{Note that a single particle enclosed by two type-1 blockages represents a frozen site too.}, while at finite $d$ they are expected to represent the largest source by far (see concluding remarks of \cref{subsec:type1distributionNumerics}). The value of $\beta=1$ agrees with the one suggested for the density of frozen sites in \cite{Morningstar_HSF} on the basis of numerical evidence and heuristic arguments, and with the linear scaling observed in \cref{fig:FESdistribution}(c) for $\nu$ close to $\nu_c$. \\

From an FES we can identify “active regions” as those subregions that are enclosed by two consecutive type-1 blockages and that contain more than 1 particle. For $d=\infty$, this definition coincides with that of active bubbles from \cref{subsec:ActiveBubbleDensityFunction}. For finite $d$ it does not, due to the possible presence of blockages beyond type-1 which give rise to frozen sites. Given the minor role that these are expected to play at the low fillings $\nu<\nu_c \le 1$, we expect that the scaling of the size of active regions also characterises the scaling associated with active bubbles, i.e.~the scaling with $(\nu_c-\nu)$ of the density $\tilde{A}^{(d,k)}(r,N,L)$ briefly introduced in \cref{subsec:ActiveBubbleDensityFunction} (where with $r$ we denote the size of the active bubble).

First, we notice that by definition the density $\rho_{a}=N_a/L$ of active regions coincides with $\rho_1 - \tilde{\rho}$, where $\tilde{\rho}$ is the density of occurrences of a single particle enclosed by two consecutive type-1 blockages (which hence constitutes a frozen site). Defining $\braket{r}=\sum_{i=1}^{N_a}r_i/N_a$ to be the average length of active regions, we can write
\begin{equation}
\label{eq:criticalavr}
1 = \sum_{\ell=k-1}^{\ell_\text{cut}} \ell \, \rho_{1,\ell} + \tilde{\rho} + (\rho_1 - \tilde{\rho}) \braket{r} \ .
\end{equation}
We notice that by definition $\tilde{\rho} \le \rho_1$. From the self-averaging argument of \cref{app:SelfAveraging} and the results in Figs.~\ref{fig:FESdistribution}(d)-(e) we know that type-1 blockages are quite uniformly distributed over the chain, which implies that $(\rho_1 - \tilde{\rho})$ must scale like $\rho_1$, i.e.~as $(\nu_c-\nu)$. Using this we obtain directly from \eqref{eq:criticalavr} that 
\begin{equation}
\label{eq:avrscaling}
    \braket{r} \propto (\nu_c - \nu)^{-\beta} \qquad \beta = 1 \ ,
\end{equation}
in agreement with the critical exponent suggested in \cite{Morningstar_HSF}. This scaling is also confirmed by the numerical results in \cref{fig:xi_length_scale}(a) for $d=2,k=6$, obtained by applying the FES picture algorithm of Appendix \ref{app:FESalgorithm} to many random initial states at different values of $\nu$, and identifying active regions from the FES obtained.

By resolving all the active regions we can also extract the probability distribution function $P(r)$ of a given active region having length $r$. This is expected to be directly proportional to $\tilde{A}^{(d,k)}(r,N,L)$ from \cref{subsec:ActiveBubbleDensityFunction}, which is associated with active bubbles. The results for a few values of $\nu$ are plotted in \cref{fig:xi_length_scale}(b) and show that $P(r)$ presents an exponential tail, from which we can extract a characteristic length scale $\xi$. As expected from the continuous nature of the phase transition, $\xi$ diverges for $\nu \to \nu_c$, reflecting the fact that the entire chain becomes a single active region beyond $\nu_c$, and also that extensively large active regions emerge exactly at $\nu=\nu_c$ (see next subsection). From \cref{fig:xi_length_scale}(c) we extract the critical scaling
\begin{equation}
    \xi \propto (\nu_c - \nu)^{-\gamma} \qquad \gamma = 2  \ . 
\end{equation}
This coincides with the exponent argued for in \cite{Morningstar_HSF}. We also note that Ref.~\cite{Pozderac_DinfHSF}, which considers a slightly different and simpler length scale $\xi$, derives for it analytically the same value $\gamma = 2$.

 \subsection{Density of blockages and transport at the critical point}
 \label{sec:blockandPCstratnuc}

  \begin{figure*}[th]
		\centering
		{\includegraphics[scale = 0.35]{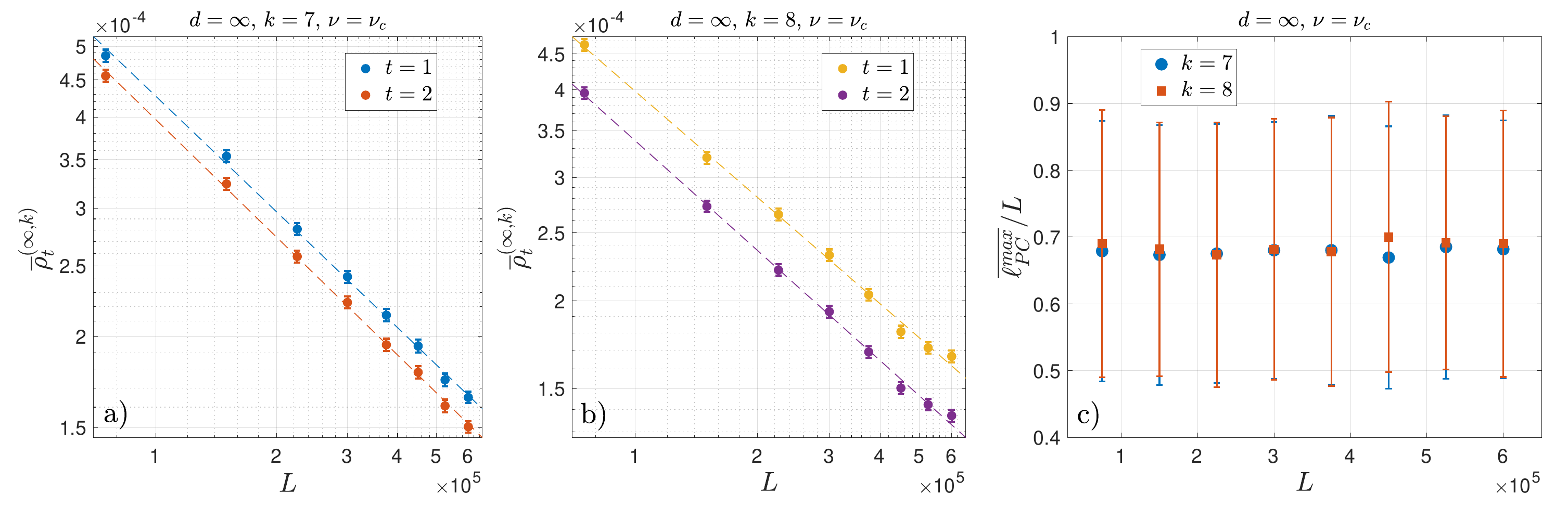}}
		\caption{Numerical results for the scaling with $L$ of several observables at $\nu_c$. (a) For $k=7$, the average densities of type-1 blockages and type-2 edges $\overline{\rho}^{(\infty,k)}_t$ (with $t=1,2$ respectively) are plotted. Averages are obtained by randomly sampling 500 states per data point, and error bars represent standard errors. The dashed lines are 2-parameter linear fits $\ln\overline{\rho}_t = -\alpha_t \ln L + \beta_t$, from which we obtain $\alpha_1 = 0.528 \pm 0.023$ and $\alpha_2 = 0.537 \pm 0.017$. (b) Analogous plots for $k=8$. The fitting slopes are $\alpha_1 =0.503  \pm 0.027$ and $\alpha_2 =0.521 \pm 0.024$. (c) Ratio of the average size of the largest particle-connected (PC) string $\overline{\ell^{\text{max}}_{\text{PC}}}$ over system size $L$. 500 states are averaged over per data point and the error bars indicate the full standard deviation so as to illustrate the spread in sizes.}
         \label{fig:CritScalingWithL}
\end{figure*}

Transport in the weakly fragmented phase of multipole-conserving systems has been established to be subdiffusive \cite{Morningstar_HSF, PhysRevResearch.2.033124, PhysRevLett.125.245303,PhysRevE.103.022142}. In the case of dipole conservation the characteristic space-time scaling (with dynamical exponent $z$) is expected to be given, in the entire weakly fragmented phase, by
\begin{equation}
\label{eq:subdiffscaling}
    x \propto t^{1/z} \qquad z=4 \ .
\end{equation}
In the strongly fragmented phase $(\nu<\nu_c)$ our results from \cref{sec:strongfrag,sec:StrongFragFurtherAnalysis} (see also \cref{sec:EEstrongfrag}) prove absence of transport over extensive length scales, which could be rephrased as $z \to \infty$. On the other end, the value of $z$ at the critical point $\nu=\nu_c$, if well-defined at all, is currently unknown. Numerical estimates from Ref.~\cite{Morningstar_HSF,Pozderac_DinfHSF} can roughly be placed within the bounds $5 \le z_c \le 10$, even though the authors acknowledge that the power-law form \eqref{eq:subdiffscaling} might be inappropriate at $\nu_c$. 

In this subsection, we make partial progress on uncovering the nature of transport at the critical point. In particular, we numerically verify that blockages, despite being strongly suppressed in number with respect to the case of $\nu<\nu_c$, are still present at $\nu=\nu_c$ and fracture the chain into dynamically disconnected subregions that cannot be coupled by transport. However, we find that typically the largest subregion devoid of blockages has extensive size. Within it, and within any other sufficiently large subregion devoid of blockages, transport can occur. \\

We have shown in \cref{eq:finalscalingrho12} that the average densities $\rho_t$ of type-1 blockages and type-2 edges vanish linearly as $\nu$ approaches $\nu_c$ \emph{in the thermodynamic limit}. We next consider how these two densities tend to zero with $L$ at $\nu_c$. Indeed, the fact that for large $L$ type-1 blockages and type-2 edges are vanishingly rare at the critical filling does not imply that they are absent, but merely indicates that their total number grows at most subextensively with $L$. We consider first the simpler case of $d=\infty$, and remark at the end of this subsection how our conclusions generalise to finite $d$. 

We numerically establish that the subextensive growth of the total number of blockages and edges is consistent with a polynomial decay in their respective densities $\rho_t$:
 \begin{equation}
 \label{eq:LmohRhot}
     \rho_t(L) = c\,  L^{-\alpha}\ \ \ \ \text{for }\nu=\nu_c,\ \ t=1,2 \ ,
 \end{equation}
where again the index $t=1,2$ distinguishes type-1 blockages from type-2 edges, $c$ is a constant and $\alpha \approx 1/2$. In \cref{fig:CritScalingWithL}(a) and (b) we plot these densities for several values of $L$ at $d=\infty$, $\nu= \nu_c$, $k=7,8$. These are obtained by mapping randomly drawn states within an $N=\nu_c L$ sector to their FES (via the efficient algorithm of \cref{app:FESalgorithm}) and identifying the location of blockages and edges from it, similarly to what was done in \cref{subsec:type1distributionNumerics}. To verify the functional form \eqref{eq:LmohRhot} we perform a 2-parameter fit $\ln\rho_t = -\alpha \ln L + \beta$ and find excellent agreement with the data. The four values of $\alpha$ are reported in \cref{fig:CritScalingWithL} and are all found to be close to $1/2$. \\

\begin{figure}
    \centering
    \includegraphics[scale=0.26]{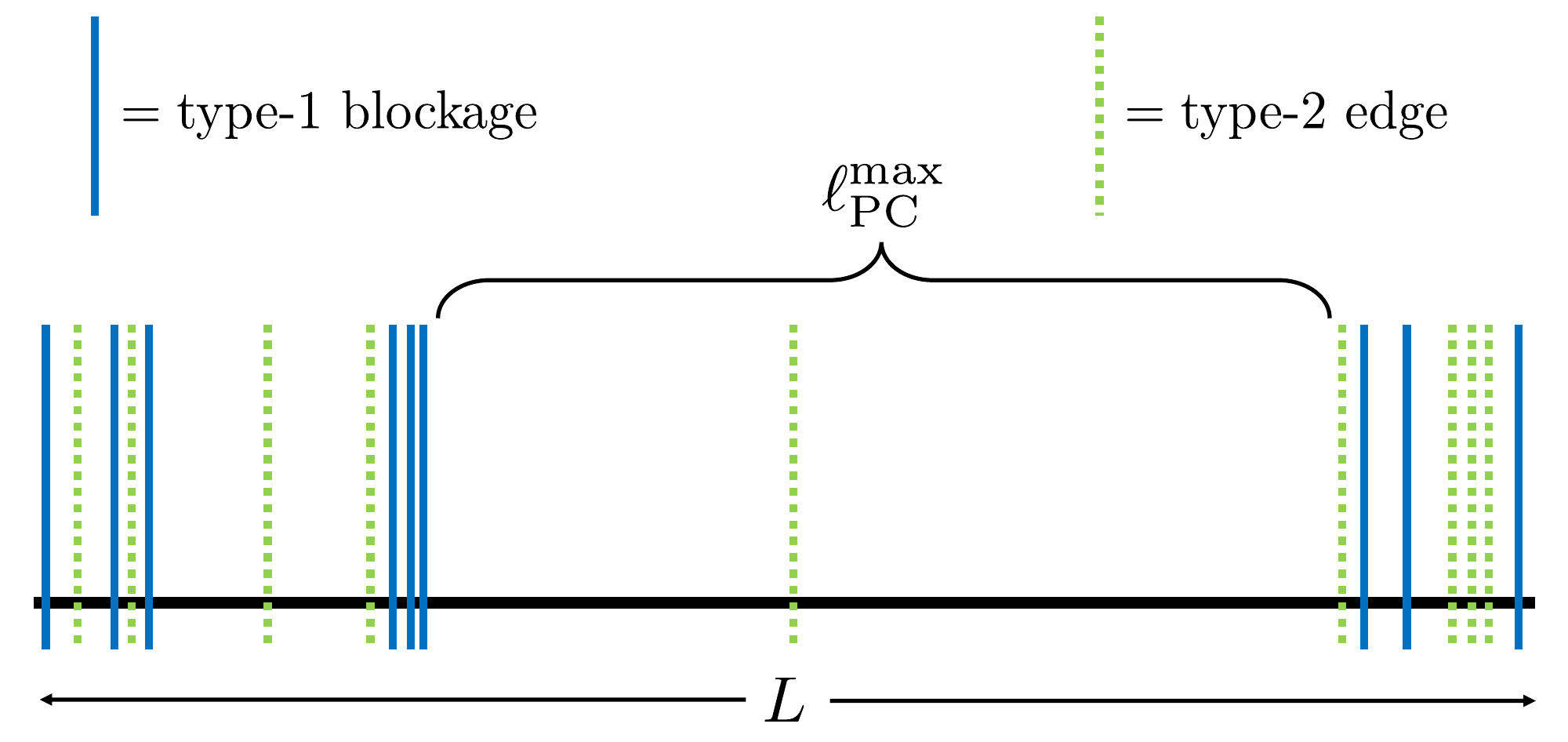}
    \caption{Schematic example of what a distribution of type-1 blockages and type-2 edges at the critical point $\nu=\nu_c$ could look like. Typically the distribution is nonuniform, giving rise to a wide range of sizes for blockage-free subregions, the largest of which has size $\ell^\text{max}_\text{PC}$.}
    \label{fig:schemdistblockcrit}
\end{figure}

A uniform distribution along the chain of these $\mathcal{O}(L^{1-\alpha})$ type-1 blockages and type-2 edges would result in a similar number of dynamically disconnected blockage-free subregions of $\mathcal{O}(L^{\alpha})$ size, within which transport can occur. However, we find that typically the distribution of blockages and edges is \emph{highly nonuniform} at $\nu_c$, see \cref{fig:schemdistblockcrit} for a schematic example. Such distributions give rise to a wide range of sizes for blockage-free subregions, which go from extensively large to finite in size (i.e.~$\mathcal{O}(L^0)$) ones. 
To show this, in \cref{fig:CritScalingWithL}(c) we plot the ratio $\overline{\ell^{\text{max}}_{\text{PC}}}/L$ for several values of $L$, with $\overline{\ell^{\text{max}}_{\text{PC}}}$ designating the average (again performed with respect to randomly drawn initial state in a fixed $N=\nu_c L$ sector) size of the \emph{largest} particle-connected (PC) string in the FES (see \cref{tab:glossary}). Indeed, for $d=\infty$, PC strings in an FES are in one-to-one correspondence with subregions of the chain that are devoid of blockages, and within which transport can occur without any hindrance. We note that the ratio is roughly fixed at a value $\overline{\ell^{\text{max}}_{\text{PC}}}/L\approx 0.7$, with a standard deviation of about 0.2 reflecting different possible arrangements of the blockages within the chain (see \cref{fig:schemdistblockcrit}). 

These results suggest the following observations.

\emph{i}. \ Typical subdivions of the chain into dynamically disconnected subregions are very different in the three regimes $\nu<\nu_c$, $\nu=\nu_c$ and $\nu>\nu_c$. For example, for $\nu<\nu_c$ the largest blockage-free subregion has at most subextensive size (see \cref{app:SelfAveraging}), while in the weakly fragmented phase typically the entire chain is devoid of blockages, as shown in \cref{sec:weakFragNumerics}. The critical case $\nu=\nu_c$ stands somewhere in between, with the largest blockage-free subregion having extensive size but still representing only a fraction of the whole chain. 

\emph{ii.} \ Due to the presence of $\mathcal{O}(L^{1-\alpha})$ blockages, space-time scalings like the one in \eqref{eq:subdiffscaling} are not well defined at the critical point, at least when transport is considered over the entire chain. On the other hand, transport within any blockage-free subregion is well-defined. Here, the determination of a characteristic space-time scaling is a problem equivalent to understanding transport within Krylov sectors that host a blockage-free FES with filling $\nu=\nu_c$. It is unclear whether the dynamical exponent $z$ associated with these instances of transport coincides with the value of $z=4$ that characterises the weakly fragmented phase. We leave this interesting question open for future work. 

We finally comment on how the previous results generalise to finite on-site dimensions $d$. Using the FES picture, similar numerical tests to those shown in \cref{fig:CritScalingWithL} were performed for randomly drawn states of $d=2$ and $d=3$ systems. These also yielded a $\rho_t(L) \propto L^{-\alpha}$ trend, with $\alpha \approx 1/2$, indicating subextensive growth in the number of type-1 blockages and type-2 edges. We also checked the ratio $\overline{\ell^{\text{max}}_{\text{PC}}}/L$, finding it to be also in these cases finite and approximately constant as a function of $L$. We remark that at finite $d$ the value of $\ell^{\text{max}}_{\text{PC}}$ obtained by the FES picture can in principle overestimate the size of the largest blockage-free subregion in the chain, due to the possible presence of finite-$d$ blockages. However, as already noticed several times, we expect the latter to be rare at the low fillings $\nu=\nu_c\le1$.

\subsection{Numerical results for the ratio of dimensions}
	\label{sec:criticalScalingRatio}
	\begin{figure*}[t]
		\centering
		{\includegraphics[scale = 0.36]{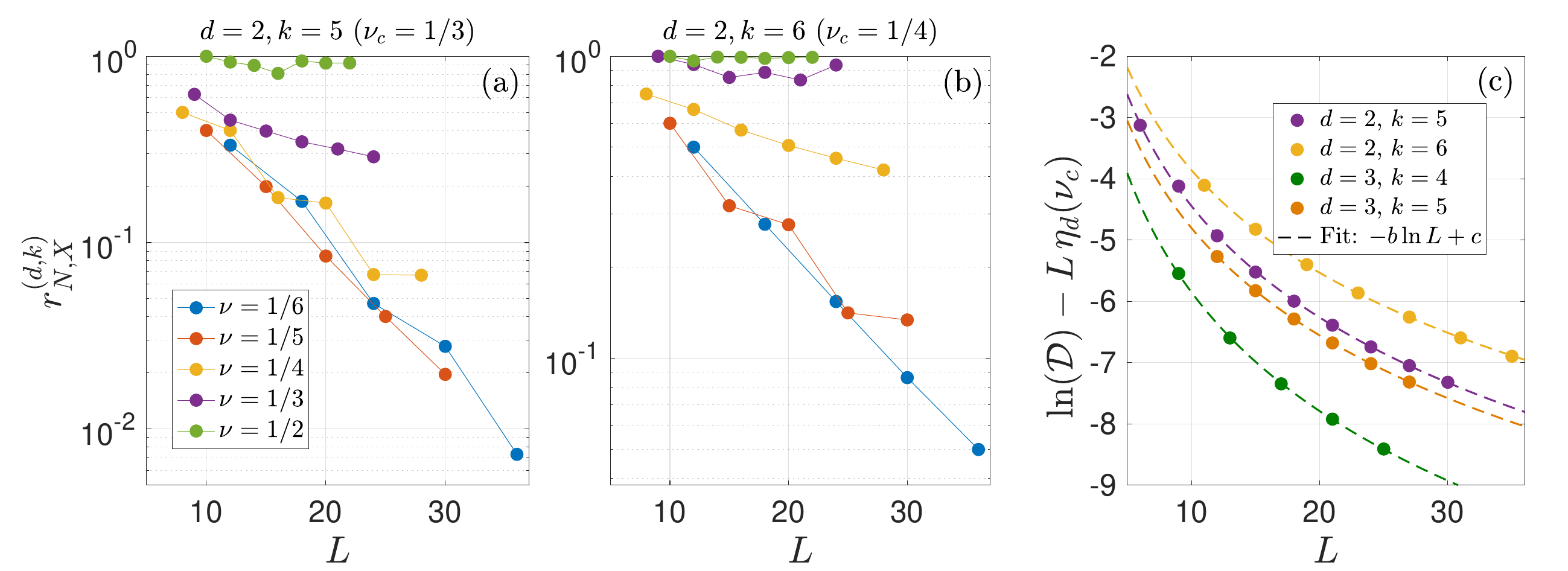}}
		\caption{(a)-(b) Exact values of $r^{(2,k)}_{N,X}$ for the largest symmetry sector $(N,X)$ at each given $N$ and locality $k=5,6$. Results are shown for a few sizes $L$ and fillings $\nu$. (c) Values of $\mathcal{D}(L)$ for $d=2,3$ and several values of $k$. The dashed lines are $2$-parameter fits to verify that, aside for polynomial corrections in $L$, $\mathcal{D}(L)$ has dependence on $L$ equal to $\exp(L\,\eta_d(\nu_c))$.\label{fig:RatioCritical}}
	\end{figure*}
     
	In \cref{sec:strongfrag,sec:StrongFragFurtherAnalysis,sec:weakFragNumerics} we presented strong analytic and numerical evidence of the existence of a universal $d$-independent strong-to-weak transition at the critical filling $\nu_c=(k-2)^{-1}$. Here we numerically determine for $d=2,3$ the dimension of the largest Krylov sector in typical symmetry sectors at different values of the filling $\nu$. This enables us to characterise the scaling with $L$ of the ratio $r^{(d,k)}_{N,X}$ as a function of $\nu$. In particular, following \cite{Pozderac_DinfHSF}, we aim to verify the presence of a critical scaling of $r^{(d,k)}_{N,X}$ at $\nu=\nu_c$.

    In Fig.~\ref{fig:RatioCritical}(a)-(b) we report exact numerical results for $r^{(2,k)}_{N,X}$ and $k=5,6$ as a function of a few small values of $L$ and a few values of $\nu$. Here $(N,X)$ is always chosen to be the largest symmetry sector for a given $N$, that is, we always have centre of mass located exactly at or just next to the middle of the chain. These values of $r^{(2,k)}_{N,X}$ were obtained by numerically partitioning the chosen $(N,X)$ sector into all its Krylov sectors, identifying the largest one, and dividing the dimension of the latter by the numerically determined $D_{N,X}^{(2)}$. We see how below $\nu_c$ the ratio $r^{(2,k)}_{N,X}$ appears to be decaying exponentially with $L$, while above $\nu_c$ it remains always close to $1$, consistently with a phase transition characterised by \cref{eq:ratio2} happening at $\nu_c$. 
    
    At the critical filling $\nu_c$ we find an intermediate behaviour, with $r^{(d,k)}_{N,X}$ still decaying to zero with $L$ but more slowly than for $\nu<\nu_c$. Following the results of \cite{Pozderac_DinfHSF} for $d=\infty$, we numerically verify that $r^{(d,k)}_{N,X}$ at $\nu=\nu_c$ decays to zero only polynomially with $L$, a feature expected to hold only at the critical point. To do so, based on the evidence from Section \ref{sec:weakFragNumerics}, we study the dimension of a Krylov sector that contains a blockage-free extended state, as the former is expected to represent the largest Krylov subspace within the $(N,X)$ sector it belongs to. In this way we avoid the need to partition the symmetry sector into all of its Krylov subsectors, and thus we manage to reach slightly higher values of $L$. We always choose $\nu \simeq \nu_c$ and $\nu_x = 1/2$ such that the blockage-free extended state has empty boundary sites and contains no sequence of $k-2$ holes, as pictured below  
	\begin{align*}
		\begin{tikzpicture}[scale=0.8, transform shape]
			\def\siteLength{0.3} 
			\def\spacing{0.45}  
			\def\particleSize{2.5pt} 
			\def\adjust{0.03}
			\foreach \i in {-1,0, 1, 3, 4, 5, 7, 8, 10, 11, 13, 14, 15, 17, 18,19} {
				\draw (\i * \spacing, 0) -- (\i * \spacing + \siteLength, 0);
			}
			\foreach \i in {0, 4, 8, 10, 14, 18} {
				\fill (\i * \spacing + \siteLength/2, 0.2) circle (\particleSize);
			}
			\node at (2 * \spacing+\siteLength/2+\adjust, 0.0) {\dots};
			\node at (6 * \spacing+\siteLength/2+\adjust, 0.0) {\dots};
			\node at (9 * \spacing+\siteLength/2+\adjust, 0.0) {\dots};
			\node at (12 * \spacing+\siteLength/2+\adjust, 0.0) {\dots};
			\node at (16 * \spacing+\siteLength/2+\adjust, 0.0) {\dots};
			\node at (2 * \spacing +\siteLength/2, -0.3) {\(k-3\)};
			\node at (6 * \spacing +\siteLength/2, -0.3) {\(k-3\)};
			\node at (12 * \spacing +\siteLength/2, -0.3) {\(k-3\)};
			\node at (16 * \spacing +\siteLength/2, -0.3) {\(k-3\)};
        \node at (20 * \spacing, 0) {.};
		\end{tikzpicture}
	\end{align*}
	We vary $L$ while retaining the general form pictured above, noting that because of the latter one obtains exactly $\nu \to \nu_c$ for $L \to \infty$. We call $\mathcal{D}(L)$ the dimension of the Krylov sector that contains such a blockage-free extended state. We compute $\mathcal{D}(L)$ by starting from the blockage-free extended state and applying a very high number of successive random gates to explore the entire Krylov sector. At each discrete time step, a gate is randomly selected from the set of all possible gates compatible with $d$, $k$ and conservation of $N$ and $X$. We then count the number of different configurations reached in the stochastic evolution up to a certain time and stop the search when this number remains constant for a sufficiently long time window. This allows us to reach slightly higher values of $L$ compared with those of Fig.~\ref{fig:RatioCritical}(a)-(b) and to study $d=3$ in addition to $d=2$. The data in Fig.~\ref{fig:RatioCritical}(c) suggest that the ratio of $\mathcal{D}$ over $D_{N,X}^{(d)}= \exp[L \eta_d(\nu) -2 \ln L + \mathcal{O}(L^0)]$ decays to zero polynomially with $L$, i.e.,  
	\begin{equation}
		\label{eq:criticalscalingnum}
		\ln \left[\frac{\mathcal{D}(L)}{D_{N,X}^{(d)}(L)}\right] = -\gamma \ln L + \lambda + o(L^0) \ ,
	\end{equation}
	where $\gamma, \lambda$ are $L$-independent constants and $\gamma>0$. Indeed, we perform a preliminary $3$-parameter fit of the form
	\begin{equation}
		\ln \mathcal{D}(L) = a L - b \ln L + c \ . 
	\end{equation}
	We always find that $a$ coincides with $\eta_d(\nu_c)$ from Section \ref{sec:sizesymmetrysec} up to a $1\%$ error in the worst case. Then we perform a second $2$-parameter fit
	\begin{equation}
		\ln \mathcal{D}(L) - L \eta_d(\nu_c) = - b \ln L + c \ , 
	\end{equation}
    shown in Fig.~\ref{fig:RatioCritical}(c), for which we always get $b > 2$. We find excellent agreement between the fit and the data, suggesting that the ratio $r^{(d,k)}_{N,X}$ has critical scaling at $\nu_c$ for arbitrary $d$.

\section{Entanglement entropy and quantum dynamics in the strongly fragmented phase}
	\label{sec:EEstrongfrag}

    In this section we present analytical and numerical results for the scaling of the entanglement entropy of eigenstates, and for quantum dynamics, at fillings $\nu<\nu_c$. As we have seen, the extensive presence of type-1 blockages proved in the previous sections implies that the chain is partitioned into independent subregions that cannot exchange quantum information. In \cref{sec:EEtype1} we show how this trivially implies that typical eigenstates in the strongly fragmented phase satisfy an area-law for entanglement entropy. In other words, the quantum many-body problem typically translates into an extensive set of few-body problems. 
    
    Contrary to frozen blockages, type-2 blockages allow the exchange of some amount of quantum information (see \cref{sec:EEtype2} below) between the two regions that they disconnect from the point of view of transport (of particles and dipole moment). This is because these regions can both interact with the active type-2 blockage, as described in \cref{sec:BlockagesFES2colour}. This is the mechanism at the basis of the “inverse quantum many-body scar” phenomenon we conjecture in \cref{sec:EEtype2}.
	
    \subsection{Type-1 blockages}
    \label{sec:EEtype1}
	
	In generic quantum chaotic systems, eigenstates at a finite energy density above the ground state possess volume-law entanglement entropy (EE)
    \begin{equation}
        S_A = -\text{Tr}_{A}(\rho_A \ln \rho_A) \propto |A| \ ,
    \end{equation}
    where $A$ is a subsystem of size $|A|$, $B$ its complement, $\rho_A = \text{Tr}_B(\ket{E_i}\bra{E_i})$ and $\ket{E_i}$ an eigenstate at energy $E_i$. 
    The EE density for $L\to\infty$ is expected to coincide with the thermodynamic entropy density at the corresponding energy, as computed from, e.g., the microcanonical or Gibbs ensembles \cite{RigolReview, Deutsch_2010, PhysRevE.86.010102, PhysRevE.87.042135, PhysRevLett.119.220603, PhysRevX.8.021026, PhysRevE.97.012140}. Volume law entanglement is also a characteristic feature of typical eigenstates at finite energy densities in integrable models \cite{PhysRevE.100.062134, PRXQuantum.3.030201}. 
    
    In this section we argue that in the class of models discussed in this work, the EE of typical eigenstates in $N$-sectors characterised by $\nu<\nu_c$ follows an \emph{area law}. 
	
	The bipartite EE of each eigenstate within the $i$-th Krylov sector of an $(N,X)$ sector can be upper bounded by $\ln \mathcal{D}_{i}^{(d,k)}(A)$, where $\mathcal{D}_{i}^{(d,k)}(A)$ is the dimension of the Krylov sector restricted to the subsystem $A$ of the bipartition \cite{Khemani_HSF,Moudgalya_2022,PhysRevLett.124.207602}, where we are assuming $\mathcal{D}_{i}^{(d,k)}(A)\le \mathcal{D}_{i}^{(d,k)}(B)$. In particular, this ensures that \emph{all} eigenstates of strongly fragmented $(N,X)$ families have EE density $s$ upper bounded in the thermodynamic limit by
	\begin{equation}
		\label{eq:sineq}
		s \le \lim_{L\to\infty}\frac{1}{L}\ln \mathcal{D}_\text{max}^{(d,k)} < \lim_{L\to\infty}\frac{1}{L}\ln D_{N,X}^{(d)} \  . 
	\end{equation}
	In quantum chaotic models, the rightmost term in the previous inequality would be expected to coincide with the EE density of highest-DoS eigenstates (i.e.~typical ones) in the chosen family of $(N,X)$ sectors (see works on generalisations of Page formula \cite{PhysRevLett.71.1291,PhysRevLett.77.1} to symmetry sectors \cite{PhysRevD.100.105010, PRXQuantum.3.030201}). Thus, the inequality in Eq.~(\ref{eq:sineq}) proves ergodicity breaking at the level of entanglement on the basis of the exponential suppression of the dimension of Krylov sectors compared to the dimension of symmetry sectors. However, given that usually $\mathcal{D}_\text{max}^{(d,k)}$ scales exponentially with $L$, the inequality in \eqref{eq:sineq} is, in principle, still compatible with volume-law EE scaling in the strongly fragmented phase. 
	It has been first remarked in Ref. \cite{Khemani_HSF} that further restrictions on the generation of entanglement can arise from local configurations that completely disconnect the regions of the chain to their left and right. In \cite{Khemani_HSF} these disconnections coincided with the “absolute blockages” discussed in Section \ref{sec:StrongFragModels}, that exist only in the very specific models with $d=2, k=4$ and $d=3,k=3$, which are strongly fragmented irrespective of the filling $\nu$. Given our definition of type-1 blockages from Section \ref{sec:FES}, we now generalise this idea to models with any $d$ and $k$. 
	
	Consider a Krylov sector $\mathcal{K}$ that contains a type-1 blockage somewhere along the chain. By the definition of Krylov sectors, the Hamiltonian leaves the vector space $\mathcal{K}$ invariant, i.e., $H \mathcal{K} \subseteq \mathcal{K}$. Call $H_\mathcal{K}$ the restriction of $H$ to $\mathcal{K}$. We can partition the chain by inserting a cut somewhere in the middle of the type-1 blockage, and obtain in this way two subregions $\alpha$ and $\beta$. By definition of type-1 blockages we have
	\begin{equation}
		\label{eq:HboldK}
		H_{\mathcal{K}} = H_\alpha \otimes \mathbb{I}_\beta + \mathbb{I}_\alpha \otimes H_\beta \ , 
	\end{equation}
	where $\mathbb{I}$ represents the identity operator and $H_\alpha$, $H_\beta$ are Hermitian operators. From \eqref{eq:HboldK} we see that the eigenstates of $H_\mathcal{K}$, which form a basis of $\mathcal{K}$, have the form
	\begin{equation}
		|n,m\rangle = |n\rangle_\alpha \otimes |m \rangle_\beta \ , 
	\end{equation}
	where $|n\rangle_\alpha$ and $|m\rangle_\beta$ are eigenstates of respectively $H_\alpha$ and $H_\beta$. More in general, if a Krylov sector contains several type-1 blockages which partition the system into $G$ regions $\alpha_i$, then all its eigenstates can be expressed as product states of active regions enclosed by the type-1 blockages
	\begin{equation}
		\label{eq:activeprodstate}
		|n_1, \ldots, n_G \rangle =  |n_1 \rangle_{\alpha_1} \otimes \ldots \otimes |n_G \rangle_{\alpha_G} \ . 
	\end{equation}
	From this it is elementary to verify that for any of these eigenstates:\\
	
	1. \ The bipartite EE associated with a single entanglement cut located within a type-1 blockage is exactly zero. \\
	
	2. \ The bipartite EE associated with a single entanglement cut that has to its left or right a type-1 blockage is upper bounded by $\ln \mathcal{D}^*$, where $\mathcal{D}^*$ is the dimension of the Krylov sector when it is restricted to the “active region” enclosed by the entanglement cut and the type-1 blockage. We notice that for an active region of size $\ell$,  $\ln \mathcal{D}^* \le \ell \ln d$.\\
	
	The previous results trivially generalise to the case in which the entanglement bipartition arises from two entanglement cuts. In \cref{sec:StrongFragFurtherAnalysis} we proved that typical Krylov sectors in any $N$-family with $\nu<\nu_c$ possess an extensive number of type-1 blockages and we showed that the latter are quite uniformly distributed along the chain. Given the implications of type-1 blockages for EE that we just established, we conclude that typical eigenstates in these families possess area-law EE, i.e.~the bipartite EE (with single entanglement cut) in these eigenstates does not grow with $L$ as the latter is increased to infinity. 

    More in general, the decomposition in \cref{eq:HboldK} and the finite density of uniformly distributed type-1 blockages imply that, in the strongly fragmented phase, eigenstate construction and quantum unitary dynamics \footnote{For example from a product state belonging to a strongly fragmented $(N,X)$ sector.} can be typically resolved into the independent action of extensively many ultralocal Hamiltonians on regions of $L$-independent size. In other words, the quantum many-body problem is transformed into a set of extensively many independent \emph{few-body} problems. The consequences of this range from eigenstate expectation values of local operators strongly deviating from the ETH prediction, to entanglement growth by quench dynamics (e.g.~starting from a simple product state) saturating to an area-law value, and to the possible presence of long-lived oscillations in the dynamics of local observables.
    
    We note that such conclusions do not hold if one considers eigenstates in weakly fragmented sectors, or dynamics starting from product states that belong to them, due to the absence of type-1 blockages in such cases.

	\subsection{Type-2 blockages and inverse quantum many-body scars}
	\label{sec:EEtype2}

    \begin{figure*}[th]
		\centering
		\includegraphics[scale=0.38]{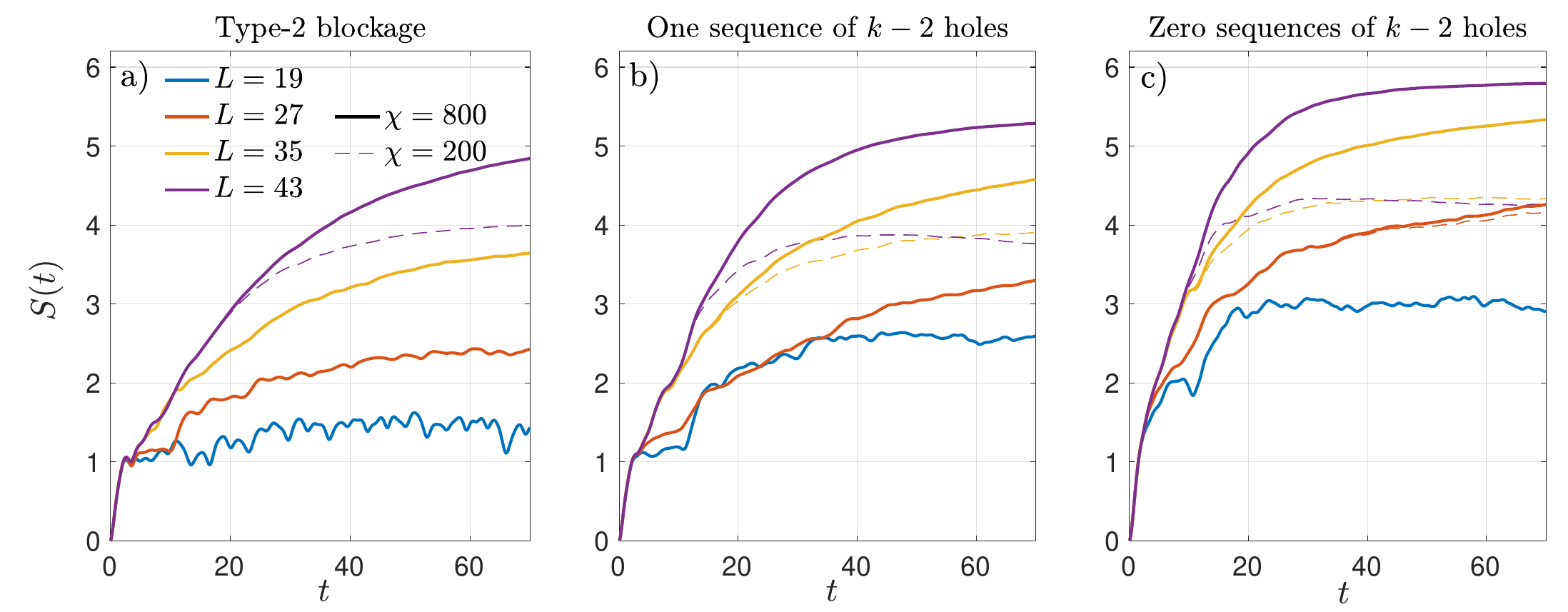}
		\caption{Time-dependence of the bipartite von Neumann entropy $S(t)$ under Hamiltonian dynamics computed using tDMRG. The Hamiltonian is characterised by $d=2$, $k=6$ and the initial product state is chosen among the set of configurations discussed in Section \ref{sec:EEtype2}, for various system sizes $L$. We report tDMRG results for bond dimensions $\chi=200$ and $800$. (a) Local disturbance coinciding with the simplest kind of type-2 blockage. (b) Local disturbance represented by a single sequence of $k-2$ holes (no blockage). (c) No disturbance. \label{fig:type2EE}}
	\end{figure*}
	We consider the effect of type-2 blockages on the EE, which turns out not to be as drastic as the one of type-1 blockages, leading us to conjecture the presence of an inverse quantum many-body scars phenomenon \cite{Srivatsa_IQMBS1,Chen_IQMBS2,Srivatsa_IQMBS3,Iversen_IQMBS4} in the family of models discussed in this work. 
	
	Given that type-2 blockages prevent transport of particles and dipole moment between the regions to their left and right, their presence along the chain typically reduces the value of $\ln \mathcal{D}_{i}^{(d,k)}(A)$ discussed in the previous subsection, thus lowering the maximal eigenstate EE reachable in principle through a bipartition into $A$ and $B$. This is however only a weak suppression of EE generation compared with the strong disconnecting effect of type-1 blockages, and it is compatible with volume-law EE scaling. One might wonder whether any type of strong suppression, similar to type-1 blockages, is produced also by active blockages. However, due to the possibility of the regions to the left and right of a type-2 blockage interacting with the particles involved in the blockage, these active blockages are not in principle expected to fully block the generation of entanglement. This is verified numerically in Fig.~\ref{fig:type2EE} using tDMRG \cite{10.21468/SciPostPhysCodeb.4} for $d=2$. We consider a chain of length $L$ in an FES, in which the boundary sites are occupied and particles are separated by exactly $k-3$ holes, with the only exception of a local disturbance in the middle of the chain. This is chosen to be a type-2 blockage formed by two type-2 edges separated by just one particle.  
	\begin{align*}
		\begin{tikzpicture}[scale=0.6, transform shape]
			\def\siteLength{0.3} 
			\def\spacing{0.45}  
			\def\particleSize{2.5pt} 
			\def\adjust{0.03}
			\foreach \i in {0, 1, 3, 4, 6, 7,9, 10, 11, 13, 14, 15, 17, 18,19,21,22,24,25,27,28} {
				\draw (\i * \spacing, 0) -- (\i * \spacing + \siteLength, 0);
			}
			\foreach \i in {0, 4, 6, 10, 14, 18,22,24,28} {
				\fill (\i * \spacing + \siteLength/2, 0.2) circle (\particleSize);
			}
			\node at (2 * \spacing+\siteLength/2+\adjust, 0.0) {\dots};
			\node at (5 * \spacing+\siteLength/2+\adjust, 0.0) {\dots};
			\node at (8 * \spacing+\siteLength/2+\adjust, 0.0) {\dots};
			\node at (12 * \spacing+\siteLength/2+\adjust, 0.0) {\dots};
			\node at (16 * \spacing+\siteLength/2+\adjust, 0.0) {\dots};
			\node at (20 * \spacing+\siteLength/2+\adjust, 0.0) {\dots};
			\node at (23 * \spacing+\siteLength/2+\adjust, 0.0) {\dots};
			\node at (26 * \spacing+\siteLength/2+\adjust, 0.0) {\dots};
			\node at (2 * \spacing +\siteLength/2, -0.3) {\(k-3\)};
			\node at (8 * \spacing +\siteLength/2, -0.3) {\(k-3\)};
			\node[text=blue] at (12 * \spacing +\siteLength/2, -0.3) {\(k-2\)};
			\node[text=blue] at (16 * \spacing +\siteLength/2, -0.3) {\(k-2\)};
			\node at (20 * \spacing +\siteLength/2, -0.3) {\(k-3\)};
			\node at (26 * \spacing +\siteLength/2, -0.3) {\(k-3\)};
            \node at (29 * \spacing, 0) {.};
		\end{tikzpicture}
	\end{align*}
	We focus on quantum Hamiltonian dynamics, with $H$ from Eq.~(\ref{eq:genericquantummodel}) characterised by $d=2$ and $k=6$, for which we choose the $h_j$'s to represent a gate-complete set as defined in Section \ref{sec:familyofmodels}. We place the entanglement cut somewhere within the type-2 blockage and calculate the von Neumann entropy $S(t)$ as a function of time. From Fig.~\ref{fig:type2EE}(a) we see that $S(t)$ is different from zero and becomes larger as $L$ is increased. This must be contrasted with the case in which the local disturbance is a type-1 blockage with entanglement cut within it, for which $S(t)=0 \ \forall \ t, \forall \ L$. In Fig.~\ref{fig:type2EE}(b) and \ref{fig:type2EE}(c) we produce the same kind of dynamics but replacing the type-2 blockage with, respectively, a single sequence of $k-2$ holes and no sequence of $k-2$ holes, as these do not give rise to blockages of any kind. These plots show a very similar evolution for $S(t)$ compared with the type-2 blockage case, confirming that the latter does not lead to any absolute hindrance to the generation of EE. Similar conclusions are reached if one places the entanglement cut just to the left or just to the right of the disturbance. 
	This supports the claim that the bipartite EE with entanglement cut lying in a subregion $A$ that hosts active blockages but no type-1 blockages can have a dependence on the size of $A$. 
	
	To conclude, in the entire phase $\nu<\nu_c$ there exist atypical Krylov sectors that contain extensively large “connected” subregions devoid of type-1 and type-2 blockages. These can give rise to eigenstates with volume-law EE, if the entanglement cuts lie within one of these subregions. Even more interesting are the rare Kyrlov sectors in the interval $(k-1)^{-1}\le \nu \le \nu_c$ which possess an extensive number of type-2 blockages uniformly distributed along the chain, but no type-1 blockages. The numerics above suggest that their eigenstates' EE scales beyond any area law, irrespective of the entanglement cut's location. It is an intriguing problem for future work to understand whether the EE of these types of eigenstates follows a volume law or not. Note that this phenomenon of rare eigenstates with beyond-area-law EE, embedded in a sea of area-law eigenstates, could be interpreted as inverse quantum many-body scars \cite{Srivatsa_IQMBS1,Chen_IQMBS2,Srivatsa_IQMBS3,Iversen_IQMBS4}. Much remains to be understood about the mechanisms responsible for inverse scars in general, and an analytic understanding of the entanglement structure of states only containing type-2 blockages could reveal a novel scenario in which they can emerge.

	\section{Models without a phase transition}
	\label{sec:StrongFragModels}
	
	We now examine special models whose families of typical symmetry sectors are always strongly fragmented, irrespective of the filling $\nu$. This occurs when the lower critical density $\nu_c$ and the upper critical density $d-1-\nu_c$ coincide, ensuring no phase transition occurs. This overlap implies the relation
	\begin{equation}
		\nu_c =  \frac{1}{k-2}=\frac{d-1}{2}
	\end{equation}
	The previous equation is satisfied only for the values $d=2,k=4$ and $d=3,k=3$. Models with these parameters are particularly remarkable in that they possess \emph{absolute blockages}, which are special local configurations of particles or holes that, whenever they occur in a state, necessarily lead to the presence of a sequence of $k-1$ frozen sites and hence of a blockage. In other words, they give rise to frozen blockages \emph{irrespective} of the particle content around them, unlike type-1 blockages that represent a global feature. These objects were first studied under the name of ``bottlenecks'' in Ref. \cite{Khemani_HSF}. It is easy to show combinatorially that local configurations of particles that represent absolute blockages appear, for any filling $\nu$, an extensive number of times in each typical state and are evenly distributed along the chain. We use this property to compute \emph{exact} active bubble densities for $d=2$, $k=4$. 
	
	\subsection{Absolute blockages}
	\label{sec:AbsoluteBlockages}

	We begin by demonstrating, using arguments similar to those in Ref. \cite{Khemani_HSF} (see also \cite{doi:10.1142/9789811231711_0009}), that for $d=2,k=4$ any sequence of 5 holes (or 5 particles by particle-hole symmetry) will necessarily contain at least $k-1=3$ frozen sites, making this an absolute blockage. In this model, only one hopping move is consistent with particle number and dipole moment conservation:
	\begin{align}
		\begin{tikzpicture}[scale=0.8, transform shape]
			\def\siteLength{0.3} 
			\def\spacing{0.45}  
			\def\particleSize{2.5pt} 
			\def\adjust{0.03}
			\foreach \i in {0,1,2, 3} {
				\draw (\i * \spacing, 0) -- (\i * \spacing + \siteLength, 0);
			}
			\definecolor{darkgreen}{rgb}{0.0, 0.5, 0.0}
			\newcommand{\getcolor}[1]{%
				\ifnum#1=0 black\fi
				\ifnum#1=1 black\fi
				\ifnum#1=2 black\fi
				\ifnum#1=3 black\fi
				\ifnum#1=4 black\fi
				\ifnum#1=5 black\fi
			}
			\foreach \i [count=\j from 0] in {1,2} {
				\fill[fill=\getcolor{\j}] (\i * \spacing + \siteLength/2, 0.2) circle (\particleSize);
			}
		\end{tikzpicture} \quad \longleftrightarrow \quad \begin{tikzpicture}[scale=0.8, transform shape]
			\def\siteLength{0.3} 
			\def\spacing{0.45}  
			\def\particleSize{2.5pt} 
			\def\adjust{0.03}
			\foreach \i in {0,1, 2,3} {
				\draw (\i * \spacing, 0) -- (\i * \spacing + \siteLength, 0);
			}
			\definecolor{darkgreen}{rgb}{0.0, 0.5, 0.0}
			\newcommand{\getcolor}[1]{%
				\ifnum#1=0 black\fi
				\ifnum#1=1 black\fi
				\ifnum#1=2 black\fi
				\ifnum#1=3 black\fi
				\ifnum#1=4 black\fi
				\ifnum#1=5 black\fi
			}
			\foreach \i [count=\j from 0] in {0,3} {
				\fill[fill=\getcolor{\j}] (\i * \spacing + \siteLength/2, 0.2) circle (\particleSize);
			}
		\end{tikzpicture}\ .
		\label{eq:k4HoppingMove}
	\end{align}
	This highly restricted moveset has important implications for the mobility of particles. Indeed, say we had a particle we wished to move two sites to the right via outward hops. For it to perform its first hop, it would need a particle to its left as in the LHS of Eq.~\eqref{eq:k4HoppingMove}. Colouring the right particle in red and the left one in blue, the first hop would look like
	\begin{align}
		\begin{tikzpicture}[scale=0.8, transform shape]
			\def\siteLength{0.3} 
			\def\spacing{0.45}  
			\def\particleSize{2.5pt} 
			\def\adjust{0.03}
			\foreach \i in {0,1,2, 3} {
				\draw (\i * \spacing, 0) -- (\i * \spacing + \siteLength, 0);
			}
			\definecolor{darkgreen}{rgb}{0.0, 0.5, 0.0}
			\newcommand{\getcolor}[1]{%
				\ifnum#1=0 blue\fi
				\ifnum#1=1 red\fi
				\ifnum#1=2 black\fi
				\ifnum#1=3 black\fi
				\ifnum#1=4 black\fi
				\ifnum#1=5 black\fi
			}
			\foreach \i [count=\j from 0] in {1,2} {
				\fill[fill=\getcolor{\j}] (\i * \spacing + \siteLength/2, 0.2) circle (\particleSize);
			}
		\end{tikzpicture} \quad \longrightarrow \quad \begin{tikzpicture}[scale=0.8, transform shape]
			\def\siteLength{0.3} 
			\def\spacing{0.45}  
			\def\particleSize{2.5pt} 
			\def\adjust{0.03}
			\foreach \i in {0,1, 2,3} {
				\draw (\i * \spacing, 0) -- (\i * \spacing + \siteLength, 0);
			}
			\definecolor{darkgreen}{rgb}{0.0, 0.5, 0.0}
			\newcommand{\getcolor}[1]{%
				\ifnum#1=0 blue\fi
				\ifnum#1=1 red\fi
				\ifnum#1=2 black\fi
				\ifnum#1=3 black\fi
				\ifnum#1=4 black\fi
				\ifnum#1=5 black\fi
			}
			\foreach \i [count=\j from 0] in {0,3} {
				\fill[fill=\getcolor{\j}] (\i * \spacing + \siteLength/2, 0.2) circle (\particleSize);
			}
		\end{tikzpicture}\ .
		\label{eq:k4HoppingMoveColour}
	\end{align}
	Hence, for the red particle to perform a second hop right, the blue particle to its left would have to travel two sites right via outward hops first. However, this requires the presence of a particle to the left of the blue particle, which must also move two sites to the right through outward hops, which in turn necessitates another particle to its left that can perform two outward hops to the right, and so on.
	
	Since this sequence never terminates,  no particle can travel two sites in a given direction via outward hops alone. Hence, for any sequence of 5 holes in a $d=2,k=4$ system, the middle 3 holes can never be occupied and are thus frozen irrespective of what surrounds them.

	For the case of $d=3,k=3$, there is again only one hopping move available, this time given by
	\begin{align}
		\begin{tikzpicture}[scale=0.8, transform shape]
			\def\siteLength{0.3} 
			\def\spacing{0.45}  
			\def\particleSize{2.5pt} 
			\def\adjust{0.03}
			\foreach \i in {0,1,2} {
				\draw (\i * \spacing, 0) -- (\i * \spacing + \siteLength, 0);
			}
			\definecolor{darkgreen}{rgb}{0.0, 0.5, 0.0}
			\newcommand{\getcolor}[1]{%
				\ifnum#1=0 black\fi
				\ifnum#1=1 black\fi
				\ifnum#1=2 black\fi
				\ifnum#1=3 black\fi
				\ifnum#1=4 black\fi
				\ifnum#1=5 black\fi
			}
			\foreach \i [count=\j from 0] in {0,2} {
				\fill[fill=\getcolor{\j}] (\i * \spacing + \siteLength/2, 0.2) circle (\particleSize);
			}
		\end{tikzpicture} \quad \longleftrightarrow \quad \begin{tikzpicture}[scale=0.8, transform shape]
			\def\siteLength{0.3} 
			\def\spacing{0.45}  
			\def\particleSize{2.5pt} 
			\def\adjust{0.03}
			\foreach \i in {0,1, 2} {
				\draw (\i * \spacing, 0) -- (\i * \spacing + \siteLength, 0);
			}
			\definecolor{darkgreen}{rgb}{0.0, 0.5, 0.0}
			\newcommand{\getcolor}[1]{%
				\ifnum#1=0 black\fi
				\ifnum#1=1 black\fi
				\ifnum#1=2 black\fi
				\ifnum#1=3 black\fi
				\ifnum#1=4 black\fi
				\ifnum#1=5 black\fi
			}
			\foreach \i [count=\j from 0] in {1} {
				\fill[fill=\getcolor{\j}] (\i * \spacing + \siteLength/2, 0.2) circle (\particleSize);
			}
			\fill (1*\spacing + \siteLength/2, 0.45) circle (\particleSize);
		\end{tikzpicture}\ .
		\label{eq:k3HoppingMove}
	\end{align}
	In this case, any sequence of 4 holes (or of 4 sites with 2 particles each) will constitute an absolute blockage, with $k-1=2$ frozen sites in the middle. The derivation can be found in Ref. \cite{Khemani_HSF}, and follows logic identical to that of the $d=2,k=4$ model.
	
	\subsection{Exact active bubble density}
	\label{sec:AB_D2K4}
	The $d=2,k=4$ model has a rich analytic structure and a number of properties that can be derived exactly. An in-depth study of the model's properties can be found in \cite{doi:10.1142/9789811231711_0009}. In this subsection, we show how the restricted particle mobility discussed in the previous subsection makes it possible  to exactly compute $ \lim_{L\rightarrow \infty} a^{(2,4)}(x,\ell,\nu L,L)$, which is the density function for individual active bubble configurations as defined in \cref{subsec:ActiveBubbleDensityFunction}. We begin by noting a few simple combinatorial results, which are rigorously derived in Appendix \ref{app:typicalstatesd2}. 
	
	Given $m$ neighbouring sites in an arbitrary state of a $d=2$ system with size $L$ and $N$ particles, the probability that those sites will contain any particular configuration of $x$ particles is given by
	\begin{equation}
		\left.\binom{L-m}{N-x}\right/\binom{L}{N},
		\label{eq:ProbFiniteL}
	\end{equation}
	where the numerator gives the number of ways to arrange the remaining $N-x$ particles in the system amongst the remaining $L-m$ sites, and the denominator gives the total number of possible states. We note this result is independent of the exact arrangement of the $x$ particles in the configuration. Thus, for example,  both particle configurations in \eqref{eq:k4HoppingMove} could occur on a given sequence of 4 sites with equal probability. Setting $N=\nu L$ and taking the thermodynamic limit, \cref{eq:ProbFiniteL} simplifies to
	\begin{equation}
		\nu^x (1-\nu)^{m-x},
		\label{eq:FiniteConfigProb}
	\end{equation}
	up to $\mathcal{O}(1/L)$ corrections. Hence, a given configuration of $x$ particles over $m$ sites will occur on average  $ L \nu^x (1-\nu)^{m-x}$ times in randomly chosen states (and so \cref{eq:FiniteConfigProb} is also the density of occurrences in a single typical state). This is proven in Appendix \ref{app:typicalstatesd2}, and has important consequences for the dynamics of the system. In particular, it implies the presence of a finite density of absolute blockages (and hence of frozen sites) in any typical state, in turn implying strong fragmentation at any filling $\nu$ by the results in Section \ref{sec:strongfrag}.

	We now use \eqref{eq:FiniteConfigProb} to compute the exact active bubble  density function $a^{(2,4)}(x,\ell,\nu L,L)$ in the thermodynamic limit. It is clear from the above discussion that this function must be of the general form
	\begin{equation*}
		\lim_{L\rightarrow \infty} a^{(2,4)}(x,\ell,\nu L,L) = \nu^x (1-\nu)^{\ell-x} (P_{\text{b}}(\nu))^2,
	\end{equation*}
	where the function $P_{\text{b}}(\nu)$ is the probability that the particles next to the active bubble, either to the right or the left, will be arranged so as to form a blockage consisting of at least 3 frozen sites. By the restricted mobility derived in Section \ref{sec:AbsoluteBlockages}, the simplest way for this to happen would be if there was a sequence of 4 holes or 4 particles next to the active bubble \footnote{Note that 4 holes or 4 particles, instead of 5, are already sufficient to obtain a blockage because by definition the particles (and holes) in an active bubble cannot propagate past the leftmost and rightmost sites of the bubble.}, resulting in the lower bound
	\begin{equation*}
		P_{\text{b}}(\nu)>\nu^4+(1-\nu)^4.
	\end{equation*}
	In \cref{app:D2K4pnh}, we make use of the restricted particle mobility to exactly enumerate all possible patterns of particles and holes which result in a blockage of 3 frozen sites, giving the final result
	\begin{equation}
		\begin{aligned}
			P_{\text{b}}&(\nu)=\\
			&(1-\nu)^3\frac{1 - 2 \nu + 2 \nu^2 - \nu^3 + \nu^4}{    1 - 2 \nu + 3 \nu^2 - 3 \nu^3 + 4 \nu^4 - 3 \nu^5 + \nu^6}\\
			&+ (\nu \leftrightarrow 1-\nu ) \ .
		\end{aligned}
	\end{equation}
	Using the result for $a^{(2,4)}$, we can directly calculate the exact bubble density function $  \lim_{L\rightarrow\infty}A^{(2,4)}(x,\nu L, L)$ by using \cref{eq:MasterD2ActiveBubble} introduced in \cref{subsec:ActiveBubbleDensityFunction}.

	\section{Conclusion and outlook}
	\label{sec:conclusion}
    
    In this work, we introduced a number of approaches for characterising the strongly and weakly Hilbert-space-fragmented phases of dipole-conserving 1D quantum systems. These allowed us to derive numerous rigorous  results concerning these phases and the ``freezing'' transition between them, as well as to develop efficient algorithms for numerically studying these models at large system sizes without relying on approximations. Our results lead to the interesting conclusion that dipole-conserving chains have the universal phase diagram depicted in \cref{fig:PhaseDiagram}, with a critical density of $\nu_c=(k-2)^{-1}$ irrespective of the on-site dimension $d$ or the precise details of the dynamics, as long as the natural assumption of gate-completeness (see \cref{sec:familyofmodels}) holds. 
    By mapping states in finite $d$ systems to their corresponding fully extended states (FESs), we were able to rigorously prove the persistence of blockages for particle fillings $\nu<\nu_c=(k-2)^{-1}$, leading to strong fragmentation in typical symmetry sectors. We were also able to analytically characterise the distribution of blockages and active bubbles in this phase, as well as to prove area-law entanglement scaling in typical eigenstates. However, the presence of active blockages leads us to conjecture the existence of an inverse quantum many-body scar phenomenon. For the weakly fragmented phase, by developing an efficient algorithm for mapping arbitrary initial states to their corresponding blockage-free extended states at $\nu>\nu_c$ and general $d$, we provided strong numerical evidence of weak fragmentation in typical symmetry sectors. Furthermore, we analytically derived some critical exponents characterising the transition and discussed some features of transport at the critical point. 

    We remark that while in this work we focused solely on open boundary conditions (OBC), which enable us to define precise structures like FESs and prove their uniqueness, we expect that the main physical conclusions drawn here (like the location of the phase-transition point or the existence of a finite density of blockages in the strongly fragmented phase) would also characterise systems with periodic boundary conditions (PBC), see e.g.~Ref.~\cite{Pozderac_DinfHSF}. It is also worth noting that OBC represents the natural choice in experimental setups on tilted lattices from which dipole-conservation emerges \cite{scherg2021observing,PhysRevLett.130.010201}. 
	
	Many of the topics explored in this paper warrant further analysis. It would be useful to better understand the nature and distribution of finite-$d$ blockages in typical states, as well as obtain expressions for the dimension of maximal Krylov sectors at any $d$. This could lead to a fully analytic proof that the critical density equals $\nu_c=(k-2)^{-1}$ at all $d$. An analytic understanding of how entanglement entropy scales in the strongly fragmented phase for states hosting only active blockages would also be valuable because it may prove the presence of inverse quantum many-body scars in the system. It would be very interesting to uncover whether a strong-to-weak phase transition can be driven by tuning the intensive centre of mass $\nu_x$ instead of the filling $\nu$, as we briefly discussed in \cref{appendix:detailsoffullproof}. Furthermore, despite our progress, a complete characterisation of transport at the critical point remains to be addressed. Throughout our work, we aimed to derive universal characteristics of a broad class of dipole-conserving models, and it would be valuable to next study these traits in the context of more specific physical realisations. This would allow for a more concrete assessment of the timescales associated with various dynamical phenomena, as well as the exploration of how violations of gate-completeness impact the location of the critical density, possibly shifting it to higher than $(k-2)^{-1}$ values. It would furthermore allow for a better understanding of how the results derived in this paper transfer to real-world systems, where fragmentation may only be approximate.
	
	There are also several natural extensions of our work. It is important to understand whether the methods introduced in this work could be generalised to the case of systems in higher spatial dimensions~\cite{Khemani_HSF,LehmannSalaPollmannRakovsky10.21468/SciPostPhys.14.6.140} as well as to systems with more complex symmetries~\cite{SalaLehmannRakovskyPollmann_ModulatedPhysRevLett.129.170601,PhysRevE.103.022142,Moudgalya_HSF} that allow Hilbert space fragmentation. Another interesting direction is to consider the inclusion of ordinary on-site and lattice symmetries, and to study the possible interplay of fragmentation with equilibrium orders mirroring the ``localisation protected quantum order" phenomenon of systems featuring many-body localisation (MBL)~\cite{LPQOPhysRevB.88.014206,Parameswaran_2018,PhysRevB.96.165136}. It would also be useful to clarify if and in what ways our results on the lattice are related to ergodicity breaking in continuum dipole-conserving systems~\cite{AP_NRFactons1_PhysRevB.109.054313,AP_NRFactons2_PhysRevB.110.024305,AP_NRFactons3_classicalfractonslocalchaos,AP_NRFactons4_phasespacefractons}.  
    Finally, we remark that the strong-to-weak transition driven by the filling represents an example of a transition driven by the degree of “connectivity” in the lattice. In this sense, it shares several qualitative features with the widely studied percolation phase transition \cite{stauffer2018introduction, PhysRevB.99.104206}. For example, both transitions are characterised by global order parameters that can be related to the size of “connected” clusters. It would be interesting to further explore analogies between the two.

	\acknowledgments
	We thank Calvin Pozderac, Brian Skinner and David Huse for encouraging us to address transport at the critical point within our formalism. We also thank Rahul Nandkishore for a suggestion regarding terminology, and the anonymous referees for useful feedback to improve the readability of the paper. 
    
	This work was supported by the Department of Physics of the University of Oxford (J.C.-H. and R.S.), the Fonds de recherche du Qu\'ebec -- Nature et technologies (J.C.-H.), the Engineering and Physical Sciences Research Council, Grant No. EP/T517811/1 (R.S.) and the European Union Horizon 2020 Research and Innovation Programme, Grant Agreement No. 804213-TMCS (A.P.).

	\appendix

	\section{Uniqueness of fully extended states}
	\label{appendix:uniqueFES}
	
In this appendix, as in the rest of this work, we restrict ourselves to systems with open boundary conditions and we assume gate-completeness. We define a configuration of $N$ particles in a system of length $L$ with $d=\infty$ and range-$k$ interactions  to correspond to a “fully extended state” (FES) if no pair of particles in the state can perform an outward hop, that is, there is no point in the state to which an outward hop gate can be applied. As already discussed in \cref{sec:BlockagesFES2colour}, it is clear that this implies that in the bulk of the system all particles must be separated by at least $k-3$ holes from their neighbours and no particles can be stacked; otherwise, it would still be possible for some particles to perform outward hops. The only exceptions to these rules are at the boundaries: an indefinite number of particles can be stacked on the leftmost and rightmost sites of the system, and there is no lower bound on the number of holes separating each of these stacks from the next closest particle.
	
	An important feature of open boundary systems is that starting from any initial configuration of particles, outward hops cannot be performed indefinitely: a sequence of outward hops must necessarily terminate after a finite number of moves, with the resulting configuration of particles corresponding to an FES. A simple proof of this fact is presented in Ref. \cite{Pozderac_DinfHSF} and makes use of the quadrupole moment: 
	\begin{equation}
		Q  =  \sum_{i=0}^{L-1} i^2\, n_i\ ,
	\end{equation}
	where as in \cref{eq:defNandX}, $n_i$ counts the number of particles on each site. It is easy to check that any outward hop performed by two particles will necessarily increase the value of $Q$ by at least 2. However, for finite $N$ and $L$, there is a clear upper bound on $Q$, given by the quadrupole moment of the particle configuration in which all particles are stacked on the right boundary site. This proves that, starting from any configuration, an FES must be reached in a finite number of moves.

	In this appendix, we prove that in a $d=\infty$ system with open boundary and any finite interaction range $k$, each particle configuration is dynamically connected to \emph{one and only one} FES. This also implies that for $d=\infty$ each Krylov sector contains a single unique FES, by which it can be characterised. This generalises the result of Ref.~\cite{Pozderac_DinfHSF}, where uniqueness was proven in the special case of $k=3$.
	
	We start by proving the following lemma:\\ 
	
	\emph{Assume that a Krylov sector contains more than one FES. Then it must be possible to map each FES in the sector to at least one other FES in the same sector by a sequence of only inward hops followed by a sequence of only outward hops.} \\
	
	In the following derivation, we for simplicity generically denote outward hop gates involving two particles both hopping one site away from each other with the letter $O$, and inward hop gates involving a pair of particles hopping one site towards each other with the letter $I$. We use subscripts to indicate the order in which we apply the gates. For example, a sequence involving a single inward hop followed by two outward hops and then two further inward hops would be designated by
\begin{equation}
    I_5 \, I_4 \, O_3 \, O_2 \, I_1 \ . 
\end{equation}
In the reasoning that will follow, the exact nature of each of these hops (i.e., where the hop is performed and the number of sites separating the hopping particles) is not important, hence permitting the above simplified notation.

Consider a given initial FES and assume that it is not the only FES within its Krylov sector. There must therefore exist a sequence of hops that will map it to a different FES. This sequence of moves necessarily starts with some number $n>0$ of consecutive inward hops $I_i$, with $ i=1,\ldots, n$. Let $O_{n+1}$ denote the first outward hop performed. The sequence of hops up to this point is hence given by
\begin{equation}
    \dots O_{n+1} I_{n} \ldots I_1 \ . 
\end{equation}
Before acting with the next hop in the sequence, we consider ``inserting the identity'': 
\begin{equation}
   \dots (\widetilde{I}_m \ldots \widetilde{I}_1)(\widetilde{O}_m \ldots \widetilde{O}_1) \, O_{n+1} I_{n} \ldots I_1 \ ,
\end{equation}
for some set $\{\widetilde{O}_i\}_{i=1}^m$ of outward hops with $\widetilde{I}_{m-i+1}=\widetilde{O}_i^{-1} \ \forall \ i$. As we established, it is always possible to reach an FES from a given configuration of particles by applying a finite number of consecutive outward hops, and so we may choose the operators $\{\widetilde{O}_i\}_{i=1}^m$ such that the state after the last outward hop $\widetilde{O}_m$ in the sequence is applied is an FES. Now, if the FES obtained after the application of $\widetilde{O}_m$ is different from the initial one, we have achieved the desired result, since the sequence $(\widetilde{O}_m \ldots \widetilde{O}_1) O_{n+1} I_{n} \ldots I_1$ maps from our initial FES to a different FES by using a sequence of purely inward then purely outward hops. If, on the other hand, the resulting FES is identical to the starting one, we have learned that the subsequence  $O_{n+1} I_n \ldots I_1$ of inward and outward hops can be replaced by the sequence of only inward hops $\widetilde{I}_m \ldots \widetilde{I}_1$. Iterating this reasoning for every further outward hop $O_i$ appearing in the overall hopping sequence, we see that at some point we will arrive at a sequence of purely inward and purely outward hops connecting our initial FES to a different FES, either by virtue of finding such a sequence of inward and outward hops at some intermediate step, or by virtue of successfully changing the original sequence of mixed inward and outward hops into an equivalent sequence of purely  inward hops followed by purely outward hops.\\
	
	We now prove the main result of this appendix, i.e., the uniqueness of FESs: \\
	
	\emph{Each Krylov sector of a $d=\infty$ system with local range-$k$ interactions and open boundaries contains a unique FES.}\\
	
	We proceed inductively. The cases of $1$ and $2$ particles are easy to verify. Assume that uniqueness holds up to FESs containing $N-1$ particles and consider an FES composed of $N$ particles. We shall prove that there is no sequence of purely inward hops followed by purely outward hops that maps this FES to a different FES. Thus, uniqueness will follow as a direct consequence of our previous lemma. In what follows, we refer to a sequence of purely inward then purely outward hops as an $O-I$ sequence. We now explore different structures of FESs on a case-by-case basis. 

\vspace{1em}
Case 1. \ Assume that in the FES there is at least one pair of particles separated by a sequence of $k-1$ or more empty sites:
    \begin{equation}
    \begin{tikzpicture}[scale=0.8, transform shape]
    \def\siteLength{0.3} 
    \def\spacing{0.45}  
    \def\particleSize{2.5pt} 
    \def\adjust{0.03}
        \foreach \i in {0, 1, 3, 4} {
            \draw (\i * \spacing, 0) -- (\i * \spacing + \siteLength, 0);
        }
        \foreach \i in {0, 4} {
            \fill (\i * \spacing + \siteLength/2, 0.2) circle (\particleSize);
        }
        \node at (-1 * \spacing+\siteLength/2+\adjust, 0.15) {\dots};
        \node at (2 * \spacing+\siteLength/2+\adjust, 0.0) {\dots};
        \node at (5 * \spacing+\siteLength/2+\adjust, 0.15) {\dots};
        \node at (2 * \spacing +\siteLength/2, -0.3) {\(\ge k-1\)};
        \node at (6 * \spacing, 0) {.};
    \end{tikzpicture}
    \end{equation}
    We apply an $O-I$ sequence with the aim of obtaining a different FES. During the application of the inward hops, the particles to the right and left of the separation of $k-1$ holes cannot interact with each other since the range-$k$ gates can only act on particles separated by $k-2$ sites or fewer. Hence, the particles on each side of the $k-1$ empty sites form in their own right independent FESs, and so by our inductive hypothesis, the subsequent sequence of outward hops can only map the particles to the left and right of the $k-1$ holes back to their initial positions. Hence, the original FES is re-obtained.
    \vspace{1em}
    
Case 2. \ Assume that in the FES there is no pair of particles separated by $k-1$ or more holes, but there are two pairs or more separated by $k-2$ holes. Assume first that two of these pairs overlap.
\begin{align}
\begin{tikzpicture}[scale=0.8, transform shape]
    \def\siteLength{0.3} 
    \def\spacing{0.45}  
    \def\particleSize{2.5pt} 
    \def\adjust{0.03}
    \foreach \i in {0, 1, 3, 4, 5, 7, 8} {
        \draw (\i * \spacing, 0) -- (\i * \spacing + \siteLength, 0);
    }
    \foreach \i in {0, 4, 8} {
        \fill (\i * \spacing + \siteLength/2, 0.2) circle (\particleSize);
    }
    \node at (-1 * \spacing+\siteLength/2+\adjust, 0.15) {\dots};
    \node at (2 * \spacing+\siteLength/2+\adjust, 0.0) {\dots};
    \node at (6 * \spacing+\siteLength/2+\adjust, 0.0) {\dots};
    \node at (9 * \spacing+\siteLength/2+\adjust, 0.15) {\dots};
    \node at (2 * \spacing +\siteLength/2, -0.3) {\(k-2\)};
    \node at (6 * \spacing +\siteLength/2, -0.3) {\(k-2\)};
    \node at (10 * \spacing, 0) {.};
\end{tikzpicture}
\label{eq:Case2FESV1}
\end{align}
If the particle in the middle does not move during the inward hops of the $O-I$ sequence, then the particles to its left cannot interact with the particles to its right, and by induction the only FES that can be reached is the original one via the same logic as in case 1. If the particle in the middle moves as part of the inward hops, then in performing an inward hop with one of its neighbouring particles, it will necessarily create a spacing of at least $k-1$ holes between itself and its other neighbouring particle. For example, if the middle particle performed an inward hop with the particle to its left, we would then have the configuration
    \begin{align}
\begin{tikzpicture}[scale=0.8, transform shape]
    \def\siteLength{0.3} 
    \def\spacing{0.45}  
    \def\particleSize{2.5pt} 
    \def\adjust{0.03}
    \foreach \i in {0, 1, 3, 4, 5, 7, 8} {
        \draw (\i * \spacing, 0) -- (\i * \spacing + \siteLength, 0);
    }
    \foreach \i in {1, 3, 8} {
        \fill (\i * \spacing + \siteLength/2, 0.2) circle (\particleSize);
    }
    \node at (-1 * \spacing+\siteLength/2+\adjust, 0.15) {\dots};
    \node at (2 * \spacing+\siteLength/2+\adjust, 0.0) {\dots};
    \node at (6 * \spacing+\siteLength/2+\adjust, 0.0) {\dots};
    \node at (9 * \spacing+\siteLength/2+\adjust, 0.15) {\dots};
    \node at (2 * \spacing +\siteLength/2, -0.3) {\(k-4\)};
    \node at (6 * \spacing +\siteLength/2, -0.3) {\(k-1\)};
    \node at (10 * \spacing, 0) {.};
\end{tikzpicture}
\end{align}
This has placed the middle particle out of range of the particle to its right, and so the two sides cannot interact during the inward hop sequence.  If on the other hand the middle particle had originally hopped to the right, we would have
    \begin{align}
\begin{tikzpicture}[scale=0.8, transform shape]
    \def\siteLength{0.3} 
    \def\spacing{0.45}  
    \def\particleSize{2.5pt} 
    \def\adjust{0.03}
    \foreach \i in {0, 1, 3, 4, 5, 7, 8} {
        \draw (\i * \spacing, 0) -- (\i * \spacing + \siteLength, 0);
    }
    \foreach \i in {0, 5, 7} {
        \fill (\i * \spacing + \siteLength/2, 0.2) circle (\particleSize);
    }
    \node at (-1 * \spacing+\siteLength/2+\adjust, 0.15) {\dots};
    \node at (2 * \spacing+\siteLength/2+\adjust, 0.0) {\dots};
    \node at (6 * \spacing+\siteLength/2+\adjust, 0.0) {\dots};
    \node at (9 * \spacing+\siteLength/2+\adjust, 0.15) {\dots};
    \node at (2 * \spacing +\siteLength/2, -0.3) {\(k-1\)};
    \node at (6 * \spacing +\siteLength/2, -0.3) {\(k-4\)};
    \node at (10 * \spacing, 0) {,};
\end{tikzpicture}
\end{align}
once again disconnecting the two regions. This again splits the state into disconnected left and right regions during the $O-I$ sequence, and we may use our inductive hypothesis to see that the same FES will be retrieved after the outward hops have been applied.

We next assume that there are no two sequences of $k-2$ holes next to each other in the FES, i.e., there is at least one sequence of $k-3$ holes separating each pair of sequences of $k-2$ holes. Consider then the two sequences of $k-2$ holes that are closest to each other: 
\begin{align}
    \begin{tikzpicture}[scale=0.8, transform shape]
    \def\siteLength{0.3} 
    \def\spacing{0.45}  
    \def\particleSize{2.5pt} 
    \def\adjust{0.03}
    \foreach \i in {0, 1, 3, 4, 5, 7, 8, 10, 11, 13, 14, 15, 17, 18} {
        \draw (\i * \spacing, 0) -- (\i * \spacing + \siteLength, 0);
    }
    \foreach \i in {0, 4, 8, 10, 14, 18} {
        \fill (\i * \spacing + \siteLength/2, 0.2) circle (\particleSize);
    }
    \node at (-1 * \spacing+\siteLength/2+\adjust, 0.15) {\dots};
    \node at (2 * \spacing+\siteLength/2+\adjust, 0.0) {\dots};
    \node at (6 * \spacing+\siteLength/2+\adjust, 0.0) {\dots};
    \node at (9 * \spacing+\siteLength/2+\adjust, 0.0) {\dots};
    \node at (12 * \spacing+\siteLength/2+\adjust, 0.0) {\dots};
    \node at (16 * \spacing+\siteLength/2+\adjust, 0.0) {\dots};
    \node at (19 * \spacing+\siteLength/2+\adjust, 0.15) {\dots};
    \node at (2 * \spacing +\siteLength/2, -0.3) {\(k-2\)};
    \node at (6 * \spacing +\siteLength/2, -0.3) {\(k-3\)};
    \node at (12 * \spacing +\siteLength/2, -0.3) {\(k-3\)};
    \node at (16 * \spacing +\siteLength/2, -0.3) {\(k-2\)};
    \node at (20 * \spacing, 0) {.};
\end{tikzpicture}
\label{eq:Case2FESV2}
\end{align}
If, during the inward hops of the $O-I$ sequence, any of the particles between the two sets of $k-2$ holes remains immobile, then we again have by our induction hypothesis that the original FES is retrieved after the outward hops. Assume that this is not the case, i.e., all particles between the two  sets of $k-2$ holes perform an inward hop at some point. Consider  the second particle from the left in \cref{eq:Case2FESV2}. If this particle performs an inward hop with the particle to its right, this will necessarily separate it by at least $k-1$ holes from the particle on its left:
\begin{align}
    \begin{tikzpicture}[scale=0.8, transform shape]
    \def\siteLength{0.3} 
    \def\spacing{0.45}  
    \def\particleSize{2.5pt} 
    \def\adjust{0.03}
    \foreach \i in {0, 1, 3, 4, 5, 7, 8, 10, 11, 13, 14, 15, 17, 18} {
        \draw (\i * \spacing, 0) -- (\i * \spacing + \siteLength, 0);
    }
    \foreach \i in {0, 5, 7, 10, 14, 18} {
        \fill (\i * \spacing + \siteLength/2, 0.2) circle (\particleSize);
    }
    \node at (-1 * \spacing+\siteLength/2+\adjust, 0.15) {\dots};
    \node at (2 * \spacing+\siteLength/2+\adjust, 0.0) {\dots};
    \node at (6 * \spacing+\siteLength/2+\adjust, 0.0) {\dots};
    \node at (9 * \spacing+\siteLength/2+\adjust, 0.0) {\dots};
    \node at (12 * \spacing+\siteLength/2+\adjust, 0.0) {\dots};
    \node at (16 * \spacing+\siteLength/2+\adjust, 0.0) {\dots};
    \node at (19 * \spacing+\siteLength/2+\adjust, 0.15) {\dots};
    \node at (2.5 * \spacing +\siteLength/2, -0.3) {\(k-1\)};
    \node at (6 * \spacing +\siteLength/2, -0.3) {\(k-5\)};
    \node at (12 * \spacing +\siteLength/2, -0.3) {\(k-3\)};
    \node at (16 * \spacing +\siteLength/2, -0.3) {\(k-2\)};
    \node at (20 * \spacing, 0) {.};
\end{tikzpicture}
\end{align}
Hence the inductive hypothesis can be applied again to show the $O-I$ sequence must yield the same FES. If this particle performs a hop instead with the particle to its left, this will create a spacing of $k-2$ or more holes between it and the particle to its right:
\begin{align}
    \begin{tikzpicture}[scale=0.8, transform shape]
    \def\siteLength{0.3} 
    \def\spacing{0.45}  
    \def\particleSize{2.5pt} 
    \def\adjust{0.03}
    \foreach \i in {0, 1, 3, 4, 5, 7, 8, 10, 11, 13, 14, 15, 17, 18} {
        \draw (\i * \spacing, 0) -- (\i * \spacing + \siteLength, 0);
    }
    \foreach \i in {1, 3, 8, 10, 14, 18} {
        \fill (\i * \spacing + \siteLength/2, 0.2) circle (\particleSize);
    }
    \node at (-1 * \spacing+\siteLength/2+\adjust, 0.15) {\dots};
    \node at (2 * \spacing+\siteLength/2+\adjust, 0.0) {\dots};
    \node at (6 * \spacing+\siteLength/2+\adjust, 0.0) {\dots};
    \node at (9 * \spacing+\siteLength/2+\adjust, 0.0) {\dots};
    \node at (12 * \spacing+\siteLength/2+\adjust, 0.0) {\dots};
    \node at (16 * \spacing+\siteLength/2+\adjust, 0.0) {\dots};
    \node at (19 * \spacing+\siteLength/2+\adjust, 0.15) {\dots};
    \node at (2 * \spacing +\siteLength/2, -0.3) {\(k-4\)};
    \node at (5.5 * \spacing +\siteLength/2, -0.3) {\(k-2\)};
    \node at (12 * \spacing +\siteLength/2, -0.3) {\(k-3\)};
    \node at (16 * \spacing +\siteLength/2, -0.3) {\(k-2\)};
    \node at (20 * \spacing, 0) {.};
\end{tikzpicture}
\end{align}
If the new spacing exactly equals $k-2$ holes, then once more, unless the two particles on either side of this new sequence of $k-2$ holes perform an inward hop, the regions on either side of them will be dynamically disconnected and the inductive hypothesis can be applied. A hop between the two particles, however, will again necessarily create a new spacing of at least $k-2$ holes to its right.

By iterating the above analysis, we see that at each step we reduce the distance between the sequence of $k-2$ holes on the right of the configuration in Eq. \eqref{eq:Case2FESV2} and the next-nearest sequence of $k-2$ holes on its left. Eventually, we must again arrive at a scenario with two neighbouring sequences of $k-2$ holes separated by a single particle, and we may use the same reasoning as in the scenario in Eq. \eqref{eq:Case2FESV1}.\\
	
	\vspace{1em}
Case 3. \ Assume that in the FES there is at most one pair of particles separated by $k-2$ holes, with all the other pairs being separated by $k-3$ holes (or less when there is overlap with the boundary). We refer to these as ``particle-connected'' (PC) FESs, and introduce them in detail in \cref{sec:ergodicFES} (see also the summary in \cref{tab:glossary}).  We now prove that two different PC FESs  must have different dipole moments $X$. Given that the dynamics considered in this work conserve $X$ and that the FESs from cases 1 and 2 have already been shown to be unique, this proves uniqueness of the PC FESs as well. 

We first consider the case of PC FESs for which particles do not overlap with the boundary sites. We denote by $i_0$ the site of the leftmost particle and assume that the $m$-th particle from the left is separated from the previous one by $k-2$ holes, with all other pairs of particles being separated by $k-3$ holes.
\begin{align}
    \begin{tikzpicture}[scale=0.6, transform shape]
    \def\siteLength{0.3} 
    \def\spacing{0.45}  
    \def\particleSize{2.5pt} 
    \def\adjust{0.03}
    \foreach \i in {-2,-1,0, 1, 3, 4, 6, 7,9, 10, 11, 13, 14, 15, 17, 18,20,21,23,24,25,26} {
        \draw (\i * \spacing, 0) -- (\i * \spacing + \siteLength, 0);
    }
    \foreach \i in {0, 4, 6, 10, 14, 18,20,24} {
        \fill (\i * \spacing + \siteLength/2, 0.2) circle (\particleSize);
    }
    \node at (-3 * \spacing+\siteLength/2+\adjust, 0.0) {\dots};
    \node at (2 * \spacing+\siteLength/2+\adjust, 0.0) {\dots};
    \node at (5 * \spacing+\siteLength/2+\adjust, 0.0) {\dots};
    \node at (8 * \spacing+\siteLength/2+\adjust, 0.0) {\dots};
    \node at (12 * \spacing+\siteLength/2+\adjust, 0.0) {\dots};
    \node at (16 * \spacing+\siteLength/2+\adjust, 0.0) {\dots};
    \node at (19 * \spacing+\siteLength/2+\adjust, 0.0) {\dots};
    \node at (22 * \spacing+\siteLength/2+\adjust, 0.0) {\dots};
    \node at (27 * \spacing+\siteLength/2+\adjust, 0.0) {\dots};
    \node at (2 * \spacing +\siteLength/2, -0.3) {\(k-3\)};
    \node at (8 * \spacing +\siteLength/2, -0.3) {\(k-3\)};
    \node at (12 * \spacing +\siteLength/2, -0.3) {\(k-2\)};
    \node at (16 * \spacing +\siteLength/2, -0.3) {\(k-3\)};
    \node at (22 * \spacing +\siteLength/2, -0.3) {\(k-3\)};
    \node at (0 * \spacing +\siteLength/2, 0.5) {\(1\)};
    \node at (24 * \spacing +\siteLength/2, 0.5) {\(N\)};
    \node at (10 * \spacing +\siteLength/2, 0.5) {\(m-1\)};
    \node at (14 * \spacing +\siteLength/2, 0.5) {\(m\)};
    \node at (28.5 * \spacing, 0) {.};
    \end{tikzpicture}
    \label{eq:Case3FES}
\end{align}
If there is no pair of particles separated by $k-2$ holes, we choose by convention to set $m=N+1$, hence $2\le m \le N+1$. The dipole moment of the configuration in Eq. \eqref{eq:Case3FES} is then
\begin{equation}
    X = N\, i_0 - m + h(k,N) \ ,
\end{equation}
where $h(k,N)=(k-2)N(N-1)/2+N+1$ does not depend on $m$ or $i_0$. We now imagine a different PC FES whose particles do not overlap with the boundary, which must possess different values of $i_0$, $m$ or both. Its dipole moment would thus read
\begin{equation}
    X' = N \, i_0' - m' + h(k,N) \ . 
\end{equation}
It is elementary to see that the equation $X=X'$ has no solution because $|i_0 - i_0'|$ is an integer and $2 \le m' \le N+1$.  Hence when there is no overlap with the boundary, we have shown that PC FESs of the type of Eq. \eqref{eq:Case3FES} are uniquely determined by $N$ and $X$.

We next allow for the possibility of overlap with the boundary. Hence we are considering PC FESs in which at most one pair of particles is separated by $k-2$ sites, and all other pairs are separated by $k-3$ sites, up to the possible exception of particles stacking on the boundary and a separation of fewer than $k-3$ sites between particles on the boundary and the next-nearest particles. An example of such a PC FES for $k=6$, $L=21$, $N=12$, $X=133$ (with the leftmost site being site 0) is given by
\begin{align}
    \begin{tikzpicture}[scale=0.8, transform shape]
    \def\siteLength{0.3} 
    \def\spacing{0.45}  
    \def\particleSize{2.5pt} 
    \def\adjust{0.03}
    \foreach \i in {0, 1, 2, 3, 4, 5, 6, 7, 8, 9, 10, 11, 12, 13, 14, 15, 16, 17, 18,19,20} {
        \draw (\i * \spacing, 0) -- (\i * \spacing + \siteLength, 0);
    }
    \foreach \i in {0, 2, 6, 11, 15, 19, 20} {
        \fill (\i * \spacing + \siteLength/2, 0.2) circle (\particleSize);
    }
    \fill (0+\siteLength/2,0.45) circle (\particleSize);
    \fill (0+\siteLength/2,0.7) circle (\particleSize);
    \fill (20*\spacing +\siteLength/2,0.45) circle (\particleSize);
    \fill (20*\spacing +\siteLength/2,0.7) circle (\particleSize);
    \fill (20*\spacing +\siteLength/2,0.95) circle (\particleSize);
    \node at (4 * \spacing +\siteLength/2, -0.3) {\(k-3\)};
    \node at (8.5 * \spacing +\siteLength/2, -0.3) {\(k-2\)};
    \node at (13 * \spacing +\siteLength/2, -0.3) {\(k-3\)};
    \node at (17 * \spacing +\siteLength/2, -0.3) {\(k-3\)};
    \node at (21 * \spacing, 0) {.};
\end{tikzpicture}
\label{eq:BFESexample}
\end{align} We shall show that the dipole moment $X$ of a PC FES uniquely determines the number of particles stacked on the right and left boundary sites. To begin with, we consider the case of the left boundary site. Say there are $N_L>0$ particles stacked on top of it. The highest dipole moment $X_\text{max}$ achievable in this case is attained when we place the remaining particles in the system as far to the right as possible, leading to the configuration 
\begin{align}
        \begin{tikzpicture}[scale=0.8, transform shape]
    \def\siteLength{0.3} 
    \def\spacing{0.45}  
    \def\particleSize{2.5pt} 
    \def\adjust{0.03}
    \foreach \i in {0,1,3,4,5,7,8} {
        \draw (\i * \spacing, 0) -- (\i * \spacing + \siteLength, 0);
    }
    \foreach \i in {0,4,8} {
        \fill (\i * \spacing + \siteLength/2, 0.2) circle (\particleSize);
    }
    \fill (0 + \siteLength/2, 0.9) circle (\particleSize);
    \node at (0 + \siteLength/2, 0.65) {\(\vdots\)};
    \node at (- 1*\siteLength, 0.55) {\(N_L\)};
    \node at (2 * \spacing+\siteLength/2+\adjust, 0.0) {\dots};
    \node at (6 * \spacing+\siteLength/2+\adjust, 0.0) {\dots};
    \node at (9 * \spacing+\siteLength/2+\adjust, 0.0) {\dots};
    \node at (2 * \spacing +\siteLength/2, -0.3) {\(k-2\)};
    \node at (6 * \spacing +\siteLength/2, -0.3) {\(k-3\)};
    \node at (10.5 * \spacing, 0) {.};
    \end{tikzpicture}
    \label{eq:Xmax}
\end{align}
In the above configuration, after the sequence of $k-2$ holes, we repeat only sequences of $k-3$ holes until either all $N$ particles are positioned or, if the right boundary is reached first, all remaining particles are stacked on the right boundary. 
Next, consider an integer $a\leq N_L$ and say we construct our PC FES with $N_L-a\geq0$ particles on the left boundary. In this case the lowest dipole moment $X'_\text{min}$ is achieved with the configuration
\begin{align}
    \begin{tikzpicture}[scale=0.8, transform shape]
    \def\siteLength{0.3} 
    \def\spacing{0.45}  
    \def\particleSize{2.5pt} 
    \def\adjust{0.03}
    \foreach \i in {-1,0,1,3,4,5,7,8} {
        \draw (\i * \spacing, 0) -- (\i * \spacing + \siteLength, 0);
    }
    \foreach \i in {-1,0,4,8} {
        \fill (\i * \spacing + \siteLength/2, 0.2) circle (\particleSize);
    }
    \fill (-1*\spacing + \siteLength/2, 0.9) circle (\particleSize);
    \node at (-1*\spacing + \siteLength/2, 0.65) {\(\vdots\)};
    \node at (-2.3*\spacing, 0.55) {\(N_L-a\)};
    \node at (2 * \spacing+\siteLength/2+\adjust, 0.0) {\dots};
    \node at (6 * \spacing+\siteLength/2+\adjust, 0.0) {\dots};
    \node at (9 * \spacing+\siteLength/2+\adjust, 0.0) {\dots};
    \node at (2 * \spacing +\siteLength/2, -0.3) {\(k-3\)};
    \node at (6 * \spacing +\siteLength/2, -0.3) {\(k-3\)};
    \node at (10.5 * \spacing, 0) {.};
    \end{tikzpicture}
    \label{eq:Xmin}
\end{align}
Here again, we only repeat sequences of $k-3$ holes between particles until all $N$ particles are placed or we reach the right boundary and the remaining particles are stacked. By comparing Eq. \eqref{eq:Xmax} and Eq. \eqref{eq:Xmin}, it is clear that $X'_\text{min} > X_\text{max}$. Thus a  PC FES  with $N_L>0$ particles on the left boundary cannot be dynamically connected to a PC FES with $N_L-a\geq0$ particles on it and vice versa.

We have thus shown that, for a PC FES, the dipole moment $X$ uniquely determines the number of particles $N_L$ on the left boundary. Likewise, an identical demonstration shows that $X$ also determines the number $N_R$ of particles on the right boundary site. What remains to be determined are the positions of the particles in the bulk of the system, but with $N_L$ and $N_R$ already known. However, the positions of the remaining particles are then also uniquely determined by the dipole moment using the same logic as for the PC FES configuration with no boundary stacking in Eq. \eqref{eq:Case3FES}. This completes the proof.\\

	A notable property of PC FESs is that, unlike the FESs discussed in cases 1 and 2, they do not have any regions that are necessarily disconnected during a sequence of inward hops. In particular, we call PC FESs whose particles span the whole system (which can only occur at fillings $N/L\geq\nu_c=(k-2)^{-1}$) ``blockage-free FESs'', and we discuss them in more detail in Section \ref{sec:ergodicFES} (see also \cref{tab:glossary}). Their particular structure is of importance to our discussion of the weakly Hilbert-space-fragmented phase in Section \ref{sec:weakFragNumerics}, where we identify dominant Krylov sectors in typical symmetry sectors of $d=\infty$ systems as those sectors that contain blockage-free FESs.

	\section{Proof of 2-colour connectivity}
	\label{sec:proof2colour}
	
	\begin{figure}[ht]
		\centering
		\begin{tikzpicture}[scale=0.85, transform shape]
			\def\siteLength{0.3} 
			\def\spacing{0.45}  
			\def\particleSize{2.5pt} 
			\def\adjust{0.03}
			\foreach \i in {0, 1, 3, 4, 5, 7, 8, 10, 11, 13, 14, 15, 17, 18} {
				\draw (\i * \spacing, 0) -- (\i * \spacing + \siteLength, 0);
			}
			
			\definecolor{darkgreen}{rgb}{0.0, 0.5, 0.0}
			\newcommand{\getcolor}[1]{%
				\ifnum#1=0 darkgreen\fi
				\ifnum#1=1 blue\fi
				\ifnum#1=2 blue\fi
				\ifnum#1=3 blue\fi
				\ifnum#1=4 blue\fi
				\ifnum#1=5 red\fi
			}
			\foreach \i [count=\j from 0] in {0, 4, 8, 10, 14, 18} {
				\fill[fill=\getcolor{\j}] (\i * \spacing + \siteLength/2, 0.2) circle (\particleSize);
			}
			\node[text=darkgreen] at (-1 * \spacing+\siteLength/2+\adjust, 0.15) {\dots};
			\node at (2 * \spacing+\siteLength/2+\adjust, 0.0) {\dots};
			\node at (6 * \spacing+\siteLength/2+\adjust, 0.0) {\dots};
			\node at (9 * \spacing+\siteLength/2+\adjust, 0.0) {\dots};
			\node at (12 * \spacing+\siteLength/2+\adjust, 0.0) {\dots};
			\node at (16 * \spacing+\siteLength/2+\adjust, 0.0) {\dots};
			\node[text=red] at (19 * \spacing+\siteLength/2+\adjust, 0.15) {\dots};
			\node at (2 * \spacing +\siteLength/2, -0.3) {\(k-2\)};
			\node at (6 * \spacing +\siteLength/2, -0.3) {\(k-3\)};
			\node at (12 * \spacing +\siteLength/2, -0.3) {\(k-3\)};
			\node at (16 * \spacing +\siteLength/2, -0.3) {\(k-2\)};
			\node[text=darkgreen] at (0 * \spacing +\siteLength/2, +0.6) {\( G_1 \)};
			\node[text=blue] at (4 * \spacing +\siteLength/2, +0.6) {\( B_1 \)};
			\node[text=blue] at (8 * \spacing +\siteLength/2, +0.6) {\( B_2 \)};
			\node[text=blue] at (10 * \spacing +\siteLength/2, +0.6) {\( B_{m-1} \)};
			\node[text=blue] at (14 * \spacing +\siteLength/2, +0.6) {\( B_m \)};
			\node[text=red] at (18 * \spacing +\siteLength/2, +0.6) {\( R_1 \)};
			
		\end{tikzpicture}
		\caption{Configuration of particles in an FES from Fig.~\ref{fig:colorscheme}(b)-(c), which leads to the formation of a type-2 blockage. Regions separated by a type-2 edge (i.e., a sequences of $k-2$ holes in an FES) host particles of different colour. We name the $m$ particles in the blue region $B_1, \ldots, B_m$, while the right-most green particle and left-most red particle are respectively $G1$ and $R1$.\label{fig:colorschemeappendix}}
	\end{figure}
	
	The proof of the 2-colour connectivity introduced in Section \ref{sec:BlockagesFES2colour} is identical in spirit to the proof discussed in case 2 of Appendix \ref{appendix:uniqueFES}. However, for completeness and in order to make explicit use of the colour scheme used in defining the $2$-colour connectivity, we present a separate proof in this appendix. 
	
	It is enough to prove the statement for a subregion of the chain containing $3$ colours, as the result trivially generalises to the whole system. We can thus refer to the particle configurations, colours, and labels from Fig.~\ref{fig:colorschemeappendix}. We start by noticing that because of uniqueness of the FES, from any configuration in the Krylov we can reach the FES by performing only outward hops. This automatically implies that any given configuration can be reached starting from the FES and performing only inward hops. Call “central” the region consisting of the two type-2 edges (i.e., sequences of $k-2$ holes ) and all the sites between them. Assume that there is a configuration of particles $\mathcal{C}$ in the Krylov sector that violates the 2-colour connectivity. This means that wherever in the central region we place a cut between two sites, during the series of inward hops from the FES to $\mathcal{C}$ there must be at least one inward hop that involves a particle on the left and one on the right of the cut. This implies that at least one inward hop involving $G_1$ and $B_1$ from Fig.~\ref{fig:colorschemeappendix} must occur before these interact with any other particle, otherwise a sequence of $k-1$ holes is created between them, preventing any future inward hop between the two. Thus, at some point during the series of inward hops, $B_1$ is moved for the first time and goes at least one site to its left. This necessarily creates a separation of $k-2$ or more holes between $B_1$ and $B_2$. Given the assumption about $\mathcal{C}$, this separation must necessarily be of $k-2$ holes, otherwise inward hops between $B_1$ and $B_2$ would be prevented and the 2-colour connectivity would be respected. We can now repeat the same reasoning used for $G_1$ and $B_1$, which were separated by $k-2$ holes, for $B_1$ and $B_2$. In this way we keep propagating to the right a sequence of $k-2$ holes. Clearly, by analogous reasoning the sequence of $k-2$ holes that originally separate $B_m$ from $R_1$ must propagate to the left. Thus, at some point two sequences of $k-2$ holes must be next to each other, with a single particle in the middle that has not been part of any inward hop yet. By the assumption about $\mathcal{C}$, the particle in the middle must perform an inward hop, and this necessarily creates a sequence of $k-1$ holes. By placing a cut anywhere between two sites that are part of this sequence of $k-1$ holes, we manage to partition the system into two independent subregions that never interact and that contain at most 2 different colours. This contradicts the assumption about $\mathcal{C}$ and proves that any configuration of particles in the Krylov sector respects the 2-colour connectivity.

	\section{Existence of particle-connected FESs and extended states for any $N$ and $X$}
	\label{app:Eoftype1FESforNandX}
	
    In Appendix \ref{appendix:uniqueFES} it is shown that for fixed $L$ and $N$ two different PC FESs (see \cref{sec:ergodicFES}, or \cref{tab:glossary} for a summary) necessarily have different dipole moments $X$. We now show the \emph{existence} of a PC FES for any value of $X$ compatible with a fixed pair of $N$ and $L$. 
	
	The configuration of $N$ particles with minimal dipole moment is the one in which all particles are stacked on the left boundary. We call its dipole moment $X^*_L$. Similarly, the configuration with maximal dipole moment has all particles stacked on the right boundary, with dipole moment $X^*_R$. We now present a simple algorithm to construct PC FESs that have any dipole moment $X$ between $X^*_L$ and $X^*_R$. Importantly, every move we perform in the following brings us from a PC FES to another PC FES, i.e., we never leave the class of PC FESs.  
	
	We fix $N$ and $L$ and imagine starting from the configuration in which all the particles are stacked on the left boundary. Thus, we have $N_L=N$, where $N_L$ is the number of particles on the left boundary. The first move takes one particle from the left stack and puts it on top of the site next to it, leading to $N_L=N-1$ and $X=X_L^*+1$. We can continue increasing the value of $X$ in steps of $1$ by moving the same particle to the right until there are $k-2$ holes between this and the left boundary. The next move is to take another particle from the left stack, that is, $N_L-1\to N_L-2$, and place it over the site next to it. Then we alternate in moving the rightmost particle and the intermediate particle of one site to the right, so as to have at most one sequence of $k-2$ holes after each of these steps. This lasts until there are $k-2$ holes between the intermediate particle and the left stack, at which point we again take a particle from the left stack and place it over the site next to it. We then repeat the process of iteratively moving of one site to the right the particles that are not on the left stack, starting from the rightmost one, arriving to the one closest to the left boundary, and then restarting from the rightmost one unless between the left boundary and the first particle to its right there are $k-2$ holes. In the latter case we remove an additional particle from the left stack. 
	In this way, we can continue to generate PC FESs that differ in dipole moment of exactly $1$ unit at each step. 
	
	At some point, the left boundary will necessarily remain without particles. The algorithm in this case runs identically to before, without the step in which we remove a particle from the left stack. The final scenario to take into consideration is when we reach the right boundary with the rightmost particle, irrespective of whether this happens before or after the left boundary has been emptied. Also in this case the algorithm remains the same, with the only modification that when between the right boundary and the first particle to its left there are no holes, the next right hop of this particle brings it onto the right boundary. Over time this increases the number of particles on the right stack, until we place all the $N$ particles on it, reaching $X^*_R$. 
	
	This, together with the results from Appendix \ref{appendix:uniqueFES}, proves that for any values of $N$ and $L$ there exists \emph{one and only one} PC FES for each value of $X$ compatible with $N$ and $L$.
	
	It is apparent that a same derivation can be used to show the existence of a particle-connected extended state (see Section \ref{sec:weakFragNumerics} or \cref{tab:glossary}) at finite $d$ for any $N$ and $X$ symmetry sector as well. In this case, we begin with the particle configuration for a given $N$ and $L$ in which all particles are stacked up on the left of the system, i.e., the leftmost $\lfloor N/(d-1)\rfloor$ sites have occupancy $d-1$, with the next site over having $N-\lfloor N/(d-1)\rfloor$ particles. The steps followed are almost identical to the PC FES case, with the main difference being that instead of new particles being taken from the leftmost site of the system, they are taken from the rightmost site of the boundary particle stack which is still nonempty. Likewise, as the particles are moved to the right of the system, instead of all stacking on the rightmost site, they will fill up the sites one by one to their maximal $d-1$ occupancy, starting from the rightmost site and progressing left. Otherwise, the algorithm used is the same, and the proof also holds at finite $d$.

	\section{Asymptotic formulas for the size of symmetry sectors}
	\label{appendix:scalingofsizeofss}
    
	The total number of possible particle configurations on a chain of $L$ sites with on-site dimension $d$ is $d^L$. 
	We start by fixing the total number of particles to a given value $N$ and, as in the main text, we refer to the total number of configurations in this symmetry sector as $D^{(d)}_N$. There are $L \,(d-1) + 1$ distinct $N$-sectors, corresponding in the thermodynamic limit to the real values in the interval $0 \le \nu \le d-1$. The fact that these are only polynomially many in $L$, while the total number of configurations scales exponentially with $L$, shows that there must be large $N$-sectors with dimension $D_N^{(d)}$ whose exponential scaling coincides with $d^L$. For $d=2$ and $d=\infty$ the expression for $D_N^{(d)}$ can be easily derived using Stirling's asymptotic formula in the limit of large $L$ on the two exact expressions
	\begin{equation}
		D_N^{(2)}(L) = \binom{L}{N} \qquad D_N^{(\infty)}(L) = \binom{N+L-1}{L-1} \ .
	\end{equation}
	We obtain for $0<\nu<d-1$
	\begin{equation}
    \label{eq:D_N_appendix}
		\ln D_N^{(d)}(L) = L \, \eta_d(\nu) - \frac{1}{2} \ln L + \mathcal{O}(L^0) \ , 
	\end{equation}
	where 
	\begin{equation}
		\begin{aligned}
			\label{eq:eta2etainf}
			\eta_2(\nu) &= -\nu \ln \nu - (1-\nu) \ln (1-\nu) \\
			\eta_\infty(\nu) & =  -\nu \ln \nu + (1+\nu) \ln (1+\nu) \ .
		\end{aligned}
	\end{equation}

We now show that the functional form of \eqref{eq:D_N_appendix} applies to all $2\le d\le \infty$. Furthermore, we prove that $\eta_d(\nu)$ is a strictly concave function for any $d$. For $d=2$ and $\infty$ this is straightforward to check by computing $\eta_d''(\nu)$ from \eqref{eq:eta2etainf} and noticing that it is negative for any $\nu$ in the allowed domain. By particle-hole symmetry and strict concavity it follows that $\eta_d(\nu)$ has a unique global maximum at the half filling $\nu^* = (d-1)/2$, where it takes the value $\eta_d(\nu^*)=\ln d$, as expected from the argument above.
   
For any finite $d$ the following identity holds
	\begin{equation}
		\label{eq:expansionforDN}
		(1+z+\ldots+z^{d-1})^L = \sum_{N=0}^{(d-1)L} D_{N}^{(d)} z^N \ . 
	\end{equation}
This can be re-written as
	\begin{equation}
		\begin{aligned}
			&(1+z+\ldots+z^{d-1})^L = \\
			&\sum_{\substack{e_0,e_1,\ldots,e_{d-1}=0 \\ [e_0+e_1+\ldots+e_{d-1}=L]}}^{L} \frac{L!}{e_0!e_1!\ldots e_{d-1}!}\prod_{n=1}^{d-1}z^{n e_n} \ . 
		\end{aligned}
	\end{equation}
	Thus, finding $D_N^{(d)}$ amounts to summing all the multinomial coefficients in the previous expression associated with sets of multiplicities $\{e_i\}$ such that
	\begin{equation}
		\label{eq:condsetsei}
		\sum_{n=0}^{d-1} e_n=L \qquad \quad  \sum_{n=0}^{d-1} n e_n=N \ . 
	\end{equation}
	For any fixed $N$ and $d$, the number of different sets $\{e_i\}$ satisfying \cref{eq:condsetsei} scales at most polynomially with $L$. This implies that if one of these sets $\{e_i\}$ has a multinomial coefficient that exponentially dominates in $L$ over the others, its exponential scaling determines the function $\eta_d(\nu)$ appearing in \cref{eq:D_N_appendix}. By particle-hole symmetry we can restrict our attention to fillings $\nu\le (d-1)/2$ without loss of generality. Using Stirling's asymptotic formula we easily get
	\begin{equation}
		\ln \left[ \frac{L!}{e_0!e_1!\ldots e_{d-1}!} \right] = -L \sum_{n=0}^{d-1}\varepsilon_n \ln \varepsilon_n + \mathcal{O}(\ln L) \ , 
	\end{equation}
	where we have defined $\varepsilon_n=e_n/L$. We can maximize by the method of Lagrange multipliers the leading term in the previous expression subject to the constraints in \cref{eq:condsetsei}, which we rewrite as 
	\begin{equation}
		\begin{aligned}
			\label{eq:varepnewverconstraints}
			h_1(\varepsilon_1, \ldots, \varepsilon_{d-1}) &= \sum_{n=0}^{d-1} \varepsilon_n - 1 = 0 \\
			h_2(\varepsilon_1, \ldots, \varepsilon_{d-1}) &= \sum_{n=0}^{d-1} n \, \varepsilon_n - \nu = 0 \ . 
		\end{aligned}
	\end{equation}
	We thus look for the stationary points of the function 
	\begin{equation}
		- \sum_{n=0}^{d-1}\varepsilon_n \ln \varepsilon_n - q \, h_1 - \tilde{q} \, h_2 \ . 
	\end{equation}
	These are given by 
	\begin{equation}
		\label{eq:varepsmaxfufo}
		\varepsilon_n^* = e^{-1-q} \, t^n \ \ \forall \ n \qquad \quad t \equiv e^{-\tilde{q}} \ .  
	\end{equation}
	Assume that $0\le t < 1$. By imposing the constraints in \cref{eq:varepnewverconstraints} on the set $\{ \varepsilon_n^* \}$ we obtain 
	\begin{align}
		e^{1+q} &= \frac{1-t^d}{1-t} \label{eq:1plusqdef} \ , \\
		\nu & = d + \frac{t}{1-t}-\frac{d}{1-t^d} \label{eq:eqdeft} \ . 
	\end{align}
	Note that in \cref{eq:eqdeft} we obtain $\nu=0$ by setting $t=0$ and $\nu\to(d-1)/2$ in the limit $t \to 1^{-}$. By continuity, there always exists a solution $0\le t<1$ to \cref{eq:eqdeft} in the domain of interest $0\le \nu < (d-1)/2$, and thus our previous assumption about the domain of $t$ is justified. We also notice that the right-hand side of \cref{eq:eqdeft} is a strictly increasing function in the interval $0 \le t < 1$ and thus the solution $t=t(\nu,d)$ of \cref{eq:eqdeft} is unique. We can now use the functional form of the maxima \eqref{eq:varepsmaxfufo}, together with \eqref{eq:1plusqdef} and \eqref{eq:eqdeft}, to obtain
	\begin{equation}
		\label{eq:implicitetadnu}
		\eta_d(\nu) = - \sum_{n=0}^{d-1} \varepsilon_n^* \ln \varepsilon_n^* = \ln \left[ \frac{1-t^d}{1-t} \right] - \nu \ln t \ . 
	\end{equation}
	This is an implicit expression for $\eta_d(\nu)$ as a function of $t(\nu,d)$, valid for any $d$. The solution $t(\nu,d)$ can be easily determined analytically in the case of $d=2$ yielding $\eta_2(\nu)$ from \cref{eq:eta2etainf}, while allowing $d \to \infty$ one recovers $\eta_\infty(\nu)$. For $d=3$ the solution $t$ is given by
	\begin{equation}
		t(\nu, 3) = \frac{1-\nu-\sqrt{1+6 \nu -3 \nu^2}}{2(\nu-2)} \ ,
	\end{equation}
	from which $\eta_3(\nu)$ is readily obtained using \cref{eq:implicitetadnu}. The strict concavity of $\eta_d(\nu)$ for $d=3$, similarly to $d=2$ and $\infty$, follows easily from the fact that $\eta_d''(\nu)<0$ for any $\nu$ in the domain $0\le \nu \le (d-1)/2$. In the general $d$ case, the value of $\eta_d(\nu)$ at a fixed $\nu$ can be found by numerically solving \eqref{eq:eqdeft} and inserting the solution $t$ in \eqref{eq:implicitetadnu}. Furthermore, by differentiating \cref{eq:eqdeft} with respect to $\nu$, we can express both $\partial_\nu t$ and $\partial^2_\nu t$ as functions of $d$, $\nu$ and $t(\nu,d)$ itself. This lets us analytically express $\eta_d''(\nu)$ as a function of the latter variables by using \eqref{eq:implicitetadnu}
	\begin{equation}
		\eta_d''(\nu) = Z(\nu,d,t(\nu,d)) \ .  
	\end{equation}
	We do not report explicitly the function $Z$, as the latter involves sums of many terms and its determination requires only simple algebraic manipulations. Then, the value of $\eta_d''(\nu)$ for any fixed $\nu$ and any generic $d$ can be obtained by numerically solving \eqref{eq:eqdeft} to obtain $t(\nu,d)$, and using the latter in the third argument of the function $Z$. By applying this procedure to different $d$ values one shows for any $d$ that the function $\eta_d''(\nu)$ is strictly negative in the entire interval $0\le \nu \le (d-1)/2$. By particle-hole symmetry the same is true for $(d-1)/2 \le \nu \le d-1$.

\begin{figure*}[th]
		\centering
		{\includegraphics[scale = 0.33]{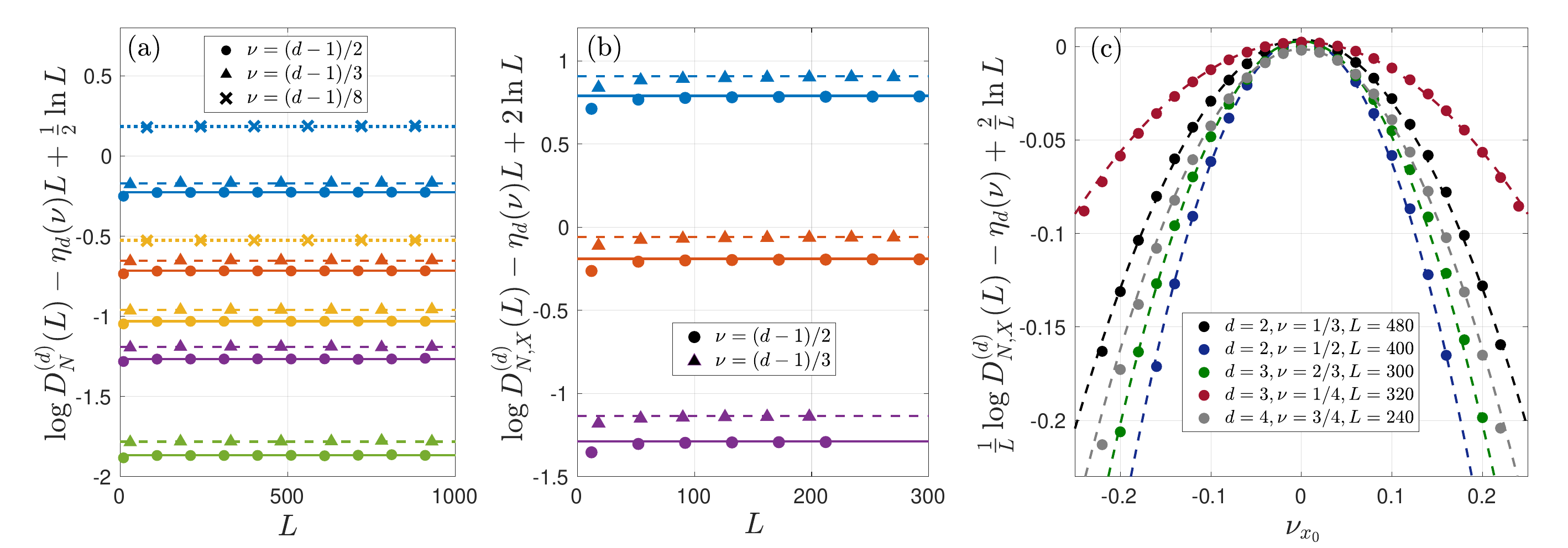}}
		\caption{(a) Scaling of $D_N^{(d)}(L)$ for $d$ equal to $2,3,4,5$ and $9$, respectively in blue, red, yellow, purple and green. For $\nu=(d-1)/2$, $\eta_d(\nu)=\ln d$, while for the others fillings $\eta_d(\nu)$ is computed analytically for $d=2,3$ and by numerically solving \cref{eq:eqdeft} for $d=4,5,9$. The horizontal lines coincide with the $L$ independent constants appearing in Eq.~(\ref{eq:generaldDN}). (b) Scaling of $D_{N,X}^{(d)}(L)$, where $x=X/N$ coincides with the centre of the chain. Colour scheme and values of $\eta_d(\nu)$ as in Figure (a). The horizontal lines are the estimated values for the $L$-independent constants appearing in Eq.~(\ref{eq:DNXscaling}) and obtained as the parameter $a$ in a two-parameter fit $a-b/x$. (c) Scaling of $D_{N,X}^{(d)}(L)$ as a function of the intensive centre of mass $\nu_{x_0}$ with origin in the centre of the chain, for fixed values of $d$, $\nu$ and $L$. The dashed lines are 2-parameter quadratic fits $a-b \, x^2$, $b>0$. These have been performed excluding a few points at the boundaries of the $\nu_{x_0}$-range displayed in the figure, so to focus on the small-$\nu_{x_0}$ regime at the basis of the quadratic expansion in \cref{eq:Lambdadnux0}.\label{fig:NXsectors}}
	\end{figure*}

    In Fig.~\ref{fig:NXsectors}(a) we verify numerically that the subleading terms in \cref{eq:D_N_appendix} are independent of $d$. Here the value of $D_N^{(d)}$ is computed numerically using Eq.~(\ref{eq:expansionforDN}) for several choices of $d$ and $\nu$. \\

	We now consider to the size $D_{N,X}^{(d)}$ of $(N,X)$ sectors. For a given $N$-family such that $0<\nu<d-1$, the number of distinct $X$-sectors scales asymptotically as $N(L-N/(d-1)) = \mathcal{O}(L^2)$. In the thermodynamic limit, these correspond to the interval $\nu/(2d-2) \le \nu_x \le 1-\nu/(2d-2) $. Given that the exponential scaling of the size of an $N=\nu L$ sector is governed by $\eta_d(\nu)$ and that there are only polynomially many $X$-sectors for each $N$, the largest $X$-sectors must have size with exponential scaling equal to $\exp(L \, \eta_d(\nu))$. Recently, Ref.~\cite{MelczerPanovaPemantle} derived a sharp asymptotic formula for $D_{N,X}^{(d)}$ in the case of $d=\infty$. Below we review this result, and show how from this we can also derive an exact asymptotic formula for $d=2$. Both cases lead to the functional form of \cref{eq:DNXscaling}
    \begin{equation}
		\begin{aligned}
        \label{eq:DNXscaling_appendix}
			\ln D_{N,X}^{(d)}(L) &= L  \Big(\eta_d(\nu)  - \Lambda_d(\nu,\nu_x) \Big) \\
			& \qquad \quad  - 2 \ln L + \mathcal{O}(L^0) \ .
		\end{aligned}
	\end{equation}
We will numerically verify that this expression is valid for any $d$. 

Similarly to Eq.~(\ref{eq:expansionforDN}), we can obtain the values of $D_{N,X}^{(d)}$ as
	\begin{equation}
		\begin{aligned}
			\label{eq:expansionforDNX}
			&\prod_{n=0}^{L-1}(1+z y^n+z^2 y^{2n} +\ldots+z^{d-1}y^{(d-1)n}) = \\
			& \qquad \qquad \qquad \qquad \sum_{N,X} D_{N,X}^{(d)} \, z^N y^X \ . 
		\end{aligned}
	\end{equation}
	In our formalism, the sharp asymptotic formula of Ref. \cite{MelczerPanovaPemantle} reads
	\begin{equation}
		\begin{aligned}
			\label{eq:DNXexact}   
			&\ln D_{N,X}^{(\infty)} (L) = L \left[r+2\, s\, \nu_x - \nu \ln \left( 1-e^{-r-s} \right)\right] \\
			&\qquad \qquad \qquad \qquad  - 2 \ln L + \mathcal{O}(L^0) \ .
		\end{aligned}
	\end{equation}
	Here $r=r(\nu,\nu_x)$ and $s=s(\nu,\nu_x)$ are $L$-independent functions of the filling $\nu$ and the intensive centre of mass $\nu_x$, which are implicitly defined as the unique positive real solutions of the following coupled equations \cite{MelczerPanovaPemantle}
	\begin{widetext}
		\begin{equation}
			\begin{aligned}
				\label{eq:randscoupled}
				\frac{1}{\nu} &= \frac{1}{s} \ln \left( \frac{e^{r+s}-1}{e^{r}-1} \right) - 1 \ , \\
				\frac{\nu_x}{\nu} &= \frac{s \ln \left( 1-e^{-r-s}\right)+\text{dilog}\left( 1-e^{-r}\right)-\text{dilog}\left( 1-e^{-r-s}\right)}{s^2} \ ,
			\end{aligned}
		\end{equation}
	\end{widetext}
	where 
	\begin{equation}
		\text{dilog}(x) \equiv \int_{1}^x dt \frac{\ln t}{1-t}  \quad \qquad |1-x|<1  \  . 
	\end{equation}
	Note that we are restricting $\nu_x$ to $\nu_x \le 1/2$ without loss of generality. For $\nu_x=1/2$ the solutions to the set \eqref{eq:randscoupled} is
	\begin{equation}
		r = \ln(1+\nu) \qquad s = 0 \ , 
	\end{equation}
	from which we obtain \cref{eq:DNXscaling} with $d=\infty$ and $\Lambda_\infty(\nu,1/2)=0$. For small negative values of $\nu_{x_0}=\nu_x-1/2$ we can study the set \eqref{eq:randscoupled} perturbatively in $\nu_{x_0}$ and obtain the following expansion for $r$ and $s$
	\begin{equation}
		\begin{aligned}
			\label{eq:randsexpanded}
			r &= \ln(1+\nu) + \frac{6 \, \nu \, \nu_{x_0}}{1+\nu}+ \mathcal{O}(\nu_{x_0}^2) \ ,  \\
			s &= -\frac{12 \, \nu \, \nu_{x_0}}{1+\nu} + \mathcal{O}(\nu_{x_0}^2) \ . 
		\end{aligned}
	\end{equation}
	Substituting this into \eqref{eq:DNXexact} yields \eqref{eq:DNXscaling_appendix} for $d=\infty$, with $\Lambda_\infty(\nu,\nu_x)$ expanded as in \cref{eq:Lambdadnux0}. Note that the analytic determination of $\lambda_\infty(\nu)$, which appears in \eqref{eq:Lambdadnux0}, requires knowledge of higher-order terms in the expansion \eqref{eq:randsexpanded}. 
	
	We now show that it is possible to derive an exact asymptotic formula for $D_{N,X}^{(2)}$ based on the knowledge of $D_{N,X}^{(\infty)}$. A known generalisation of the binomial theorem, see e.g. \cite{TAKACS1986123}, gives
	\begin{equation}
		\begin{aligned}
			\label{eq:genbintheorem}
			&\prod_{n=0}^{L-1}(1+z y^n) = \\
			& \ \ \ \sum_{N=0}^L \sum_{j=0}^{N(L-N)} D_{N,\,j}^{(\infty)}(L-N) \, z^N \, y^{j+N(N+1)/2} \ .
		\end{aligned}
	\end{equation}
	Comparing \eqref{eq:genbintheorem} with \eqref{eq:expansionforDNX} we realize that
	\begin{equation}
		\label{eq:equivDNX2inf}
		D_{N,X}^{(2)}(L) = D_{N,\widetilde{X}}^{(\infty)}(\widetilde{L}) \ ,
	\end{equation}
	where $\widetilde{L}=L-N$, $\widetilde{X}=X-N(N+1)/2$ and $N,X$ belong to the domain allowed by $d=2$. Similarly to the definitions of $\nu$ and $\nu_{x_0}$ we can define $\tilde{\nu}=N/\widetilde{L}$ and $\tilde{\nu}_{x_0} = \widetilde{X}/(N \widetilde{L})-1/2$. In the range $0<\nu<1$ of interest we have 
	\begin{equation}
		\tilde{\nu} = \frac{\nu}{1-\nu} \qquad \tilde{\nu}_{x_0} = \frac{\nu_{x_0}}{1-\nu} + \mathcal{O}(L^{-1}) \ . 
	\end{equation}
	Using the previous two equations together with \eqref{eq:equivDNX2inf} we arrive at the asymptotic expression \eqref{eq:DNXscaling_appendix} for $d=2$. In particular, for small $\nu_{x_0}$ values we obtain
	\begin{equation}
		\begin{aligned}
			&\ln D_{N,X}^{(2)}(L) = \ln D_{N,\widetilde{X}}^{(\infty)}(\widetilde{L}) \\
			& \ \ \ = L \left[ (1-\nu) \, \eta_{\infty}\left( \tilde{\nu} \right) - \frac{\lambda_{\infty}\left( \tilde{\nu} \right)}{1-\nu}\nu_{x_0}^2 + o(\nu_{x_0}^2) \right] \\
			& \qquad \qquad \qquad \quad - 2 \ln L + \mathcal{O}(L^0) \ . 
		\end{aligned}
	\end{equation}
	From this we see that
	\begin{equation}
		\lambda_2(\nu) = \frac{\lambda_{\infty}\left( \tilde{\nu} \right)}{1-\nu} \ .
	\end{equation}
	It is elementary to verify using Eq.~(\ref{eq:eta2etainf}) that 
	\begin{equation}
		(1-\nu) \, \eta_{\infty}(\tilde{\nu}) = \eta_2(\nu) \ ,
	\end{equation}
	as required by Eq.~(\ref{eq:DNXscaling_appendix}). \\

    Finally, we numerically verify that the functional form \eqref{eq:DNXscaling_appendix} and the expansion \eqref{eq:Lambdadnux0} hold for generic $d$ values. The asymptotic value of $D_{N,X}^{(d)}(L)$ for a dipole moment with centre of mass $X/N$ coinciding with the centre of the chain is numerically studied in Fig.\ref{fig:NXsectors}(b) for a few finite $d$ values, and coincides with the one reported in Eq.~(\ref{eq:DNXscaling_appendix}). Fig.~\ref{fig:NXsectors}(c) studies the dependence of $D_{N,X}^{(d)}$ on $\nu_{x_0}$, for small values of the latter, at fixed $L$ and for different finite $d$ values. The 2-parameter quadratic fits match the data very well, confirming that $\ln D_{N,X}^{(d)}/L-\eta_d(\nu)$ is at leading order in $\nu_{x_0}$ a quadratic function of the latter for generic $d$, as expressed by \cref{eq:Lambdadnux0}. The coefficient in front of the quadratic term of the fit is an estimate for the value of $\lambda_d(\nu)$ from \cref{eq:Lambdadnux0}.

	\section{Details of the full proof and strong fragmentation in atypical $(N,X)$ families}
	\label{appendix:detailsoffullproof}

In Section \ref{sec:detailscase3} we present the technical details involved in the treatment of “Case 3” in the general proof of \cref{sec:strongfrag2}.  In Section \ref{eq:strongfragatypical} we briefly discuss fragmentation in atypical families of $(N,X)$ sectors, i.e., those families having centre of mass $X/N$ sufficiently far from the centre of the chain. 

\subsection{Details of the proof for Case 3}
\label{sec:detailscase3}

To address Case 3 from \cref{sec:strongfrag2} we first need to generalise the inequality between $\eta_d(\nu)$ and $g_{d,\nu}(p,s)$ of \cref{eq:bipartitionchain} to the case in which we take into account the dipole moment. Consider a symmetry sector of the chain characterised by $N= \nu L$ and $X=\nu_x \nu L^2$ and create a bipartition into two subregions of size $sL$ and $(1-s)L$. Say that the first of the two subregions has $N_1 = p L$ particles and dipole moment $X_1 = x\, p\, s\, L^2$. The second subregion has $N_2$, $X_2$ such that $N_1+N_2=N$ and $X_1+X_2=X$. Define $\tilde{g}_{d,\nu,\nu_x}(p,x,s)$ as
	\begin{equation}
		\begin{aligned}
			&\tilde{g}_{d,\nu,\nu_x}(p,x,s) = \lim_{L\to\infty} \frac{1}{L}\ln \widetilde{W}_d \ , \\
			& \widetilde{W}_d  =  D_{N_1,X_1}^{(d)}(sL) \, D_{N_2,X_2}^{(d)}(L-sL) \ ,
		\end{aligned}
	\end{equation}
	where $\widetilde{W}_d$ is the total number of particle configurations compatible with the bipartition. Given that everyone of these $\widetilde{W}_d$ configurations is also a configuration of the global $(N,X)$ sector chosen, using Eq.~(\ref{eq:DNXscaling}) and taking the thermodynamic limit one arrives to
	\begin{equation}
		\label{eq:gitildeNXineq}
		\eta_d(\nu)-\Lambda_d(\nu,\nu_x) \, \ge \, \tilde{g}_{d,\nu,\nu_x}(p,x,s) \ ,
	\end{equation}
	for any $p, x$ and $s$ in their respective domains. The inequality (\ref{eq:gitildeNXineq}) trivially generalises to the case in which we partition the system into more than $2$ subregions. Note that identifying the global maxima of $\tilde{g}_{d,\nu,\nu_x}$ requires a more precise determination of the function $\Lambda_d(\nu,\nu_x)$ from \cref{sec:sizesymmetrysec} compared with the numerical one performed in Appendix \ref{appendix:scalingofsizeofss}.\\

    As we have seen in \cref{sec:strongfrag2}, Case 3 implies, in addition to having a zero fraction $f$ of frozen sites being part of type-1 blockages, that any extensive subregion of the chain has local filling approaching the global $\nu$ for large $L$ values. We now exploit a trivial property of the FES picture starting from any state of the system $S$ with on-site dimension $d$: \\
    
    \emph{If an extensive subregion of the chain is enclosed by two blockages of either type-1 or type-2 of $o(L)$ size, then for asymptotically large $L$ values this region has identical local filling in the FES of the auxiliary system $\tilde{S}$ and in the chosen state of the system $S$.} \\
    
    This provides us with an algorithm to partition the system as in Fig.~\ref{fig:type2partition}, i.e.~into $G$ subregions $\mathcal{A}_i$ that are separated by type-2 blockages $\mathcal{B}_i$ of $L$-independent size. The first subregion $\mathcal{A}_1$ must necessarily have size $\ell_1$ scaling as $\mathcal{O}(L^{\gamma_1})$, with $0 \le \gamma_1 < 1$. Indeed, assume for the sake of contradiction that the first $L$-independent in size type-2 blockage is encountered only extensively far from from the left boundary, and so $\ell_1 = \mathcal{O}(L)$. This means that either $\mathcal{A}_1$ does not possess type-2 edges at all or that those present are at least an $\mathcal{O}(h(L))$ apart from each other, with $h(L)$ a function that diverges for $L \to \infty$. This would however imply that for large $L$ the local filling in $\mathcal{A}_1$ is greater or equal to $\nu_c$, which is not possible given the condition of having local filling equal to $\nu$, i.e.~lower than $\nu_c$, in any extensive subregion of the chain. Exactly the same reasoning can be applied to all the remaining $\mathcal{A}_i$ regions. In particular, after each blockage $\mathcal{B}_{i-1}$ we try to look for the next type-2 blockage of $L$-independent size at a fixed distance $\ell = \mathcal{O}(L^0)$ to the right of $\mathcal{B}_{i-1}$, with $\ell \gg 1$. If no such blockage is found at such $\ell$ distance, we keep scanning to the right, until we find $\mathcal{B}_{i}$ at a distance $\ell_i$ from $\mathcal{B}_{i-1}$. From the argument above we are guaranteed that for any subregion $\mathcal{A}_i$ the size $\ell_i \ge \ell$ is subextensive, i.e., it scales as $\ell_i = \mathcal{O}(L^{\gamma_i})$ for some $0 \le \gamma_i < 1$.
	We can now generalise Eq.~(\ref{eq:dimABCbound}) from \cref{sec:strongfrag2} to the multipartite case as (to simplify the notation we now indicate $D^{(d)}_{N_{\mathcal{A}},X_{\mathcal{A}}}$ simply as $D^{(d)}_{\mathcal{A}}$)
	\begin{equation}
		\label{eq:dimABmultipartite}
		\mathcal{D}^{(d,k)}_\text{max} <
		D_{\mathcal{A}_1+\mathcal{B}_1}^{(d)} \left[\prod_{i=2}^{G-1} D_{\mathcal{B}_{i-1}+\mathcal{A}_{i}+\mathcal{B}_i}^{(d)} \right]  D_{\mathcal{B}_{G-1}+\mathcal{A}_{G}}^{(d)} \, . 
	\end{equation}
	Using the analog of \eqref{eq:QboundDmax} in the multipartite case we get 
	\begin{equation}
		\label{eq:multipartQbound}
		\mathcal{D}^{(d,k)}_\text{max} <
		Q^{G-1} \left[\prod_{i=1}^{G-1} D_{\mathcal{A}_{i}+\mathcal{B}_i}^{(d)} \right]  D_{\mathcal{A}_{G}}^{(d)} \ ,
	\end{equation}
	where $Q$ is again an $L$-independent positive constant. The right-hand side of the previous inequality involves the total number of configurations in a partition of the chain into $G$ regions that cannot exchange particles and dipole moment with each other. Using Eq.~(\ref{eq:DNXscaling}) together with the generalisation of Eq.~(\ref{eq:gitildeNXineq}) to the multipartite case, we obtain
	\begin{equation}
		\label{eq:finalbound_appendix}
		r^{(d,k)}_{N,X} < L^2 \prod_{i=1}^G \frac{\widetilde{Q}}{\ell_i^2} \ . 
	\end{equation}
	Here $\widetilde{Q}$ is an $L$-indepdendent positive constant that has absorbed the $Q$ factors from (\ref{eq:multipartQbound}) and the $\mathcal{O}(L^0)$ constants arising from the use of (\ref{eq:DNXscaling}) on every $D_{\mathcal{A}_i+\mathcal{B}_i}^{(d)}$ and on $D_{N,X}^{(d)}$ from (\ref{eq:ratio1}). Importantly, we choose the $L$-independent distance $\ell$ to be much larger than the size of any $\mathcal{B}_i$. In this way we can always guarantee that $\ell_i^2 \ge \ell^2 > \widetilde{Q}$. We see that if $G$ is extensively large, i.e.~if after applying the subdivision algorithm we find among the set of all $\mathcal{A}_i$ regions a subset of extensively many sizes $\ell_i=\mathcal{O}(L^0)$, then $r^{(d,k)}_{N,X}$ in \cref{eq:finalbound_appendix} decays to zero exponentially with $L$. The fact that $G=\mathcal{O}(L)$ is guaranteed by our requirement of having an extensive number of type-2 edges in the system (given that Case 3 requires a zero fraction of type-1 blockages). Indeed, assume by contradiction that $G=\mathcal{O}(L^{\gamma_g})$, with $\gamma_g < 1$. The total number of type-2 edges in the system are obtained by summing the following contributions: 
	\begin{enumerate}
		\item Two edges per $\mathcal{B}_i$ blockage and at most  $\mathcal{O}(L^0)$ edges per each $\mathcal{A}_i$ region with size $\ell_i=\mathcal{O}(L^0)$.
		\item At most $\ell_i/h(L)$ edges in each $\mathcal{A}_i$ region of size $\ell_i=\mathcal{O}(L^{\gamma_i})$ with $0<\gamma_i<1$. Here $h(L)$ is again a function that diverges for $L\to\infty$. 
	\end{enumerate}
	Summing these contributions under the assumption of a subextensive $G$ yields a subextensive number of type-2 edges in the chain, which contradicts our requirements of having an extensive number of them. Hence $G=\mathcal{O}(L)$ and $r_{N,X}^{(d,k)}$ decays exponentially with $L$.

\subsection{Strong fragmentation in atypical families}
\label{eq:strongfragatypical}
    
	In this appendix we discuss fragmentation in atypical families of $(N,X)$ sectors, i.e., those families having intensive centre of mass $\nu_x=X/(NL)$ such that $\nu_{x_0}=\nu_x - 1/2$ is finite for $L \to \infty$. Even together, all these families contain only an exponentially vanishing fraction of all configurations associated with a given $N=\nu L$ sector (see \cref{sec:sizesymmetrysec}). As clear from the analysis of Sections \ref{sec:strongfrag1} and \ref{sec:strongfrag2}, the presence of a nonzero density of type-1 or type-2 blockages for $\nu<(k-2)^{-1}$ holds regardless of the value of $\nu_{x_0}$. Thus, the physics of atypical families of $(N,X)$ sectors characterised in the thermodynamic limit by $\nu_{x_0}\neq 0$ and $\nu<(k-2)^{-1}$ is still dominated by an extensive number of sites involved in blockages. 
	
	Proving strong fragmentation in atypical families (according to the ratio of dimensions $r_{N,X}^{(d,k)}$) by simple arguments similar to those employed for typical families requires the determination of the function $\Lambda_d(\nu,\nu_{x})$ appearing in Eq.~(\ref{eq:DNXscaling}), a task that goes beyond the numerical results discussed in Appendix \ref{appendix:scalingofsizeofss}. However, consider the domain $\nu<(k-1)^{-1}$, where necessarily a fraction $f>0$ of all sites is involved in type-1 blockages. Call $\Delta_{d,\nu}(f)$ the difference
	\begin{equation}
		\Delta_{d,\nu}(f) = g_{d,\nu}(0, f) - \eta_d(\nu) < 0 \ , 
	\end{equation}
    where $g_{d,\nu}$ has been defined in \cref{eq:bipartitionchain}. 
	If $\nu_{x_0}$ is different from zero in the thermodynamic limit, Eq.~(\ref{eq:useofg1}) can be replaced by
	\begin{equation}
		\label{eq:useofg3}
		\lim_{L\to\infty} \frac{1}{L} \ln r^{(d,k)}_{N,X} \le \Delta_{d,\nu}(f) + \Lambda_d(\nu,\nu_x)  \ . 
	\end{equation}
	This trivially proves that all the atypical families with $|\nu_{x_0}|$ different from zero but sufficiently small and $\nu<(k-1)^{-1}$ are strongly fragmented, with ratio $r^{(d,k)}_{N,X}$ decaying to zero exponentially with $L$. This is because, due to \cref{eq:Lambdadnux0}, there always exists a nonzero value of $|\nu_{x_0}|$ below which the left-hand side of \eqref{eq:useofg3} is negative. 
	
	More in general, for $\nu<\nu_c=(k-2)^{-1}$ and without requiring explicit knowledge of $\Lambda_d(\nu,\nu_x)$, if one can prove that in the largest Krylov subsectors of an atypical family of symmetry sectors, one of the extensively many blockages partitions the system into two extensive subregions, each characterised by a number of particles $N$ and dipole moment $X$ that are compatible with an exponentially large number of configurations according to the scaling in Eq.~(\ref{eq:DNXscaling}), then application of Eq.~(\ref{eq:gitildeNXineq}) would ensure that the ratio $r^{(d,k)}_{N,X}$ decays to zero at least as $L^{-2}$, as seen from Eq.~(\ref{eq:finalbound}) when one sets $G=2$. Note that a polynomial decay of $r^{(d,k)}_{N,X}$ is enough to rule out weak fragmentation, i.e.~the existence of a dominant Krylov sector that occupies a measure-1 fraction of its symmetry sector for $L\to\infty$. Similar reasoning can also be applied to cases with $G>2$ by making use of the disconnecting property of multiple blockages.
	
	Finally, we notice that it is possible that in addition to weak-to-strong fragmentation transitions induced by tuning $\nu$, similar transitions might be induced by varying $\nu_{x_0}$. In particular, in families with $\nu_{x_0}$ significantly different from $0$ and filling $\nu < \nu_c$, it might be possible that the dipole polarisation constrains the majority of particles into a subregion of the chain with a local density of particles larger than $\nu_c$, thus locally suppressing the presence of blockages. This mechanism might be at the basis of a transition from strong to weak fragmentation within $N$-families characterised by $\nu<\nu_c$. It might also open the possibility to the coexistence of weak fragmentation (as defined by \eqref{eq:ratio2}) and a finite density of frozen empty sites. We leave the study of this interesting problem for future work. \\
	
	\section{Calculation of densities of various particle configurations}
	\label{app:DSandP}
	
	In this appendix, we compute several functions of use for Section \ref{sec:StrongFragFurtherAnalysis}.

    \begingroup
    \renewcommand{\addcontentsline}[3]{}
	\subsection{Calculation of $b_1^{(2,k)}(N,L)$ and $b^{(\infty,k)}_1(N,L)$}
	\label{appendix:DS}
    
	Here we compute  an exact expression for the functions $b_1^{(2,k)}(N,L)$ and $b_1^{(\infty,k)}(N,L)$ which for $d=2$ and $d=\infty$ respectively give the total number of ways to arrange $N$ particles over $L$ sites such that they satisfy the no-propagation constraints $i^0_n\geq(n-1)(k-2)$ derived in \cref{sec:nopropcondition}. For $d=2$, we have
	\begin{equation}
		b_1^{(2,k)}(N,L) =  \sum_{i_{N}=(N-1)(k-2)}^{L-1} \dots \sum_{i_2=(k-2)}^{i_3-1} \, \sum_{i_1=0}^{i_2-1} 1  \ .
	\end{equation}
	
	This sum is identical to
	\begin{equation}
		b_1^{(2,k)}(N,L) = \tilde{b}_1^{(2,k)}(N,L) =  \sum_{i_{N}=N (k-2)}^{y} \dots \sum_{i_2=2(k-2)}^{i_3-1} \, \sum_{i_1=k-2}^{i_2-1} 1 
	\end{equation}
	when $y=L + k-3$. For simplicity in some of the following calculations, we consider this sum instead. We shall prove that 
	\begin{equation}
		\tilde{b}_1^{(2,k)}(N,L) = \frac{y-N(k-2)+1}{y-N+1}\binom{y}{N} 
	\end{equation}
	inductively in $N$. The case $N=1$ is trivial. Assume that the formula holds up to $N-1$, and consider it for $N$. We then find
	\begin{widetext}
		\begin{equation}
			\begin{aligned}
				\tilde{b}_1^{(2,k)}(N,L) &= \sum_{i_N = N(k-2)}^{y} D'_k(N-1,i_N-1) =\sum_{i_N = N(k-2)}^{y} \frac{i_N -(N-1)(k-2)}{i_N -N}\binom{i_N-1}{N-1}\\
				&= \sum_{i_N = N(k-2)}^{y} \left[\binom{i_N}{N-1} -(k-2)\binom{i_N-1}{N-2}\right] =\left(\sum_{i_N = N-1}^{y}-\sum_{i_N = N-1}^{N(k-2)-1}\right) \left[\binom{i_N}{N-1} -(k-2)\binom{i_N-1}{N-2}\right]\\
				& = \binom{y+1}{N}-(k-2)\binom{y}{N-1} = \frac{y-N(k-2)+1}{y-N+1}\binom{y}{N} \ , 
			\end{aligned}
		\end{equation}
	\end{widetext}
	where in the before-last step we made use of the hockey stick identity. Setting $y=L+k-3$ we obtain the desired result:
	\begin{equation}
		b_1^{(2,k)}(N,L)=\binom{L+(k-2)}{N}\frac{L-(N-1)(k-2)}{L+(k-2)}
		\label{eq:b1eq}
	\end{equation}
	
	As concerns the $d=\infty$ case, due to stacking we have
	\begin{equation}
		b_1^{(\infty,k)}(N,L) = \sum_{i_{N}=(N-1)(k-2)}^{L-1} \dots \sum_{i_2=(k-2)}^{i_3} \, \sum_{i_1=0}^{i_2} 1 .
	\end{equation}
	Following identical logic to the previous derivation, one can then show
	\begin{equation}
		b_1^{(\infty,k)}(N,L)= \frac{L - (N - 1) (k - 2)}{L + k - 2} \binom{L + k + N - 3}{ N}
	\end{equation}

	\subsection{Lower bound on the density of particular particle configurations}
	\label{appendix:lowerboundonP}
	We next compute a function which has several applications in the main text, in particular in lower bounding the density of type-1 blockages and type-2 edges in \cref{subsec:MeanB1B2andp1} and the density of active bubbles in \cref{subsec:ActiveBubbleDensityFunction}. In particular, we compute the thermodynamic limit of the function
	\begin{widetext}
		\begin{equation}
			P^{(2,k)}(x,m,N,L)=\frac{\sum_{L_R=0}^{L-m}\sum_{N_R=N_\text{min}}^{N_\text{max}} b_1^{(2,k)}(N_R,L_R)b_1^{(2,k)}(N-N_R-x,L-L_R-m)}{L\binom{L}{N}}.
			\label{eq:PkformulaAppendix}
		\end{equation}
	\end{widetext}
	For a particular local configuration of $x$ particles over $m$ sites in a $d=2$ system, this sum equals the average density of occurrences of that configuration among all states with $N$ particles and $L$ sites, such that the particles to the right and left of the configuration satisfy the no-propagation condition of \cref{eq:NoPropCond}. The average is carried out over an entire $N$ sector. The bounds on the sum over $N_R$ are given by 
	\begin{equation}
		N_\text{min} = \text{max} \left( 0, N- \left\lfloor \frac{L-L_R-m+k-3}{k-2} \right\rfloor \right) \ ,
	\end{equation}
	\begin{equation}
		N_\text{max} = \text{min} \left( N, \left\lfloor \frac{L_R+k-3}{k-2} \right\rfloor \right) . 
	\end{equation}
	The sum on the right-hand side of Eq.~(\ref{eq:PkformulaAppendix}) involves polynomially many in $L$ terms. We will show that a subset of the summands exponentially dominates in $L$ over the others, and thus the leading order of the total sum coincides with the sum of this subset. Using \cref{eq:b1eq} we rewrite the product of $b_1^{(2,k)}$'s in Eq.~(\ref{eq:PkformulaAppendix}) as
	\begin{equation}
		\begin{aligned}
			\label{eq:productofDks}
			&b_1^{(2,k)}(N_R,L_R)b_1^{(2,k)}(N-N_R-x,L-L_R-m)  \\
			& \ \ \ \ = \binom{L_R}{N_R}\binom{L-L_R}{N-N_R} \, \alpha(k,x,m,N,L,N_R,L_R) \ .
		\end{aligned}
	\end{equation}
	It is easy to check that $\alpha(k,x,m,N,L,N_R,L_R)$ is $\mathcal{O}(L^0)$ and thus the only exponential dependence on $L$ comes from the product of the two binomials. We introduce the intensive quantities  
	\begin{equation}
		\ell_R = \frac{L_R}{L} \qquad \quad \nu_R = \frac{N_R}{L} \ , 
	\end{equation} 
	and define
	\begin{equation}
		\begin{aligned}
			M_{\nu,\ell_R}(\nu_R) &= \binom{L_R}{N_R}\binom{L-L_R}{N-N_R}\Big/ \binom{L}{N} \\
			& = \binom{\ell_R \, L}{\nu_R \, L}\binom{(1-\ell_R)L}{(\nu-\nu_R)L} \Big/ \binom{L}{\nu\,L} \ .
		\end{aligned}
	\end{equation}
	Using Stirling's formula for the asymptotic value of the binomials, we find that $M_{\nu,\ell_R}(\nu_R)$ is given by a product
	\begin{equation}
		\label{eq:MnuellRnuR}
		M_{\nu,\ell_R}(\nu_R) = \left[M_{\nu,\ell_R}^{(1)}(\nu_R)\right]^L \, M_{\nu,\ell_R}^{(2)}(\nu_R) \, \,  L^{-1/2} \ ,
	\end{equation}
	where both $M_{\nu,\ell_R}^{(1)}$ and $M_{\nu,\ell_R}^{(2)}$ are $\mathcal{O}(L^0)$. It is easy to show that $M_{\nu,\ell_R}^{(1)}$ has only one global maximum for $\nu_R^* = \nu \, \ell_R$, where it takes the value
	\begin{equation}
		M_{\nu,\ell_R}^{(1)}(\nu_R^*) = 1 \ . 
	\end{equation}
	Taylor expanding $M_{\nu,\ell_R}^{(1)}$ in a small neighborhood around $\nu_R^*$ we get
	\begin{equation}
		\begin{aligned}
			\label{eq:expandingM1nuellR}
			& \left[M_{\nu,\ell_R}^{(1)}(\nu_R)\right]^L = \\
			&  \ \ \ \exp \left\{ -L \left[\frac{(\nu_R-\nu_R^*)^2}{2 \sigma^2} + O\left((\nu_R-\nu_R^*)^3\right) \right]\right\} \ ,
		\end{aligned}
	\end{equation}
	with
	\begin{equation}
		\sigma^2 = \ell_R (1-\ell_R) \, \nu (1-\nu) \ . 
	\end{equation}
	In the thermodynamic limit we can transform the sum over $N_R$ into an integral 
	\begin{equation}
		L^{-1/2} \sum_{N_R} = L^{1/2} \frac{1}{L} \sum_{N_R=N_\text{min}}^{N_\text{max}} \to \, \, L^{1/2} \int_{\nu_R^{\text{min}}}^{\nu_R^{\text{max}}} d \nu_R \ ,
	\end{equation}
	where the factor $L^{-1/2}$ comes from Eq.~(\ref{eq:MnuellRnuR}). From \eqref{eq:expandingM1nuellR} it is clear that, aside from corrections that go to zero faster than any power law in $L$, only a small neighborhood around $\nu_R^*$ of width of the order of $1/\sqrt{L}$ contributes to the integral. In this neighborhood we can neglect the $O\left((\nu_R-\nu_R^*)^3\right)$ term from Eq.~(\ref{eq:expandingM1nuellR}). Thus, by defining
	\begin{equation}
		\tilde{\alpha}_{k,x,m,\nu,\ell_R}(\nu_R) = \alpha(k,x,m,\nu \, L,L,\nu_R \, L,\ell_R \, L) \ ,
	\end{equation}
	we can rewrite Eq.~(\ref{eq:PkformulaAppendix}) as
	\begin{equation}
		\begin{aligned}
			\label{eq:valueforWk}
			&P^{(2,k)}(x,m,\nu L,L) = \\
			&\int_0^1 d\ell_R\int_{-\infty}^{\infty} d \nu_R \, \tilde{\alpha}_{k,x,m,\nu,\ell_R}(\nu_R) \, M_{\nu,\ell_R}^{(2)}(\nu_R) \, \frac{e^{-\frac{(\nu_R-\nu_R^*)^2}{2 \sigma^2/L} }}{1/\sqrt{L}} \\
			&\qquad \qquad \qquad \qquad \qquad + \varepsilon(L) \ ,
		\end{aligned}
	\end{equation}
	where $\varepsilon(L)$ goes to zero faster than any power law in $L$. Notice that in the previous expression we have extended the integral boundaries to $\pm \infty$ and absorbed the associated error inside $\varepsilon(L)$. Given that only a $\mathcal{O}(1/\sqrt{L})$ region around $\nu_R^*$ contributes to the integral in \eqref{eq:valueforWk}, we can expand $\tilde{\alpha}_{k,x,m,\nu,\ell_R} \, M_{\nu,\ell_R}^{(2)}$ in powers of $(\nu_R-\nu_R^*)$:
	\begin{equation}
		\begin{aligned}
			\label{eq:expandingAtimesM}
			&\tilde{\alpha}_{k,x,m,\nu,\ell_R}(\nu_R) \, M_{\nu,\ell_R}^{(2)}(\nu_R) = \\
			&\frac{1}{\sqrt{2 \pi \sigma^2}}\nu^x (1-\nu)^{m+2-2(k-1)} (1-\nu/\nu_c)^2 \\
            &\quad \quad + \mathcal{O}(\nu_R-\nu_R^*) \ . 
		\end{aligned}
	\end{equation}
	The odd powers in this expansion do not contribute to the final result as they vanish within the remaining Gaussian integral, while each $(\nu_R-\nu_R^*)^{2n}$ with $n\ge1$ contributes an order $\mathcal{O}(L^{-n})$. Thus we only need to worry about the leading order in Eq.~(\ref{eq:expandingAtimesM}), which brings us to 
	\begin{equation}
		\begin{aligned}
			P^{(2,k)}(x,m,&\nu L,L) = \\
			 \int_0^1 d\ell_R & \nu^x (1-\nu)^{m+2-2(k-1)} (1-\nu/\nu_c)^2\\
             &\int_{-\infty}^{\infty} d \nu_R  \, \frac{e^{-\frac{(\nu_R-\nu_R^*)^2}{2 \sigma^2/L} }}{\sqrt{2 \pi \sigma^2/L}}  + \mathcal{O}(1/L) \ . 
		\end{aligned}
	\end{equation}
	Given that the Gaussian integral in the previous equation is properly normalised, carrying the integral through we obtain
	
	\begin{equation}
    \begin{aligned}
		\lim_{L\rightarrow\infty}P^{(2,k)}(x,m&,\nu L,L) =\\
        &\nu^x (1-\nu)^{m+2-2(k-1)} (1-\nu/\nu_c)^2
        \label{eq:P2formula}
    \end{aligned}
	\end{equation}
	
	This can be used to obtain the type-1 blockage density of  \cref{eq:lowerbavdend2} by setting $m=2k-3$ and $x=1$, the type-2 edge density of \cref{eq:LowerBound_t2_d2} by setting $x=2$ and $m=3k-6$, and the active bubble configuration density of  \cref{eq:AB_d2} by setting $m=\ell+2(k-1)$.
	
	The corresponding function for $d=\infty$ is given by
	\begin{widetext}
		\begin{equation}
			P^{(\infty,k)}(x,m,N,L)=\frac{\sum_{L_R=0}^{L-m}\sum_{N_R=N_\text{min}}^{N_\text{max}} b_1^{(\infty,k)}(N_R,L_R)b_1^{(\infty,k)}(N-N_R-x,L-L_R-m)}{L\binom{L+N-1}{N}}.
			\label{eq:PkinfformulaAppendix}
		\end{equation}
	\end{widetext}
	
	The thermodynamic limit expression is derived using an identical approach, and found to be
	\begin{equation}
		\begin{aligned}
			\lim_{L\rightarrow\infty}P^{(\infty,k)}(x,m&,\nu L,L) =\\
			&\nu^{x} (1+\nu)^{-(x+m+2-2(k-1))}(1-\nu/\nu_c)^2 \ .
            \label{eq:PInfformula}
		\end{aligned}
	\end{equation}
	From this we obtain the type-1 blockage density of  \cref{eq:LowerBound_pb1_dinf} by setting $x=1$ and $m=2k-3$, the type-2 edge density of \cref{eq:LowerBound_t2_dinf} by setting $x=2$ and $m=3k-6$, and the active bubble configuration density of  \cref{eq:AB_dinf} by setting $m=\ell+2(k-1)$.
    \endgroup

    \section{Distribution of blockages in typical states}
	\label{app:SelfAveraging}
	
	It is ubiquitous in statistical mechanics to find that properties averaged over an ensemble of states match the properties of individual states in the typicality class that dominates the average. Hence, we expect in the present case the density of type-1 blockages and of type-2 edges in an individual typical state for any $d$ to approach the corresponding average density in the chosen $N=\nu L$ sector in the thermodynamic limit. We show this explicitly by making use of self-averaging \cite{Wiseman_SelfAveraging} arguments.

	Given the regime $\nu<\nu_c$, we know from Section \ref{sec:strongfrag} that any state must host an extensive number of sites that are part of either type-1 blockages or type-2 edges. We restrict our attention to typical states, which as proved in Section \ref{sec:sizesymmetrysec} have, in any extensive subregion, local filling approaching the global $\nu$ for $L \to \infty$. Given these premises, we can imagine applying the FES picture to any given typical state and then applying to the resulting FES the same subdivision algorithm used in \cref{sec:strongfrag2,sec:detailscase3} and depicted in Fig.~\ref{fig:type2partition}. In this way, we partition the chain into $G$ dynamically disconnected subregions $\mathcal{A}_i$, with sizes $\ell_i \ge \ell \gg 1$ such that $\ell_i=\mathcal{O}(L^{\gamma_i})$ for some $0\le \gamma_i<1$, that are separated by blockages $\mathcal{B}_i$ of $L$-independent size. The only two differences with respect to the algorithm applied in Section \ref{sec:strongfrag2} are
	\begin{enumerate}
		\item Now the $\mathcal{B}_i$'s can be either type-1 or type-2 blockages. In this regard notice that a type-1 blockage of any size can be counted as a blockage of $L$-independent size, by considering only a subpart of it as a blockage, if necessary. 
		\item Starting from blockage $\mathcal{B}_{i-1}$, the search for the next blockage is only stopped when we find an $L$-independent blockage $\mathcal{B}_{i}$ such that the region $\mathcal{A}_i$ has local filling sufficiently close to the global one $\nu$. This is always possible to achieve while retaining $\gamma_i<1$, given that from typicality any extensively large region enclosed by two blockages of $L$-independent size must have filling approaching $\nu$ in the thermodynamic limit. 
	\end{enumerate}
	Also, here we are choosing $\ell$ much larger than the size of any $\mathcal{B}_i$ blockage, so that the dynamical overlap associated with the presence of type-2 blockages among the $\mathcal{B}_i$'s is negligible. Furthermore, we notice that by construction $G$ diverges for $L\to\infty$. Having obtained $G$ dynamically disconnected subregions all with local filling close to the global $\nu$, we can compute the density of both type-1 blockages and type-2 edges in each of the $G$ many $\mathcal{A}_i$ subregions. Over asymptotically large chains, these can be regarded as many independent realisations of a random variable whose expected value coincides with the average density of type-1 blockages or type-2 edges calculated over the entire $N=\nu L$ sector in \cref{subsec:MeanB1B2andp1}. By the central limit theorem we thus expect the global density of type-1 blockages and type-2 edges, in any asymptotically large typical state for $\nu<\nu_c$ and any $d$, to match those obtained by computing the average densities over the entire $N$ sector. The previous argument based on the subdivision algorithm and self-average also proves that type-1 blockages and type-2 edges must be quite uniformly distributed along the chain in any typical state. 
	
	Note that combining the results just derived with the lower bounds on the average density of type-1 blockages and type-2 edges from \cref{subsec:MeanB1B2andp1}, we prove for $d=2$ and $d=\infty$ that individual typical states possess both a nonvanishing density of type-1 blockages and of type-2 edges for $\nu<\nu_c$. We expect the same to be true for any other value of $d$ and numerically show it in \cref{subsec:type1distributionNumerics}. \\
	
	\section{Lower bounds on $A^{(2,k)}(x,N,L)$ and $A^{(\infty,k)}(x,N,L)$}
	\label{app:activebubbleDistLowerBounds}
	
	In this section, we proceed to explicitly compute some lower bounds on the $d=2$ and $d=\infty$ active bubble density functions, $A^{(2,k)}(x,\nu L,L)$ and $A^{(\infty,k)}(x,\nu L,L)$, as defined in Section \ref{subsec:ActiveBubbleDensityFunction}, for various values of $k$ and $x$. We begin with the simplest case of $A^{(\infty,k)}(x,\nu L,L)$, which we lower bound for $k=7$ and from $x=2$ to $x=6$. We have that
	\begin{equation}
		\begin{aligned}
			A^{(\infty,7)}(x,&\nu L,L) =\\
			&\sum_{\ell=1+5(x-1)}^{1+6(x-1)} m^{(\infty,7)}(x,\ell) a^{(\infty,7)}(x,\ell,N,L),
		\end{aligned}
	\end{equation}
	The upper summation bound is obtained from the $d$-independent formula
    \begin{equation}
		\ell_{\text{max}}(x,k) = 1+(k-1)(x-1),
        \label{eq:l2max}
	\end{equation}
    which corresponds to a particle configuration where all $x$ particles are separated from their neighbours by $k-2$ holes. The lower summation bound is obtained from the general $d=\infty$ analytic expression
    \begin{equation}
        \ell_{\infty,\text{min}}(x,k) = 1+(k-2)(x-1) \ ,
    \end{equation}
    which corresponds to an active bubble configuration with $x$ particles each separated by $k-3$ holes from their neighbours. A lower bound on $a^{(\infty,7)}(x,\ell,N,L)$ in the thermodynamic limit is provided in \cref{eq:AB_dinf}. Hence, all that remains to be calculated are the multiplicities $m^{(\infty,7)}(x,\ell)$. Keeping in mind the definition of active bubble from \cref{subsec:ActiveBubbleDensityFunction} (see summary in \cref{tab:glossary}), this can be done methodically with the aid of a computer as follows. For each value of $\ell$ in the summation, $m^{(\infty,7)}(x,\ell)$ will equal the sum of the dimensions of all local Krylov sectors with a corresponding sub-FES consisting of $\ell-(1+5(x-1))$ pairs of particles separated by $k-2$ holes and $1+6(x-1)-\ell$ pairs of particles separated by $k-3$ holes. For example, $m^{(\infty,7)}(x,\ell=1+5(x-1))$ will equal the dimension of the local Krylov sector that contains the sub-FES in which all pairs of particles are separated by $k-3$ holes. On the other hand, $m^{(\infty,7)}(x,\ell=2+5(x-1))$ will equal the sum of the dimensions of the $x-1$ local Krylov sectors defined by the sub-FESs with exactly one pair of particles separated by $k-2$ holes; and so on. The dimensions of the local Krylov sector associated with a given sub-FES can be determined by starting from that FES, then computing all possible states that can be reached from the FES via a series of inward hop operators. This necessarily maps out the whole Krylov sector, as by the results in Appendix \ref{appendix:uniqueFES} any state in a Krylov sector of a $d=\infty$ system can be reached by starting from the unique FES and applying inward hops. We present all the multiplicity values derived in this fashion in \cref{tab:d_inf_k_7}, where the boxes filled are those between the bounds of $\ell_{\text{min}}$ and $\ell_{\text{max}}$.
	\begin{widetext}
    \begin{center}
    \begin{table}[h!]
    \caption{Active bubble configuration multiplicities for $d = \infty$, $k = 7$.}
    \label{tab:d_inf_k_7}
    \begin{tabular}{ |p{1cm}||p{2.5cm}|p{2.5cm}|p{2.5cm}|p{2.5cm}|p{2.5cm}|p{2.5cm}|  }
        \hline\hline
        \multicolumn{7}{|c|}{$m^{(\infty,7)}(x,\ell)$} \\
        \hline
        & $\ell= 1+5(x-1)$ & $\ell= 2+5(x-1)$ & $\ell= 3+5(x-1)$ &
          $\ell= 4+5(x-1)$ & $\ell= 5+5(x-1)$ & $\ell= 6+5(x-1)$ \\
        \hline
        $x=2$ & 3 & 4 & & & & \\
        \hline
        $x=3$ & 18 & 42 & 7 & & & \\
        \hline
        $x=4$ & 131 & 462 & 108 & 19 & & \\
        \hline
        $x=5$ & 1111 & 5268 & 1446 & 408 & 40 & \\
        \hline
        $x=6$ & 10462 & 62185 & 18688 & 6723 & 1077 & 97 \\
        \hline
    \end{tabular}
    \end{table}
    \end{center}
    \end{widetext}

	We next consider, for $k=5$, the $d=2$ multiplicity function $m^{(2,5)}(x,\ell)$. Although a given active bubble configuration cannot necessarily attain its corresponding sub-FES for $d$ finite via a series of inward and outward hops, one can show via an exhaustive enumeration that for $x=2$ to $x=5$ this is the case. Hence the local Krylov sectors for these values of $x$ can be determined by starting from the corresponding sub-FES and mapping out all states attainable via a combination of inward and outward hops. This then yields the results in \cref{tab:d_2_k_5}.
	\begin{widetext}
		\begin{center}
        \begin{table}[h!]
        \caption{Active bubble configuration multiplicities for $d = 2$, $k = 5$.}
        \label{tab:d_2_k_5}
			\begin{tabular}{ |p{1cm}||p{2.5cm}|p{2.5cm}|p{2.5cm}|p{2.5cm}| p{2.5cm}| }
				\hline\hline
				\multicolumn{6}{|c|}{$m^{(2,5)}(x,\ell)$} \\
				\hline
				& $\ell= 1+5(x-1)$& $\ell= 2+5(x-1)$ &$\ell= 3+5(x-1)$ &$\ell= 4+5(x-1)$ &$\ell= 5+5(x-1)$ \\
				\hline
				$x=2$   & 2 & 2 & & & \\
				\hline
				$x=3$&5 &12 &3 & & \\
				\hline
				$x=4$& 15&62 &27 &5 &\\
				\hline
				$x=5$ &56 &318 &180 &62 &8 \\
				\hline
			\end{tabular}
        \end{table}
		\end{center}
	\end{widetext}
	
	\section{Algorithm for efficiently mapping to the FES}
	\label{app:FESalgorithm}

     The simplest approach for mapping a given initial state to its corresponding FES is to apply a series of outward hop gates to nearby pairs of particles in the state until no more outward hops are possible. However, this becomes computationally expensive at the large system sizes considered in this paper. In this appendix, we present a more efficient algorithm for mapping from an initial state at any $d$ to its corresponding FES in the FES picture (see \cref{tab:glossary}). This algorithm remains efficient even for system sizes $L\sim 10^6$ and densities $\nu$ close to the critical density $\nu_c$.
     
     We note that in this appendix we take the leftmost site of the chain to have index 1 (instead of $0$ as in the rest of the paper), and the rightmost to have index $L$. The algorithm makes use of the following property. We define a ``blockage-free group'' to be a collection of occupied sites separated by $k-4$ holes or fewer from each other. Occupied sites with no other nearby occupied sites are considered in and of themselves blockage-free groups; in particular, in systems with $k=3$, all occupied sites are individually blockage-free groups. We then have that a blockage-free group embedded in an otherwise empty system (with a large number of holes on either side) will, when acted on by a series of outward hops, expand into a particle-connected (PC) string (see \cref{tab:glossary}). Furthermore, if we had an (otherwise empty) system containing two blockage-free groups, and those two groups were separated by $k-4$ sites or fewer at some point during their expansion, then the two groups would merge and collectively form one PC string. Both of these results follow immediately from the 2-colour connectivity discussed in Section \ref{sec:BlockagesFES2colour} and the uniqueness result of Appendix \ref{appendix:uniqueFES}. The algorithm for expanding to an FES is then as follows.
   
	\begin{itemize}
		\item Divide the state into blockage-free groups.
		\item For each blockage-free group, use the number of particles it contains as well as its local dipole moment to compute the leftmost and rightmost sites it would occupy if, as described above, it were to be expanded to a PC string (or a PC string plus a stack of particles on the boundaries if the expanding particles overlap with them).
		\item Merge any pair of blockage-free groups for which the corresponding PC strings are separated by $k-4$ sites or less, and compute the bounds of the new resulting PC string. Repeat this merging procedure until no further merging can occur.
		\item Use the particle number and dipole moment of each final blockage-free group to uniquely determine the location of each of the particles in the corresponding PC string, and use these results to construct the FES of the entire system.
	\end{itemize}
	
	We next show how to calculate the leftmost and rightmost sites of the PC string corresponding to a blockage-free group, as well as the positions of the particles in the PC string. By definition, the PC string will have at most one pair of particles separated by $k-2$ holes and all others separated by $k-3$ holes. Say that the blockage-free group has $n$ particles and a dipole moment $p$, and the particles do not overlap with the boundaries during their expansion. Let $i_1$ denote the leftmost site of the PC string and let $i_f$ denote the rightmost. Furthermore, let $j\in\{1,\dots,n\}$ denote the index of the particle to the right of which there is a spacing of $k-2$ holes (so if there are $k-2$ holes between the first and second particles, $j=1$). If there are no spacings of $k-2$ holes, then $j=n$. In this case, the local dipole moment equals
	\begin{equation}
		\begin{aligned}
			p &= n\, i_1 + \sum_{m=1}^{n-1}m(k-2) + (n-j)\\
			&=n\, i_1 + n(n-1)(k-2)/2 + (n-j).
		\end{aligned}
	\end{equation}
	From this equation one finds that
	\begin{equation}
		\begin{aligned}
			i_1 &= \lfloor(p-n(n-1)(k-2)/2)/n\rfloor\\
			j &= n(i_1+1) + n(n-1)(k-2)/2 -p\\
			i_f &= i_1 + (n-1)(k-2) +(1-\delta_{j,n})
		\end{aligned}
	\end{equation}
	We next consider the case where the blockage-free group, when expanded, overlaps with exactly one of the boundaries, such that the particles form a PC string as well as a stack of particles on that boundary site. We assume the overlap to be with the left boundary (with the derivation for the right boundary following identical logic). Let $m_l$ denote the number of particles piled up on the left boundary, $i_1$ the site of the leftmost particle \textit{not on the boundary} of the PC string, $j\in\{1,\dots,n-m_l\}$ the index, among the particles \textit{not on the boundary}, of the particle to the right of which there are $k-2$ holes (with $j=n-m_l$ if there is none) and $i_f$ the site of the rightmost particle. In this case, the dipole moment equals
	\begin{equation}
        \begin{aligned}
		p &= m_l +(n-m_l)\, i_1 + (n-m_l)(n-m_l-1)\frac{(k-2)}{2} \\
        & \qquad \qquad \qquad + (n-m_l-j) \ .
         \end{aligned}
	\end{equation}
	Note that in the above equation $2\leq i_1\leq k$. We proceed as follows: we know that of the possible configurations of the remaining $n-m_l$ particles, the one with the smallest dipole moment is given by $i_1=2$ and $j=n-m_l$. Hence, if we set $i_1$ and $j$ to these values in the above equation, the solution for $m_l$ rounded up to the next integer (if the solution is not already an integer) will give the correct $m_l$ value. Indeed, by the proof of uniqueness of the FES in Appendix \ref{appendix:uniqueFES}, if $m_l$ were lower or higher than this value, then it would be impossible to arrange the particles not on the boundary into a PC string such that the particles collectively had a dipole moment $p$. With $m_l$ determined, the remaining variables can be solved, as before, by determining the positions of the particles in the PC string.
	
	We finally consider the case where the expanded blockage-free group in fact spans the whole system, with particles on both of the two boundaries. In this case, let $m_l$ be the numbers of particles on the left boundary, $m_c$ the number of particles on neither boundary, $i_1$ the site of the leftmost particle not on a boundary, $i_f$ the site of the rightmost particle not on a boundary, and $j\in\{1,\dots,m_c\}$ the index of the particle to the right of which a sequence of $k-2$ holes are situated (with $j=m_c$ either if there is none or if the $k-2$ holes are right next to the right or left boundary). Noting this implies there are $n-m_l-m_c$ particles on the right boundary, we then have that
	\begin{equation}
        \begin{aligned}
		p &= m_l +m_c\, i_1 + m_c(m_c-1)\frac{(k-2)}{2}\\ 
        & \qquad \qquad \qquad+ (m_c-j) + (n-m_l-m_c)L.
        \end{aligned}
	\end{equation}
	We know that $m_c\leq\lceil (L-2)/(k-2)\rceil$, and that the smallest dipole moment for a given $m_l$ is achieved when $m_c$ saturates this bound and $i_1=2$, $j=m_c$. Plugging these values into the above equation, solving for $m_l$, and rounding up to the nearest integer gives the correct value of $m_l$. The value of $m_r$, the number of particles on the right boundary, can be determined using an identical method, following which $m_c$, $i_1$, $j$, and $i_f$ can be determined.

	\section{Proof of the particle-connected CS-to-ES algorithm}
	\label{app:ProofCStoES}
	
    In this appendix, we derive the algorithm described in \cref{sec:CStoES} for mapping a particle-connected (PC) contracted state to a PC extended state (see \cref{tab:glossary} for definitions) and vice versa. The algorithm is composed of a series of pairwise particle hopping gates which are applied to the initial state. In what follows, we denote an outward hop gate, which sends a particle on site $i+1$ and a particle on site $j-1$ to sites $i$ and $j$, respectively, by $U^+_{i,j}$, and an inward hop gate sending particles from sites $i$ and $j$ to sites $i+1$ and $j-1$ by $U^-_{i,j}$. A gate is only applied in the following algorithm if it is compatible with the constraints of on-site dimension $d$ and system length $L$. The algorithm for mapping a PC contracted state to a PC extended state for a system with onsite dimension $d$ and range-$k$ interactions is as follows.
	\begin{itemize}
    \item For site $i\in\{0,\dots, L-3\}$ in increasing order, if the sites $i$ and $i+k-1$ are within the system (i.e., are within $\{0,\dots, L-1\}$) and both have occupancy at most $d-2$ and all sites in-between them each have occupancy $d-1$, then apply the gate $U^+_{i,i+k-1}$. Repeat the loop over sites until it is no longer possible to apply this gate.
    \item For $\ell\in\{k-4,\dots, 0\}$ in decreasing order, for site $i\in\{0,\dots, L-3\}$ in increasing order, if the sites $i$ and $i+\ell+2$ are within the system and both have occupancy at most $d-2$ and all sites in-between each have occupancy $d-1$, apply the gate $U^+_{i,i+\ell+2}$ (where $U^+_{i,i+2}$ for $\ell=0$ is an outward hop of two particles on a  same site); and if the gate cannot be applied, then if the sites $i$ and $i+\ell+3$ are within the system and both have occupancy at most $d-2$ and all sites in-between each have occupancy $d-1$, apply  $U^+_{i,i+\ell+3}$ instead.
    \item For $\ell\in\{0,\dots, k-4\}$ in increasing order, for site $i\in\{0,\dots, L-3\}$ in increasing order, apply the gate $U^+_{i,i+\ell+2}$ if possible, and if not, apply the gate $U^+_{i,i+\ell+3}$ if possible.
\end{itemize}

For the case of $d=2$, since no stacking can occur, $\ell$ ranges from $k-4$ to $1$ in the second step and from $1$ to $k-4$ in the third. 	An algorithm for mapping from the PC extended state to the contracted state is obtained by replacing outward hops $U^+_{i,j}$ in the above algorithm with inward hops $U^-_{i,j}$, and by replacing on-site particle occupancies with their particle-hole conjugate (so, for example, the first step involves applying operators $U^-_{i,i+k-1}$ when sites $i$ and $i+k-1$ each contain one particle or more and all sites in-between are empty).

When the operations listed above are applied to a PC contracted state, the different steps have the following effects. If the PC contracted state contains a sequence of $k-2$ sites each with occupancy $d-1$, the first step of the algorithm shifts the position of that sequence towards the boundaries of the system until either that sequence is no longer present (which happens if some of the particles involved hop onto sites containing fewer than $d-2$ particles), or otherwise a stack of particles forms at the boundary. Likewise, the second step of the algorithm also ``unstacks'' particles by shifting them toward the boundaries until no sites (apart from particle stacks at the boundaries)  have occupancy greater than 1. The third step then expands the spacings between the remaining particles not in the boundary stacks until only a PC string remains in the bulk and hence a PC extended state is obtained overall.

    In presenting the proof, we focus on the case of $d=2$ for simplicity, and we discuss at the end how to generalise to higher $d$. We note that a PC contracted state in a $d=2$ system consists of a series of clusters of particles, where we define a cluster to be a group of occupied contiguous sites with a hole (or the system's boundary) on either side. A lone particle with no neighbouring particles also constitutes a cluster. We also note that, from the structure of PC contracted states, at most one of these clusters contains $k-2$ particles; the rest all contain $k-3$ particles, up to the possible exception of the leftmost and rightmost clusters, which may contain fewer particles (or more if there is overlap with the boundary and hence stacking of particles).

Let us label the clusters in a given PC contracted state by an index $j=1,\dots,N_C$, where $N_C$ is the initial total number of particle clusters. As the algorithm is carried out, new clusters may form to the left and right of the existing ones. When a new cluster is formed to the left, we take the range of $j$ to extend downward, so a first new leftward cluster will have $j=0$, a second new one will have $j=-1$, and so on. Likewise, when new clusters are formed to the right, the upper range of the index $j$ extends beyond $N_C$. We denote the number of particles in each cluster by $c_j$. When after a series of hops $c_j=0$, this indicates there are no particles left in that cluster (hence the holes that were originally on either side of the cluster become adjacent); we continue to label the ``empty cluster'', however, to preserve notational consistency.

Let us consider the first step in the algorithm. If there is no cluster of size $k-2$, this step does not have an impact. If there is such a cluster, then say it is at index $j$, such that $c_j=k-2$, and that the leftmost particle in the cluster is at site $i+1$. After the outward hop gate $U^+_{i,i+k-1}$ is applied to it, the following changes in cluster size occur:
\begin{align*}
    c_j&=k-4,\\
    c_{j+1}&=c_{j+1}+1,\\
    c_{j-1}&=c_{j-1}+1.
\end{align*}
Indeed, the $j$th cluster loses two particles, and the clusters to its left and right  both gain one (or new clusters are started if they are not there already). If the new value of $c_{j+1}$ satisfies $c_{j+1}<k-2$, then no further hops are applied until the algorithm loops back round to the first site. If $c_{j+1}=k-2$, on the other hand, then the outward hop operator is applied again, resulting in
\begin{align*}
    c_{j+1}&=k-4,\\
    c_{j+2}&=c_{j+2}+1,\\
    c_{j}&=k-3.
\end{align*}
Again, a further hop operator is only applied if $c_{j+2}=k-2$; these outward hops continue until an outward hop occurs in which no cluster of size $k-2$ is formed, or else a cluster of size $k-2$ (or more) is formed which overlaps with the boundary (in which case the particles in that cluster just contribute to the particle stack at the boundary). At this point, the algorithm loops back to the first site.

When the algorithm loops back to the first site, then at the start $c_{j-1}\leq k-2$, and all other clusters (up to the possible exception of those on the boundary) have size less than $k-2$. If $c_{j-1}< k-2$, then the first step of the algorithm terminates. If $c_{j-1}= k-2$, then the same series of hops is performed again, with the end result that either $c_{j-2}=k-2$ or else no cluster away from the boundaries has size $k-2$. This continues until  no clusters of size $k-2$ not on the boundary are left, which must eventually occur due to the finite size of the system.

We then consider the second step of the algorithm. For each value of $\ell$, at the beginning of the loop, $\ell+1$ is the size of the largest clusters present, and at the end of the loop, $\ell$ is the size of the largest clusters present (disregarding boundaries). To demonstrate this, we note that it is straightforward to show, following identical logic to the last step, that by the end of each loop over all sites the number of clusters of size $\ell+1$ is less than or equal to what it was before the loop over all sites; and furthermore that if it is equal, then the index of the leftmost cluster of size $\ell+1$ has decreased by 1. Hence, repeated application of the loop over all sites eventually results in the absence of any clusters of size $\ell+1$ away from the boundaries. The loop over the values of $\ell$ itself eventually results in all clusters away from the boundaries having a size of at most 1, and hence in all particles not in boundary stacks being separated by at least one (and at most 2) holes from their neighbours.

The third step of the algorithm follows identical logic in reverse. For each value of $\ell$, at the start all particles are separated by at least $\ell$ and at most $\ell+1$ holes from their neighbours, and by the end they are all separated by at least $\ell+1$ and at most $\ell+2$ holes. By the end of the algorithm, all particles are separated by at least $k-3$ and at most $k-2$ holes. However, it can readily be shown that there can only be one instance of $k-2$ holes, since if there were two or more, then following logic identical to that in step 2 of the uniqueness proof in \cref{appendix:uniqueFES}, it would be impossible via a series of solely inward hops to arrive at a PC contracted state; but it must be possible since the sequence of outward hops we applied in our algorithm can be reversed.

In the case of general finite $d$, the ``clusters'' become groups of neighbouring sites each containing $d-1$ particles. Using steps essentially identical to the ones above, one shows that the first step of the algorithm removes any cluster of $k-2$ sites, and the second step removes any clusters of even one site because it results in a state where no sites aside from boundary particle stacks have occupancy $d-1$. The third step then completes the expansion to the PC extended state. The proof of the general $d$ algorithm is almost identical to that presented above for $d=2$.

	\section{Characterising typical states for $d=2$}
	\label{app:typicalstatesd2}
	As seen in \cref{eq:generaldDN}, the logarithm of the total number of particle configurations in a family of $N=\nu L$ sectors for $d=2$ is given by
	\begin{equation}
		\label{eq:Dns2remind}
		\ln D_{N}^{(2)}(L) = \ln \binom{L}{N} = L \, \eta_2(\nu) + \mathcal{O}(\ln L) \ ,
	\end{equation}
	with $\eta_2(\nu)$ defined in \cref{eq:eta2etainf}.
	In each of these configurations, call $N_n$ the number of particles that are separated by exactly $n\ge 0$ holes from the first particle to their right (for the right-most particle in the system $n$ is the number of holes to its right up to the right-boundary), and call $f_n=N_n/N$ their fractions over the total number of particles. In this Appendix we characterise the fractions $f_n$ in typical states of $d=2$ systems at fixed fillings $0 < \nu < 1 = d-1$. 
	
	We start by noticing that counting all the configurations in an $N$ sector is equivalent to counting all the possible ways, compatible with the size $L$, of assigning right-sequences of $n$ holes to each of the $N$ particles in the system, i.e., 
	\begin{equation}
		\label{eq:NnandDN2}
		\sum_{\{N_n\}} \frac{N!}{N_0!N_1!\ldots N_{\Gamma-1}!} = D_N^{(2)}(L) \ ,
	\end{equation}
	where $\Gamma-1=L-N$ and the sum is over all sets $\{N_n \, | \, n=0,\ldots,\Gamma-1\}$ that satisfy the constraints
	\begin{equation}
		\label{eq:setNnconstraints}
		\sum_{n=0}^{\Gamma-1} N_n = N  \qquad
		\sum_{n=0}^{\Gamma-1} n N_n = L - N \ . 
	\end{equation}
	The value of $\Gamma=L-N$ has been derived from the previous constraints by setting $N_0=N-1$ and $N_{\Gamma-1}=1$. A first question is: how many different sets $\{N_n\}$ satisfy \eqref{eq:setNnconstraints}? We notice that the equations in \eqref{eq:setNnconstraints} are identical to those in \cref{eq:defNandX} for $d=\infty$, via the identifications
	\begin{equation}
		L \to \Gamma \equiv L', \quad N \to N \equiv N', \quad X \to L-N \equiv X' \, . 
	\end{equation}
	This means that the number of distinct sets $\{N_n\}$ satisfying \eqref{eq:setNnconstraints} coincides with $D_{N',X'}^{(\infty)}(L')$. Given that
	\begin{equation}
		\nu_{x}'=\frac{X'}{N'L'} = \frac{1}{\nu L} = \mathcal{O}(L^{-1}) \ ,
	\end{equation}
	$D_{N',X'}^{(\infty)}(L')$ must scale with $L$ much more slowly than $D_{N,X}^{(\infty)}(L)$ for typical $(N,X)$ sectors with $\nu_x=1/2$. However, we cannot calculate $D_{N',X'}^{(\infty)}$ by direct application of the formulas for $D_{N,X}^{(\infty)}$ discussed in Section \ref{sec:sizesymmetrysec} and Appendix \ref{appendix:scalingofsizeofss}, as those were derived under the assumption of $0<\nu_x<1$, which guarantees that $\ln D_{N,X}^{\infty} = \mathcal{O}(L)$. This suggests that the leading order of $\ln D_{N',X'}^{(\infty)}(L')$ should be $\mathcal{O}(L^\gamma)$ with $0<\gamma<1$. Calculating numerically $D_{N',X'}^{(\infty)}(L')$ as a function of a few values of $L$ we obtain $\gamma=1/2$.
	
	Given that only $O\left( \exp{c \sqrt{L}}\right)$ sets $\{N_n\}$ appear in the sum in \eqref{eq:NnandDN2}, with $c$ an $L$-independent constant, if a restricted class of these sets is associated with multinomials that exponentially dominate in $L$ over all the others, this class alone determines the fraction $f_n$ in typical states. 
	The asymptotic scaling of 
	\begin{equation}
		\label{eq:N0NGammaMulti}
		\ln \left[\frac{N!}{N_0!N_1!\ldots N_{\Gamma-1}!}\right] = L \, M_{\{N_n\}} + \mathcal{O}(\ln L)
	\end{equation}
	can be studied exactly with the same method based on Stirling's asymptotic formula and employed in Appendix \ref{appendix:scalingofsizeofss} for the multinomials 
	\begin{equation}
		\frac{L!}{e_0!e_1!\ldots e_{d-1}!} \ , 
	\end{equation}
	which were subject to the constraints in \eqref{eq:condsetsei}, which are functionally identical to \eqref{eq:setNnconstraints}. In particular, we can directly use here the results from Appendix \ref{appendix:scalingofsizeofss} via the identifications
	\begin{equation}
		\begin{aligned}
			\frac{L!}{(\varepsilon_0 L)!\ldots (\varepsilon_{d-1}L)!} \ \ &\to \ \ \frac{N!}{(f_0 N)!\ldots (f_{\widetilde{\Gamma}-1} N)!} \ , \\
			\sum_{n=0}^{d-1} \varepsilon_n = 1 \ \ &\to \ \ \sum_{n=0}^{\widetilde{\Gamma}-1} f_n = 1 \ , \\
			\sum_{n=0}^{d-1} n \varepsilon_n = \nu \ \ &\to \ \ \sum_{n=0}^{\widetilde{\Gamma}-1} n f_n = \frac{1}{\nu}-1 \ , 
		\end{aligned}
	\end{equation}
	where $\widetilde{\Gamma}<\Gamma$ is a new cutoff to be determined a posteriori by requiring that $N_n\ge1 \ \forall \ n = 0, \ldots, \widetilde{\Gamma}$. The previous inequality is necessary to justify the application of Stirling's formula, which is fairly accurate even for $N_n$ as small as $1$, but certainly not for $N_n=0$. From the results of Appendix \ref{appendix:scalingofsizeofss} we automatically deduce that the maximum of $M_{\{N_n\}}$ from \cref{eq:N0NGammaMulti} is given by
	\begin{equation}
		M_{\{f_n^* N\}} =  \nu \ln \left( \frac{1-t^{\widetilde{\Gamma}}}{1-t} \right) - (1-\nu) \ln t  \ ,
	\end{equation}
	\begin{equation}
		f_n^* = \frac{N_n^*}{N} =\frac{1-t}{1-t^{\widetilde{\Gamma}}} \, t^n \ ,
	\end{equation}
	where $0<t<1$ is the unique solution of 
	\begin{equation}
		\label{eq:eqforttildeGamma}
		\frac{1}{\nu}-1 = \widetilde{\Gamma} + \frac{t}{1-t}-\frac{\widetilde{\Gamma}}{1-t^{\widetilde{\Gamma}}} \ .
	\end{equation}
	We now assume that $\widetilde{\Gamma}$ is an increasing function of $L$ which diverges in the thermodynamic limit and we will check this at the end. This allows to determine the solution $t$ of \cref{eq:eqforttildeGamma} in terms of the expansion
	\begin{equation}
		t = 1-\nu +O\left[(1-\nu)^{\widetilde{\Gamma}}\right] \ , 
	\end{equation}
	which is consistent with $0<t<1$. From this we obtain
	\begin{equation}
		M_{\{f_n^* N\}} = \eta_2(\nu) +O\left[(1-\nu)^{\widetilde{\Gamma}}\right]  \ ,
	\end{equation}
	\begin{equation}
		\label{eq:fnstarfinal}
		f_n^* = \nu (1-\nu)^n +O\left[(1-\nu)^{\widetilde{\Gamma}}\right] \ .
	\end{equation}
	The fact that the leading order of the maximum $L\, M_{\{f_n^* N\}}$ coincides with the leading order of $\ln D_{N}^{(2)}(L)$, see \cref{eq:Dns2remind}, was expected from \cref{eq:NnandDN2}. To determine $\widetilde{\Gamma}$ we now impose that $N_{\widetilde{\Gamma}-1}^*= 1$, so to ensure $N_n^* \ge 1 \ \forall \ n=0, \ldots, \widetilde{\Gamma}-1$. This implies
	\begin{equation}
		\frac{1}{N} = f_{\widetilde{\Gamma}-1}^* = O\left[(1-\nu)^{\widetilde{\Gamma}}\right] \quad \to \quad \widetilde{\Gamma} = \mathcal{O}(\ln N) \ .
	\end{equation}
	This proves that $\widetilde{\Gamma}$ is an increasing function of $L$ which diverges in the thermodynamic limit, consistent with our previous assumption. Note that the typicality class which includes an infinitesimal neighborhood around the maxima $f_n^*$ contains a fraction of all configurations that tends to $1$ exponentially fast with $L$. \\
	
	The previous results can be interpreted in a simple way through the lens of the central limit theorem. Imagine generating a random configuration of particles and holes in the chain of size $L$ by drawing a particle on each site with probability $\nu$ and a hole with probability $1-\nu$. For large $L$ the central limit theorem guarantees that, in any state generated in this way, the global filling is very close to $\nu$, with fluctuations around this value of order $\mathcal{O}(1/\sqrt{L})$. This means that for asymptotically large $L$ this class of randomly generated states reflects the average properties over the set of $D_{\nu L}^{(2)}(L)$ states with filling $\nu$ on a chain of size $L$. 
	We can then ask what the expectation value is of $N_n$ in the randomly generated set of states. This is given by the probability of having (after any particle) $n$ consecutive holes followed by another particle, times the total number of particles in the system
	\begin{equation}
		\langle N_n \rangle = \langle N \rangle (1-\nu)^n \nu \ = L \nu^2 (1-\nu)^n \ .
	\end{equation}
	The previous result coincides with \cref{eq:fnstarfinal}, as expected from the fact that our typicality class dominates in any average over the entire set of states. Exploiting this equivalence between average properties over sets of randomly drawn states and properties of single states in a typicality class, we can generalise our previous results to the following statement: \\
	
	In typical states of $d=2$ systems with global filling $\nu$ there are 
	\begin{equation}
		L \nu^p (1-\nu)^h + o(L) 
	\end{equation}
	occurrences of $p+h$ consecutive sites hosting $p$ particles and $h$ holes arranged in a fixed chosen configuration. 
	
	\section{Deriving $P_{\text{b}}(\nu)$ for $d=2,k=4$}
	\label{app:D2K4pnh}
	We aim to derive an exact expression for the function $P_{\text{b}}(\nu)$ introduced in \cref{sec:AB_D2K4}, which for $k=4$ and $d=2$ gives the probability that, starting from a given site (and assuming the preceding 2 sites are part of an active bubble), a pattern of particles and holes will arise such that the first three sites in the pattern are frozen sites. In the following we implicitly make use of the conclusions drawn in \cref{sec:AbsoluteBlockages}, e.g.~that models with $k=4$ and $d=2$ have extremely restricted particle mobility, and that they are strongly fragmented irrespective of the filling $\nu$.
	
	We begin by noting that the first three sites in the pattern must either all be holes or particles. This is because the preceding 2 sites are part of an active bubble, and so they must consist of one particle and one hole, with the respective positions of the particle and hole swapping as the system evolves with time. Hence, if the first three sites of the pattern contained both particles and holes, it is easy to see this would allow a hopping move to occur via interaction with a particle in the active bubble, and so the 3 sites would not all be frozen.
	
	In view of this result, we proceed to decompose our function as
	\begin{equation*}
		P_{\text{b}}(\nu) = (1-\nu)^3 p_{\text{b}}(\nu) + \nu^3 p_{\text{b}}(1-\nu),
	\end{equation*}
	where $p_{\text{b}}(\nu)$ is the probability that starting from a given site, a pattern will arise such that the first site in the pattern is either empty or, if it contains a particle, the particle will be unable to perform an outward hop. Hence, if the three sites next to the active bubble are holes, $p_{\text{b}}(\nu)$ is the probability that the following sites will be configured such that those 3 holes are frozen; and likewise, if the three sites are particles, then $p_{\text{b}}(1-\nu)$ gives the probability that those three particles will be frozen.
	
	To make the computation of $p_{\text{b}}(\nu)$ more intuitive to follow, instead of expressing it as a sum over powers of $\nu$ and $(1-\nu)$, we express it as a sum over sequences of ``0'' (indicating a hole) and ``1'' (indicating a particle), with the rule $0\rightarrow(1-\nu)$ and $1\rightarrow\nu$. By definition, we immediately have that
	\begin{equation}
		p_{\text{b}}(\nu) = 0 + 1\dots,
	\end{equation}
	where by $\dots$ we mean that further terms need to be specified in the sequence to ensure that the starting particle in it cannot hop outward. If the first 1 is followed by a 0 and then another 1, we find ourselves in the same situation: the first particle in the sequence cannot hop outward only if the newly added particle cannot hop outward. This pattern repeats itself if we add on indefinitely many further terms of the form $01$, and so
	\begin{equation}
		p_{\text{b}}(\nu) = 0 + 1\left(\sum_{i=0}^{\infty}(01)^i\right)\dots
	\end{equation}
	The next term we could add to the sequence is either a 00 or a 11 (since a 10 would allow the first particle in the sequence to perform an outward hop). A 00 already achieves the goal of preventing an outward hop. A 11 on the other hand requires that more of the sequence be specified to ensure an outward hop cannot happen. Hence we have
	\begin{equation}
		p_{\text{b}}(\nu) = 0 + 1\left(\sum_{i=0}^{\infty}(01)^i\right)[00+11\dots].
	\end{equation}
	If we have a 11 term, we must ensure that the second of these two particles cannot perform an inward hop with the remaining particles in the sequence. If we add a 01 to the sequence, then we have the same situation: we must ensure that the latest particle added to the sequence cannot perform an inward hop. Thus we have
	\begin{equation}
		p_{\text{b}}(\nu) = 0 + 1\left(\sum_{i=0}^{\infty}(01)^i\right)\left[00+11\left(\sum_{i=0}^{\infty}(01)^i\right)\dots\right].
	\end{equation}
	If we add on a further 1 after this, then the inward hop cannot occur and we are done. If we add on a 00, then we must add on a third 0, as a 001 would allow an inward hop with the before-last particle. Thus 
	\begin{equation}
    \begin{aligned}
		p_{\text{b}}(\nu) =& 0 + 1\left(\sum_{i=0}^{\infty}(01)^i\right)\\
        & \times\left[00+11\left(\sum_{i=0}^{\infty}(01)^i\right)[1+000\dots]\right].
    \end{aligned}
	\end{equation}
	For the final term, we note that our aim is achieved only if the last sequence 000 is followed by a fourth hole 0, or else by a particle 1 that cannot perform an outward hop. We therefore have
	\begin{equation}
    \begin{aligned}
		p_{\text{b}}(\nu) =& 0 + 1\left(\sum_{i=0}^{\infty}(01)^i\right)\\
        &\times \left[00+11\left(\sum_{i=0}^{\infty}(01)^i\right)[1+000p_{b}(\nu)]\right].
    \end{aligned}
	\end{equation}
	Substituting $(1-\nu)$ for 0 and $\nu$ for 1 in the above equation and solving it, we obtain our desired result:
	\begin{equation}
		p_{\text{b}}(\nu)=\frac{1 - 2 \nu + 2 \nu^2 - \nu^3 + \nu^4}{    1 - 2 \nu + 3 \nu^2 - 3 \nu^3 + 4 \nu^4 - 3 \nu^5 + \nu^6} \ .
	\end{equation}
	
\bibliography{apssamp}
	
\end{document}